\documentclass[10pt]{article}%
\usepackage{amsmath}
\usepackage{graphicx}%
\usepackage{amsfonts}%
\usepackage{amssymb}
\usepackage{theorem}
\theorembodyfont{\upshape}
\newtheorem{theorem}{Theorem}[section]

\newtheorem{condition}[theorem]{Condition}

\newtheorem{corollary}[theorem]{Corollary}

\newtheorem{definition}[theorem]{Definition}

\newtheorem{proposition}[theorem]{Proposition}
\newtheorem{remark}[theorem]{Remark}

\newenvironment{proof}[1][Proof]{\textbf{#1.} }{\ \rule{0.5em}{0.5em}}
\numberwithin{equation}{section}

\begin{document}

\title{Spectral Theory of Time Dispersive and Dissipative Systems}
\author{A. Figotin\\Irvine, CA
\and J. H. Schenker\\ETH, Zurich}
\date{28 April, 2004, revised 27 August, 2004, corrected 12 December, 2004}
\renewcommand{\thefootnote}{}
\maketitle 
\renewcommand{\thefootnote}{\arabic{footnote}}

\begin{abstract}
We study linear time dispersive and dissipative systems. Very
often such systems are not conservative and the standard spectral
theory can not be applied. We develop a mathematically consistent
framework allowing (i) to constructively determine if a given time
dispersive system can be extended to a conservative one; (ii) to
construct that very conservative system -- which we show is
essentially unique. We illustrate the method by applying it to the
spectral analysis of time dispersive dielectrics and the damped
oscillator with retarded friction. In particular, we obtain a
conservative extension of the Maxwell equations which is
equivalent to the original Maxwell equations for a dispersive and
lossy dielectric medium.
\end{abstract}
\tableofcontents

\section{Introduction}

In this paper we describe a mathematical framework for a spectral theory of
linear time dispersive and dissipative (lossy) media, e.g., dielectric media.
Here is a concise formulation of the setup. We consider a linear system
(medium) whose state is described by a time dependent \emph{generalized
velocity} $v\left(  t\right)  $ taking values in a Hilbert space $H_{0}$ with
scalar product $\left(  \cdot,\cdot\right)  $. The evolution of $v$ is
governed by a linear equation incorporating \emph{retarded friction}%
\begin{equation}
m\partial_{t}v\left(  t\right)  =-\mathrm{i}Av\left(  t\right)  -\int
_{0}^{\infty}a\left(  \tau\right)  v\left(  t-\tau\right)  \,d\tau+f\left(
t\right)  ,\ \label{ax3}%
\end{equation}
where $m>0$ is a positive mass operator in $H_{0}$, $A$ is a self-adjoint
operator in $H_{0}$, $f\left(  t\right)  $ is a time dependent external
\emph{generalized force}, and $a\left(  t\right)  $, $t\geq0$, is an operator
valued function which we call the \emph{operator valued friction retardation
function} \cite[Section 1.6]{KuboToda2}, or just \emph{friction function}. The
names ``generalized velocity''\ and ``generalized force''\ are justified when
we interpret the real part of the scalar product $\operatorname{Re}\left\{
\left(  v\left(  t\right)  ,f\left(  t\right)  \right)  \right\}  $ as the
work done by $f\left(  t\right)  $ per unit of time at instant $t$. Hence, the
total work $\mathcal{W}$ done by $f\left(  t\right)  $ is%
\begin{equation}
\mathcal{W}=\int_{-\infty}^{\infty}\operatorname{Re}\left\{  \left(  v\left(
t\right)  ,f\left(  t\right)  \right)  \right\}  \,dt.\label{work1}%
\end{equation}
By the same token, the first two terms on the right hand side of (\ref{ax3}),
namely%
\begin{equation}
-\mathrm{i}Av\left(  t\right)  -\int_{0}^{\infty}a\left(  \tau\right)
v\left(  t-\tau\right)  \,d\tau,
\end{equation}
are interpreted as the force which $v$ exerts on itself at time $t$. This self
forcing depends on $v$ through two terms: (i) the instantaneous term
$-\mathrm{i}Av\left(  t\right)  $ and (ii) the ``time dispersive''\ integral
term $\int_{0}^{\infty}a\left(  \tau\right)  v\left(  t-\tau\right)  \,d\tau$
involving values $v\left(  t^{\prime}\right)  $ for $t^{\prime}<t$. \ The
integral term is interpreted as a (retarded) friction force whose special form
reflects two fundamental requirements: (i) time homogeneity and (ii)
causality. \

If we rescale the variables according to the formulas%
\begin{equation}
\tilde{v}=\sqrt{m}v,\ \Omega=\frac{1}{\sqrt{m}}A\frac{1}{\sqrt{m}%
},\ \tilde{a}=\frac{1}{\sqrt{m}}a\frac{1}{\sqrt{m}},\ \tilde{f}=\frac{1}%
{\sqrt{m}}f, \label{vax1}%
\end{equation}
then the equation (\ref{ax3}) reduces to the special form with $m$ the
identity operator, i.e.%
\begin{equation}
\partial_{t}\tilde{v}\left(  t\right)  =-\mathrm{i}\Omega\tilde{v}\left(
t\right)  -\int_{0}^{\infty}\tilde{a}\left(  \tau\right)  \tilde{v}\left(
t-\tau\right)  \,d\tau+\tilde{f}\left(  t\right)  . \label{vax2}%
\end{equation}
We refer to $\Omega=\frac{1}{\sqrt{m}}A\frac{1}{\sqrt{m}}$ as the frequency
operator in $H_{0}$ since its spectrum gives the resonant frequencies of the system.

We call the system conservative, or non-dispersive, if the friction function
vanishes, $\ $i.e.,
\begin{equation}
m\partial_{t}v\left(  t\right)  =-\mathrm{i}Av\left(  t\right)  +f\left(
t\right)  . \label{ax1}%
\end{equation}
In this case, it is natural to define the internal energy of the system at
time $t$ to be $\frac{1}{2}\left(  v(t),mv(t)\right)  $, since it follows from
(\ref{ax3}) that
\begin{equation}
\frac{d}{dt}\frac{1}{2}\left(  v(t),mv(t)\right)  =\operatorname{Re}\left\{
\left(  v\left(  t\right)  ,f\left(  t\right)  \right)  \right\}  ,
\label{energy1}%
\end{equation}
where $\operatorname{Re}\left\{  \left(  v\left(  t\right)  ,f\left( t\right)
\right)  \right\}  $ is the instantaneous rate of work (by assumption). In the
non-conservative (dispersive) case, we continue to interpret $\frac{1}{2}\left(
v(t),mv(t)\right)  $ as the internal energy but
note that (\ref{energy1}) is not true in general. Instead,%
\begin{multline}
\frac{d}{dt}\frac{1}{2}\left(  v(t),mv(t)\right) \\
= \operatorname{Re}\left\{ \left(  v\left(  t\right)  ,f\left(
t\right)  \right)  \right\} -\operatorname{Re}\left\{  \left(
v\left(  t\right)  ,\int_{0}^{\infty }a\left(  \tau\right) v\left(
t-\tau\right)  \,d\tau\right)  \right\}  .
\label{avm1}%
\end{multline}
We interpret the second term of this expression as the instantaneous rate of
\textquotedblleft work done by the system on itself,\textquotedblright\ or
more properly the negative rate of energy dissipation due to friction.

In all physical models of which we are aware the time dispersive term $a$
arises as a phenomenological description of a linear coupling between the
system and some other ``hidden''\ degrees of freedom. Our belief in the
conservation of energy suggests an equivalent description with the hidden
system described by a vector $w$ in a (different) Hilbert space $H_{1}$, such
that the extension%
\begin{equation}
V(t)=\left[
\begin{array}
[c]{c}%
v(t)\\
w(t)
\end{array}
\right]  \label{avm2}%
\end{equation}
evolves according to a conservative equation like (\ref{ax1}).

The main result of this paper is that under a physically reasonable hypothesis
it is possible to construct a conservative extension of the original system
(\ref{ax3}) in the following form
\begin{align}
m\partial_{t}v\left(  t\right)   &  =-\mathrm{i}Av\left(  t\right)
-\mathrm{i}\Gamma w\left(  t\right)  +f\left(  t\right)  ,\ m>0,\ A\text{ is
self-adjoint,}\label{avm3}\\
m_{1}\partial_{t}w\left(  t\right)   &  =-\mathrm{i}\Gamma^{\dagger}v\left(
t\right)  -\mathrm{i}A_{1}w\left(  t\right)  ,\ m_{1}>0,\ A_{1}\text{ is
self-adjoint,} \label{avm4}%
\end{align}
where $w\in H_{1}$, the Hilbert space of ``hidden'' variables, $m_{1}$ and
$A_{1}$ describe respectively the mass operator and the generator of internal
dynamics on $H_{1}$, and $\Gamma:H_{1}\rightarrow H_{0}$ is a coupling
operator between the ``hidden'' and ``observable'' variables. While it may be
conceptually clear that such extensions \emph{should }exist for reasonable
linear models,\ it is not immediately obvious how to determine from the given
equation (\ref{ax3}) if such an extension \emph{does} exist. Nonetheless, we
show there is a natural intrinsic condition -- non-positivity of the total
work done by the friction force -- which is both necessary and sufficient for
such an extension. Furthermore, we shall see that the Hilbert space $H_{1}$ of
``hidden variables'' as well as the operators $\Omega_{1}=$ $\frac{1}%
{\sqrt{m_{1}}}A_{1}\frac{1}{\sqrt{m_{1}}}$ and $\Gamma$ in (\ref{avm3}%
)-(\ref{avm4}) are essentially uniquely determined by the friction function.

It would be interesting and natural to study (\ref{avm3}), (\ref{avm4}), and
thus (\ref{ax3}),  with a random force $f(t)$ as a fluctuation-dissipation
model similar to the Langevin equation \cite{KuboToda2}. However, the
importance of the subject and the efforts needed to conduct such a study are
worthy of a separate publication.

For a homogeneous dielectric medium described by a scalar-valued frequency
dependent electric susceptibility $\hat{\chi}\left( \omega\right)  $, and
magnetic permeability $\mu=1$, the conservative extension takes the form (in
common notations)%
\begin{equation}\label{avm5}
\begin{split}
  \partial_{t}\mathbf{H}\left(  \mathbf{r},t\right)   &  =-\nabla\times
\mathbf{E}\left(  \mathbf{r},t\right)  ,\\
\partial_{t}\mathbf{E}\left(  \mathbf{r},t\right)   &  =\nabla\times
\mathbf{H}\left(  \mathbf{r},t\right) \\
& \quad -\int_{-\infty}^{\infty}\sqrt
{m_{\hat{\chi}}n_{\hat{\chi}}\left( \sigma\right)
}\mathbf{\Psi}\left(
\mathbf{r},t,\sigma\right)  \,d\sigma-4\pi\mathbf{J}\left(  \mathbf{r}%
,t\right)  , \\
m_{\hat{\chi}}\partial_{t}\mathbf{\Psi}\left(
\mathbf{r},t,\sigma\right)   &
=-\mathrm{i}m_{\hat{\chi}}\sigma\mathbf{\Psi}\left(
\mathbf{r},t,\sigma
\right)  +\sqrt{m_{\hat{\chi}}n_{\hat{\chi}}\left(  \sigma\right)  }%
\mathbf{E}\left(  \mathbf{r},t\right)  ,
\end{split}
\end{equation}
where the field $\mathbf{\Psi}\left(  \mathbf{r},t,\sigma\right)  $ describes
\textquotedblleft hidden\textquotedblright\ variables, which one may view as a
\textquotedblleft string of dipoles\textquotedblright\ with string coordinate
$\sigma\in\mathbb{R}$ attached at every space point $\mathbf{r}$. The
parameters $n_{\hat{\chi}}\left(  \sigma\right)  $ and $m_{\hat{\chi}}$ are
related to the electric susceptibility $\hat{\chi}\left(  \omega\right)  $ as
follows
\begin{equation}\label{avm6}
    \begin{split}
      n_{\hat{\chi}}\left(  \sigma\right)  =4\operatorname{Im}\left\{  \sigma
\hat{\chi}\left(  \sigma\right)  \right\}  \geq0,\ m_{\hat{\chi}}^{-1}%
=\int_{-\infty}^{\infty}n_{\hat{\chi}}\left(  \sigma\right)
\,d\sigma
,\\
4\pi\omega\hat{\chi}\left(  \omega\right)  =\lim_{\eta\rightarrow+0}%
\int_{-\infty}^{\infty}\frac{n_{\hat{\chi}}\left(  \sigma\right)  }%
{\sigma-\omega-\mathrm{i}\eta}\,d\sigma,\
\operatorname{Im}\zeta>0.
    \end{split}
\end{equation}
In the second equation of (\ref{avm5}), the term $\int_{-\infty}^{\infty}%
\sqrt{m_{\hat{\chi}}n_{\hat{\chi}}\left(  \sigma\right)  }\mathbf{\Psi}\left(
\mathbf{r},t,\sigma\right)  \,d\sigma$ is $4\pi\times$ the displacement
current, and thus the electric polarization is related to $\mathbf{\Psi}$ by
\begin{equation}
\mathbf{P}\left(  \mathbf{r},t\right)  =\frac{1}{4\pi}\int_{-\infty}^{t}%
\int_{-\infty}^{\infty}\sqrt{m_{\hat{\chi}}n_{\hat{\chi}}\left(
\sigma\right)  }\mathbf{\Psi}\left(  \mathbf{r},\tau,\sigma\right)  \,d\sigma
d\tau\mathbf{,} \label{avm8}%
\end{equation}
assuming $\mathbf{P}\equiv0$ at $t=-\infty$. In addition, the fields
$\mathbf{H}\left(  \mathbf{r},t\right)  $, $\mathbf{D}\left(  \mathbf{r}%
,t\right)  =\mathbf{E}\left(  \mathbf{r},t\right)  +\mathbf{P}\left(
\mathbf{r},t\right)  $ are required to be divergence free, i.e.%
\begin{multline}\label{avm7}
      \nabla\cdot\mathbf{H}\left(  \mathbf{r},t\right)  =0, \\
\nabla\cdot \mathbf{E}\left(  \mathbf{r},t\right)
+\int_{-\infty}^{t}\int_{-\infty
}^{\infty}\sqrt{m_{\hat{\chi}}n_{\hat{\chi}}\left(  \sigma\right)  }%
\nabla\cdot\mathbf{\Psi}\left(  \mathbf{r},\tau,\sigma\right)
\,d\sigma
d\tau=0. %
    \end{multline}
A system of extended Maxwell equations similar to (\ref{avm5}) were proposed in
\cite{Tip}.

In many physical models there is an instantaneous contribution to the
friction. Thus we assume throughout that $a\left(  t\right)  $, $t\geq0$ is of
the form
\begin{equation}
a\left(  t\right)  =\alpha_{\infty}\delta\left(  t\right)  +\alpha\left(
t\right)  \text{, where }\alpha\left(  t\right)  \text{ is continuous for
}t\geq0, \label{ax3a}%
\end{equation}
with $\alpha_{\infty}$ self-adjoint (any anti self-adjoint piece can be
incorporated in $-\mathrm{i}A$). The representation (\ref{ax3a}) explicitly
distinguishes the instantaneous component $\alpha_{\infty}\delta\left( t\right)
$ of the\ friction function from the retarded component described by
$\alpha\left(  t\right)  $. It is possible to extend some of the results below
to $\alpha(t)$ which are operator valued measures or distributions, but to
simplify the exposition such extensions -- which are physically somewhat
esoteric anyway -- will not be considered. \ On the other hand, the
instantaneous component $\alpha_{\infty}\delta\left(  t\right)  $ is completely
natural and occurs in many examples.

To state the central condition of this paper, let us consider the total work
$\mathcal{W}_{\operatorname*{fr}}$ done by the friction force, which may be
written as follows
\begin{equation}
\mathcal{W}_{\operatorname*{fr}}=-\frac{1}{2}\int_{-\infty}^{\infty}%
\int_{-\infty}^{\infty}\left(  v\left(  t\right)  ,a_{e}\left(  t-\tau\right)
v\left(  \tau\right)  \right)  \,dtd\tau, \label{ava1}%
\end{equation}
where%
\begin{equation}
a_{e}\left(  t\right)  =2\alpha_{\infty}\delta\left(  t\right)  +\left\{
\begin{array}
[c]{ccc}%
\alpha\left(  t\right)  & \text{if} & t>0\\
\operatorname{Re}\left\{  \alpha\left(  +0\right)  \right\}  & \text{if} &
t=0\\
\alpha^{\dagger}\left(  -t\right)  & \text{if} & t<0
\end{array}
\right.  \text{ , }-\infty<t<\infty. \label{ava2}%
\end{equation}
Notice that $a_{e}\left(  -t\right)  =a_{e}^{\dagger}\left(  t\right)  $. The
\emph{precise condition which distinguishes systems with conservative
extensions} \emph{is that }$\mathcal{W}_{\operatorname*{fr}}$\emph{ should be
non-positive}, that is,
\begin{equation}
\int_{-\infty}^{\infty}\int_{-\infty}^{\infty}\left(  v\left(  t\right)
,a_{e}\left(  t-\tau\right)  v\left(  \tau\right)  \right)  \,dtd\tau
\geq0,\text{ } \label{ava4}%
\end{equation}
for every function $v(t)$, signifying the ultimate conversion of mechanical
energy into heat -- the transport of energy from relevant (observable) degrees
of freedom to hidden ones. We refer to (\ref{ava4}) as the \emph{power
dissipation condition}.

Scalar functions $a_{e}\left(  t\right)  $ which satisfy (\ref{ava4}), known as
\emph{positive definite} functions, are familiar from Bochner's theorem on the
Fourier transform of a finite positive measure. \ In fact, one construction of
a conservative extension to (\ref{ax3}) is based on Theorem
\ref{tOperatorBoch}, an operator valued generalization of Bochner's theorem.

It is often useful to work in the frequency domain, and the power dissipation
condition (\ref{ava4}) may be formulated there as well. To do so we formally
apply the Fourier transform defined by%
\begin{equation}
v\left(  t\right)  =\frac{1}{2\pi}\int_{-\infty}^{\infty}\mathrm{e}%
^{-\mathrm{i}\omega t}\hat{v}\left(  \omega\right)  d\omega,\ \hat{v}\left(
\omega\right)  =\int_{-\infty}^{\infty}\mathrm{e}^{\mathrm{i}\omega t}v\left(
t\right)  \,dt \label{ax1b}%
\end{equation}
to recast the time evolution equation (\ref{ax3}) as%
\begin{equation}
\omega m\hat{v}\left(  \omega\right)  =\left[  A-\mathrm{i}\hat{a}\left(
\omega\right)  \right]  \hat{v}\left(  \omega\right)  +\mathrm{i}\hat
{f}\left(  \omega\right)  , \label{ax4}%
\end{equation}
Here%
\begin{equation}
\hat{a}\left(  \omega\right)  =\alpha_{\infty}+\hat{\alpha}\left(
\omega\right)  ,\ \hat{\alpha}\left(  \omega\right)  =\int_{0}^{\infty}%
\alpha\left(  \tau\right)  \mathrm{e}^{\mathrm{i}\omega\tau}\,d\tau
\label{ava4b}%
\end{equation}
is a formal object, and may not be defined pointwise since we do not assume
integrability of $\alpha(\tau)$. However, if it \emph{is} defined pointwise,
we see -- in view of Bochner's theorem -- that the power dissipation condition
(\ref{ava4}) becomes%
\begin{equation}
\hat{a}_{e}\left(  \omega\right)  \geq0\text{ for all }\omega, \label{ava4a}%
\end{equation}
where%
\begin{equation}
\hat{a}_{e}\left(  \omega\right)  =2\operatorname{Re}\left\{  \hat{a}\left(
\omega\right)  \right\}  =\hat{a}\left(  \omega\right)  +\hat{a}^{\dagger
}\left(  \omega\right)  . \label{ava3}%
\end{equation}
That is, $\hat{a}_{e}\left(  \omega\right)  $ is a positive semi-definite
operator,
\begin{equation}
\left(  v,\hat{a}_{e}\left(  \omega\right)  v\right)  \geq0\text{ for all
}v\in\mathcal{H}.
\end{equation}
We refer to (\ref{ava4a}) also as the power dissipation condition. If the
Fourier transform is not defined pointwise the situation is a bit more
delicate, but in effect $\hat{a}_{e}\left(  \omega\right)  d\omega$ is a
non-negative operator valued measure (see Theorem \ref{tOperatorNev2}.)

In many examples $a(t)$ is Hermitian for every $t$. In those cases the real and
imaginary parts of $\hat{a}\left(  \omega\right)  $,
\begin{equation}
\operatorname{Re}\left\{  \hat{a}\left(  \omega\right)  \right\}  =\frac{1}%
{2}\left[  \hat{a}\left(  \omega\right)  +\hat{a}^{\dagger}\left(
\omega\right)  \right]  ,\text{ }\operatorname{Im}\left\{  \hat{a}\left(
\omega\right)  \right\}  =\frac{1}{2\mathrm{i}}\left[  \hat{a}\left(
\omega\right)  -\hat{a}^{\dagger}\left(  \omega\right)  \right]  ,
\end{equation}
are respectively even and odd functions of $\omega$ given by the $\cos$ and
$\sin$ transforms of $a$,%
\begin{equation}
\begin{split}
  \operatorname{Re}\left\{  \hat{a}\left(  \omega\right)  \right\}
=\alpha_{\infty}+\int_{0}^{\infty}\alpha\left(  \tau\right)
\cos\left( \omega\tau\right)  \,d\tau,\\ \operatorname{Im}\left\{
\hat{a}\left( \omega\right)  \right\}
=\int_{0}^{\infty}\alpha\left(  \tau\right) \sin\left(
\omega\tau\right)  \,d\tau.
\end{split}
\end{equation}

In the important special case of a conservative system (\ref{ax1}), equation
(\ref{ax4}) reduces to%
\begin{equation}
\omega m\hat{v}\left(  \omega\right)  =A\hat{v}\left(  \omega\right)
+\mathrm{i}\hat{f}\left(  \omega\right)  . \label{ax1a}%
\end{equation}
When the external force vanishes ($f=0$), a time harmonic solution $v\left(
t\right)  =V_{\omega}\mathrm{e}^{-\mathrm{i}\omega t}$ with $\hat{v}\left(
\omega^{\prime}\right)  =$ $V_{\omega}\delta(\omega-\omega^{\prime})$ is
obtained for $V_{\omega}$ which solves the spectral problem%
\begin{equation}
\omega mV_{\omega}=AV_{\omega}. \label{ax2}%
\end{equation}
This eigenvalue problem is part of the standard spectral theory of
self-adjoint operators, and its analysis is instrumental to the study of the
non-dispersive evolution (\ref{ax1}). In particular, we remind the reader that
the eigenvectors $V_{\omega}$ -- which may lie in a proper extension of the
Hilbert space -- are $m$-orthogonal for different $\omega$ and form a basis of
the Hilbert space $\mathcal{H}$.

In contrast, a time harmonic solution $v\left(  t\right)  =v_{\omega
}\mathrm{e}^{-\mathrm{i}\omega t}$ to (\ref{ax3}) satisfies%
\begin{equation}
\omega mv_{\omega}=\hat{A}\left(  \omega\right)  v_{\omega},\text{ with
}\hat{A}\left(  \omega\right)  =A-\mathrm{i}\hat{a}\left(  \omega\right)  .
\label{ax4a}%
\end{equation}
In non-trivial examples, the operator $\hat{A}\left( \omega\right)  $ is
typically non self-adjoint. Consequently the spectral theory for $\hat{A}\left(
\omega\right)  $ may be rather complicated, even in the simplest examples, with
a finite dimensional Hilbert space, $m$ the identity matrix, and $\hat{A}\left(
\omega\right)  $ a finite square matrix. For instance, $\hat{A}\left(
\omega\right)  $ may not be diagonalizable as it may have nontrivial blocks in
Jordan form. Consequently, the genuine eigenvectors of $\hat{A}\left(
\omega\right)  $ may not form a basis. In many problems of interest, including
continuum dielectric media, the Hilbert space is infinite dimensional and the
operator $\hat{A}\left(  \omega\right)  $ is unbounded, in addition to being
non self-adjoint. Thus, in general it seems to be very difficult to analyze the
eigenvalue problem (\ref{ax4a}). There are some results on the completeness of
so called root vectors for dissipative linear operators, \cite[Chapter
V]{GoKr}, but the conditions of those statements -- for instance, compactness
of the imaginary part of the operator -- are too restrictive and are often not
satisfied in problems of interest.

Compounding these difficulties, even if the complicated spectral analysis of
non-self-adjoint $\hat{A}\left(  \omega\right)  $ were somehow addressed for
fixed $\omega$, another problem would arise since $\hat{A}\left( \omega\right)
$ depends on the spectral parameter $\omega$. Thus there is no obvious relation
between solutions to (\ref{ax4a}) for different values of $\omega$. In
particular, the vectors $v_{\omega}$ need not be orthogonal and it is not clear
that they span the Hilbert space.

However, for physically meaningful examples $\hat{A}\left(  \omega\right)  $
is not an arbitrary $\omega$ dependent non self-adjoint operator, but rather
one with certain properties which allow some kind of spectral analysis. The
relevant properties, as we will see below, are analyticity and a dissipation
condition for complex $\omega$ in the upper half-plane.

The focus of this article is the construction of a spectral theory for a wide
class of dispersive and dissipative systems -- described by (\ref{ax3}),
(\ref{ax4}), or (\ref{ax2b}) below -- under the assumption of a suitable power
dissipation condition. As indicated above, our approach is based on the
observation that dispersion and dissipation are caused in most physical models
by coupling with degrees of freedom which are ``hidden'' from observation.
Consequently, we consider a spectral theory for a dispersive/dissipative
system to be a realization of the system as a proper projection of a
conservative extension to which the standard spectral theory may be applied.
As we show, \emph{the power dissipation condition is necessary and sufficient
for such a conservative extension and determines the (minimal) conservative
extension uniquely.}

The paper is organized as follows.\ In section 2 we recall the analysis of a
general conservative system of the form (\ref{avm3}), (\ref{avm4}), which we
take as an abstract model for a system with ``hidden variables.'' In particular
we show that positive power dissipation holds for any system with a
conservative extension. We also describe the \emph{admittance operator
}reformulation of (\ref{ax3}) and (\ref{avm3}), (\ref{avm4}). In section 3, we
discuss the main mathematical results which demonstrate that positive power
dissipation is \emph{equivalent} to the existence of a conservative extension.
There are two approaches to constructing a conservative extension, proceeding
by either the time (\ref{ax3}) or the frequency (\ref{ax4}) representations and
based on the classical Bochner's Theorem \ref{tBoch} and Herglotz-Nevanlinna
Theorems \ref{tNev2}, \ref{tNev1} respectively. In Section 4, we summarize in
concise form the schemes by which one may construct a conservative extention\
of a given dispersive/dissipative system including (i) the space of ``hidden''
variables; (ii) a self-adjoint operator describing the internal dynamics of the
``hidden'' variables; (iii) an operator coupling the ``hidden'' variables with
the ``observable'' variables. In Section 5, we give examples of conservative
extensions for general scalar dispersive systems and homogeneous isotropic
dielectrics, obtaining an extended conservative system which is equivalent to
Maxwell's equations for a dispersive and lossy dielectric medium.\ In
forthcoming work we shall describe in greater detail the application of the
techniques presented here to dispersion in dielectric media. In section 6, we
discuss dissipation, or loss of energy, a phenomenon which can arise in systems
with dispersion. We present sufficient conditions on $a\left(  t\right)  $ for
solutions $v\left(  t\right)  $ to (\ref{ax3}) to exhibit dissipation, that is
$\lim_{t\rightarrow\infty}\left\|  v\left( t\right)  \right\|  =0$. Finally,
Section 7 is devoted to proofs of results in Section 3 and related
constructions.

\section{Modelling hidden degrees of freedom}

In this section we introduce and analyze abstract models for hidden degrees of
freedom, showing in particular that the power dissipation condition holds for
the truncation of any conservative system. In addition, we recall the
equivalent description of a linear system in terms of its admittance operator
and discuss the admittance for truncations.

\subsection{Abstract model: preliminary analysis}

Consider a conservative system with degrees of freedom (variables) divided
into two classes: the \emph{observable variables}, denoted $v$, and the
\emph{hidden variables}, denoted $w$. For instance, $v$ might describe the
polarization of a dielectric medium at different space points, while
components of $w$ account for microscopic degrees of freedom which give rise
to the material relations. We assume that $v$ and $w$ take values respectively
in Hilbert spaces $H_{0}$ and $H_{1}$, and that the combination%
\begin{equation}
V=\left[
\begin{array}
[c]{c}%
v\\
w
\end{array}
\right]  \in\mathcal{H}=H_{0}\oplus H_{1},\ v\in H_{0},\ w\in H_{1},
\label{xfa1}%
\end{equation}
describes a \emph{conservative} system governed by the following evolution
equation%
\begin{equation}
\mathcal{M}\partial_{t}V\left(  t\right)  =-\mathrm{i}\mathcal{A}V\left(
t\right)  +F\left(  t\right)  , \label{xfa2}%
\end{equation}
where $\mathcal{M}>0$ and $\mathcal{A}$ are self-adjoint operators in
$\mathcal{H}$. Furthermore we assume -- and this is key -- that the external
force $F\left(  t\right)  $ is of the form%
\begin{equation}
F\left(  t\right)  =\left[
\begin{array}
[c]{c}%
f\left(  t\right) \\
0
\end{array}
\right]  =: T^\dagger f\left(  t\right)  , \label{xfa4}%
\end{equation}
indicating that we may excite directly only the degrees of freedom
corresponding to the observable variables $v$.

In \eqref{xfa4} we have introduced the operator:
\begin{equation}
T=\left[
\begin{array}
[c]{cc}%
I_{H_{0}} & 0
\end{array}
\right]  :\mathcal{H}\rightarrow H_{0},\ TT^{\dag}=I_{H_{0}},\ T^{\dag
}T=P_{H_{0}}, \label{xfa4a}%
\end{equation}
where $I_{H_{0}}$ and $P_{H_{0}}$ are respectively the identity operator in
$H_{0}$ and the orthogonal projection onto $\operatorname{ran}T^{\dag}%
=H_{0}\oplus\left\{  0\right\}  $ in $\mathcal{H}$ . Notice that
$T^{\dag }:H_{0}\rightarrow\mathcal{H}$ is an isometric injection
from $H_{0}$ into
$\mathcal{H}$, and we can recast the equations (\ref{xfa2}), (\ref{xfa4}) as%
\begin{equation}
\mathcal{M}\partial_{t}V\left(  t\right)  =-\mathrm{i}\mathcal{A}V\left(
t\right)  +T^{\dag}f\left(  t\right)  ,\ V\left(  t\right)  \in\mathcal{H}%
,\ f\left(  t\right)  \in H_{0}. \label{MAT}%
\end{equation}
The operator $T$ is an example of an \emph{isometric truncation}:

\begin{definition}
Given two Hilbert spaces $\mathcal{H}$ and $H_{0}$, a bounded linear operator
$T:\mathcal{H}\rightarrow H_{0}$ is called an \emph{isometric truncation} of
$\mathcal{H}$ to $H_{0}$ if $TT^{\dag}=I_{H_{0}}$.
\end{definition}

\noindent Observe that when $H_{0}\subseteq\mathcal{H}$ the
orthogonal projection $P_{H_{0}}$ of $\mathcal{H}$ onto $H_{0}$ is
an isometric truncation. \emph{The conservative system as
described by the equation (\ref{MAT}) with }$\mathcal{M}>0$\emph{,
$\mathcal{A}$} \emph{self adjoint, and }$T$\emph{ an isometric
truncation is of the most general form we consider in this paper.}

We are specifically interested in conservative extensions to (\ref{ax3}) of
the form (\ref{avm3})-(\ref{avm4}), corresponding to the special case of
equation (\ref{MAT}) with operators $\mathcal{M}$ and $\mathcal{A}$ in the
following block-matrix form%
\begin{multline}
\mathcal{M}    =\left[
\begin{array}
[c]{cc}%
m & 0\\
0 & I_{H_{1}}%
\end{array}
\right]  ,\text{ }m>0,\\
\mathcal{A}    =\left[
\begin{array}
[c]{cc}%
A & \Gamma\\
\Gamma^{\dagger} & \Omega_{1}%
\end{array}
\right]  ,\ A^{\dagger}=A,\ \Omega_{1}^{\dagger}=\Omega_{1},\
\Gamma :H_{1}\rightarrow H_{0},\label{xfa3}
\end{multline}
where $I_{H_{1}}$ is the identity operator in $H_{1}$. In physical models with
unbounded operators, the block decomposition (\ref{xfa3}) may be somewhat
formal, since there remains the question of specifying the domain of each
operator. In general we consider $A$ and $\Omega_{1}$ which are self adjoint
with domains $\mathcal{D}(A)$ and $\mathcal{D}(\Omega_{1})$ respectively. For
the time being, we take the operator $\Gamma$ -- which evidently provides the
coupling between the observable and hidden degrees of freedom -- to be a
\emph{bounded} map from $H_{1}$ to $H_{0}$. In this case, the natural domain
for $\mathcal{A}$ is the direct sum $\mathcal{D}(A)\oplus\mathcal{D}%
(\Omega_{1})$, and with this choice the operator is self-adjoint. We consider
unbounded $\Gamma$ below to describe systems with non-vanishing instantaneous
friction, e.g., a damped oscillator. In that case, $\Gamma$ will be a map from
$\mathcal{D}\left(  \Omega_{1}\right)  $ to $H_{0}$, and we shall have to
specify the domain of $\mathcal{A}$ carefully.

We can represent a formal solution to (\ref{xfa2}) -- using the rescaling
(\ref{vax1}) -- by the formula:%
\begin{multline}\label{xfa5}
V\left(  t\right)    =\int_{0}^{\infty}\mathcal{M}^{-\frac{1}{2}%
}e^{-\mathrm{i}\mathcal{A}_{m}\tau}\mathcal{M}^{-\frac{1}{2}}F\left(
t-\tau\right)  \,d\tau\\
 =\int_{-\infty}^{t}\mathcal{M}^{-\frac{1}{2}}e^{-\mathrm{i}\mathcal{A}%
_{m}\left(  t-\tau\right)  }\mathcal{M}^{-\frac{1}{2}}F\left(  \tau\right)
\,d\tau,\ \mathcal{A}_{m}=\mathcal{M}^{-\frac{1}{2}}\mathcal{AM}^{-\frac{1}%
{2}},
\end{multline}
assuming the system was \emph{at rest} with $V(t)=0$, $F(t)=0$ in distant
past.\ In particular, we note that
\begin{equation}
V\left(  t\right)  =\mathcal{M}^{-\frac{1}{2}}e^{-\mathrm{i}\mathcal{A}_{m}%
t}\mathcal{M}^{-\frac{1}{2}}F_{0},\text{ }t>0,\text{ } \label{xfa6a}%
\end{equation}
for a pulse force $F\left(  t\right)  =F_{0}\delta\left(  t\right)  $, and
\begin{equation}
V\left(  t\right)  =\mathcal{M}^{-\frac{1}{2}}e^{-\mathrm{i}\mathcal{A}_{m}%
t}\left[  \int_{-\infty}^{\infty}e^{-\mathrm{i}\mathcal{A}_{m}\tau}%
\mathcal{M}^{-\frac{1}{2}}F\left(  \tau\right)  \,d\tau+o\left(  1\right)
\right] \label{xfa6}%
\end{equation}
for $t \rightarrow \infty$ if, say, $\int_{-\infty}^{\infty}\left\|  \mathcal{M}^{-\frac{1}{2}}%
F(\tau)\right\|  <\infty$.

To obtain an effective equation for the evolution of $v$, let us look at the
block form of the evolution equation (\ref{xfa2}) with $\mathcal{M}$,
$\mathcal{A}$ and $F$ respectively satisfying (\ref{xfa3}) and (\ref{xfa4}),%
\begin{align}
m\partial_{t}v\left(  t\right)   &  =-\mathrm{i}Av\left(  t\right)
-\mathrm{i}\Gamma w\left(  t\right)  +f\left(  t\right)  ,\label{xfa7a}\\
\partial_{t}w\left(  t\right)   &  =-\mathrm{i}\Gamma^{\dagger}v\left(
t\right)  -\mathrm{i}\Omega_{1}w\left(  t\right)  . \label{xfa7b}%
\end{align}
Note that $-\mathrm{i}\Gamma^{\dagger}v\left(  t\right)  $ plays the role of an
external force in eq. (\ref{xfa7b}). Thus we can solve for $w$ as in eq.
(\ref{xfa5}), obtaining%
\begin{equation}
w\left(  t\right)  =-\mathrm{i}\int_{0}^{\infty}e^{-\mathrm{i}\Omega_{1}\tau
}\Gamma^{\dagger}v\left(  t-\tau\right)  \,d\tau. \label{xfa8}%
\end{equation}
Plugging this into eq. (\ref{xfa7a}) yields%
\begin{equation}
m\partial_{t}v\left(  t\right)  =-\mathrm{i}Av\left(  t\right)  -\int
_{0}^{\infty}\Gamma e^{-\mathrm{i}\Omega_{1}\tau}\Gamma^{\dagger}v\left(
t-\tau\right)  \,d\tau+f\left(  t\right)  . \label{xfa9}%
\end{equation}

The dispersive evolution equation (\ref{xfa9}) is of the form (\ref{ax3}) with
friction function%
\begin{equation}
a(t)=\Gamma e^{-\mathrm{i}\Omega_{1}t}\Gamma^{\dagger},\text{ for }t>0,
\label{xfa9a}%
\end{equation}
and describes the evolution of the observable variable $v$. We note in
particular that the instantaneous friction $\alpha_{\infty}$ vanishes -- since
$a(0)=\Gamma\Gamma^{\dagger}$ is finite -- and the extended friction function
$a_{e}\left(  t\right)  $ is%
\begin{equation}
a_{e}\left(  t\right)  =\Gamma e^{-\mathrm{i}\Omega_{1}t}\Gamma^{\dagger
},\text{ for }-\infty<t<\infty. \label{xfa10}%
\end{equation}
Notice that for $a_{e}\left(  t\right)  $ of this form we have%
\begin{equation}
\int_{-\infty}^{\infty}\int_{-\infty}^{\infty}\left(  v\left(  t\right)
,a_{e}\left(  t-\tau\right)  v\left(  \tau\right)  \right)  \,dtd\tau
=\left\Vert \int_{-\infty}^{\infty}e^{\mathrm{i}\Omega_{1}t}\Gamma^{\dagger
}v\left(  t\right)  \,dt\right\Vert ^{2}, \label{xfa11}%
\end{equation}
readily implying that the extended friction function $a_{e}\left(  t\right)
=\Gamma e^{\mathrm{i}\Omega_{1}\tau}\Gamma^{\dagger}$ satisfies the power
dissipation condition (\ref{ava4}). \emph{The central point of this work is
that }$\Omega_{1}$\emph{ and }$\Gamma$\emph{ needed to satisfy (\ref{xfa9a})
can be re-constructed from the friction function }$a(t)$\emph{.}

\subsection{Unbounded coupling and instantaneous friction}

Let us briefly consider how to define the abstract model (\ref{xfa3}) with
unbounded $\Gamma$, a necessary step in the description of systems with
non-vanishing instantaneous friction ($\alpha_{\infty}\neq0$ in (\ref{ax3a})).
The following discussion is by nature a bit technical; we direct the reader to
the case of the damped oscillator in Section 5 for an explicit example which
may provide clarification.

Suppose $\Gamma$ is given as a map $\Gamma:\mathcal{D}(\Omega_{1})\rightarrow
H_{0}$ which is $\Omega_{1}$-bounded, i.e.%
\begin{equation}
\left\Vert \Gamma\phi\right\Vert ^{2}\leq C\left\Vert \left(  \Omega
_{1}+\mathrm{i}I_{H_{1}}\right)  \phi\right\Vert ^{2},\text{ } \label{Gbound}%
\end{equation}
where $I_{H_{1}}$ is the identity operator and $C<\infty$. We continue to take
$A$ and $\Omega_{1}$ to be self adjoint on their respective domains
$\mathcal{D}(A)$ and $\mathcal{D}(\Omega_{1})$ as above. The operator
$\Gamma\Phi_{R}$, with%
\begin{equation}
\Phi_{R}=\frac{R}{\sqrt{\Omega_{1}^{2}+R^{2}I_{H_{1}}}}\text{ for }R>0,
\end{equation}
is a bounded map from $H_{1}$ to $H_{0}$, with $\left\Vert \Gamma\Phi
_{R}\right\Vert \lesssim R$ as $R\rightarrow\infty$.\ We denote the adjoint of
this map by $\Phi_{R}\Gamma^{\dagger}$, although $\Gamma^{\dagger}$ has not
been defined,\footnote{It may be defined as a map from $H_{0}$ to a proper
extension of $H_{1}$, namely the space $\mathcal{D}\left(  B\right)  ^{\ast}$
of conjugate linear functionals on $\mathcal{D}\left(  B\right)  $, but we do
not use this fact here.} and define $\mathcal{A}$ as the limit%
\begin{equation}
\mathcal{A}\left[
\begin{array}
[c]{c}%
v\\
w
\end{array}
\right]  :=\lim_{R\rightarrow\infty}\left[
\begin{array}
[c]{c}%
Av+\Gamma\Phi_{R}w\\
\Phi_{R}\Gamma^{\dagger}v+\Omega_{1}\Phi_{R}w
\end{array}
\right]  , \label{Alimitdefn}%
\end{equation}
on the domain $\mathcal{D}\left(  \mathcal{A}\right)  $ of vectors $\left[
v,w\right]  ^{T}$ such that the limit exits.

As things stand, it is not clear if the resulting operator is self-adjoint, or
even that $\mathcal{D}\left(  \mathcal{A}\right)  $ is dense. To proceed we
require an additional assumption -- (\ref{assumedlimit}) below -- which
guarantees self-adjointness. To state that condition consider the map%
\begin{equation}
S:=\Gamma\left(  \Omega_{1}+\mathrm{i}I_{H_{1}}\right)  ^{-1}, \label{SgB}%
\end{equation}
which is bounded and satisfies
\begin{equation}
\left(  \Omega_{1}-\mathrm{i}I_{H_{1}}\right)  \Phi_{R}S^{\dagger}v=\Phi
_{R}\Gamma^{\dagger}v\text{ for any }v\in H_{0}.
\end{equation}
Therefore the limit
\begin{multline}
\lim_{R\rightarrow\infty}\Phi_{R}\Gamma^{\dagger}v+\Omega_{1}\Phi_{R}%
w \\ =\lim_{R\rightarrow\infty}\left(
\Omega_{1}-\mathrm{i}I_{H_{1}}\right) \Phi_{R}\left(
\frac{\Omega_{1}}{\Omega_{1}-\mathrm{i}I_{H_{1}}}w+S^{\dagger
}v\right)
\end{multline}
exists if and only if
\begin{equation}
\frac{\Omega_{1}}{\Omega_{1}-\mathrm{i}I_{H_{1}}}w+S^{\dagger}v\in
\mathcal{D}\left(  \Omega_{1}\right)  ,
\end{equation}
which is equivalent to saying that $w=\phi-S^{\dagger}v$ with $\phi
\in\mathcal{D}\left(  \Omega_{1}\right)  $. Thus $\left[  v,w\right]  ^{T}%
\in\mathcal{D}\left(  \mathcal{A}\right)  $ if and only if $w=\phi-S^{\dagger
}v$ with $\phi\in\mathcal{D}\left(  \Omega_{1}\right)  $ and%
\begin{equation}
Gv:=\lim_{R\rightarrow\infty}\Gamma\Phi_{R}S^{\dagger}v=\lim_{R\rightarrow
\infty}\Gamma\frac{1}{\Omega_{1}-\mathrm{i}I_{H_{1}}}\Phi_{R}\Gamma^{\dagger
}v\text{ } \label{assumedlimit}%
\end{equation}
exists. We \emph{require }of $\Gamma$ and $\Omega_{1}$ that the limit
(\ref{assumedlimit}) exists for every $v\in\mathcal{D}\left(  A\right)  $, and
defines an $A$ bounded operator $G$ with $A$ bound less than one -- i.e.,
there are $\delta<1$ and $\beta_{\delta}>0$ such that%
\begin{equation}
\left\Vert Gv\right\Vert \leq\delta\left\Vert \left(  A+\mathrm{i}%
\beta_{\delta}I_{H_{0}}\right)  v\right\Vert . \label{assumedbound}%
\end{equation}
Under these assumptions
\begin{equation}
  \begin{split}
    \mathcal{D}\left(  \mathcal{A}\right)   &  =\left\{  \left[
\begin{array}
[c]{c}%
v\\
w
\end{array}
\right]  :v\in\mathcal{D}\left(  A\right)  \text{ and
}\lim_{R\rightarrow
\infty}\Phi_{R}\Gamma^{\dagger}v+\Omega_{1}\Phi_{R}w\text{
exists}\right\}
\\
&  =\left\{  \left[
\begin{array}
[c]{c}%
v\\
\phi-Sv
\end{array}
\right]  :v\in\mathcal{D}\left(  A\right)  \text{ and }\phi\in\mathcal{D}%
\left(  \Omega_{1}\right)  \right\}  ,
  \end{split}
  \label{Adomain}
\end{equation}
and the operator $\mathcal{A}$ is self-adjoint:

\begin{proposition}
\label{pSelfAdjoint}Let $A$, $\Omega_{1}$ be self adjoint operators on the
Hilbert spaces $H_{0}$, $H_{1}$ with domains $\mathcal{D}\left(  A\right)  $,
$\mathcal{D}\left(  \Omega_{1}\right)  $ respectively. Suppose that
$\Gamma:\mathcal{D}\left(  \Omega_{1}\right)  \rightarrow H_{0}$ is such that
(\ref{Gbound}) holds and the limit (\ref{assumedlimit}) exists for all
$v\in\mathcal{D}\left(  A\right)  $, defining an operator $G$ for which the
bound (\ref{assumedbound}) holds. \ If $\mathcal{A}$ is defined by
(\ref{Alimitdefn}) on the domain $\mathcal{D}\left(  \mathcal{A}\right)
\subset H_{0}\oplus H_{1}$ specified in (\ref{Adomain}), then $\mathcal{A}$ is self-adjoint.
\end{proposition}

The proof of this proposition is elementary. It is obvious that $\mathcal{A}$
is symmetric, so to prove self-adjointness we need only to show that
$\mathcal{A}^{\dagger}V=\pm\mathrm{i}V$ implies $V=0$, which is an easy
exercise.

For the operator $\mathcal{A}$ defined in this way, the evolution equations
(\ref{xfa7a})-(\ref{xfa7b}) imply that%
\begin{equation}
\partial_{t}\Phi_{R}w\left(  t\right)  =-\mathrm{i}\Phi_{R}\Gamma^{\dagger
}v\left(  t\right)  -\mathrm{i}\Omega_{1}\Phi_{R}w\left(  t\right)  ,
\end{equation}
and thus%
\begin{equation}
\Phi_{R}w\left(  t\right)  =-\mathrm{i}\int_{0}^{\infty}e^{-\mathrm{i}%
\Omega_{1}\tau}\Phi_{R}\Gamma^{\dagger}v\left(  t-\tau\right)  \,d\tau.
\end{equation}
Therefore%
\begin{equation}
m\partial_{t}v\left(  t\right)  =-\mathrm{i}Av\left(  t\right)  -\lim
_{R\rightarrow\infty}\int_{0}^{\infty}\Gamma e^{-\mathrm{i}\Omega_{1}\tau}%
\Phi_{R}^{2}\Gamma^{\dagger}v\left(  t-\tau\right)  \,d\tau+f\left(  t\right)
.
\end{equation}

While this equation is formally similar to (\ref{xfa9}), we note that the
resulting friction function%
\begin{equation}
a_{e}(t)=\lim_{R\rightarrow\infty}\Gamma\frac{e^{-\mathrm{i}\Omega_{1}t}%
}{\frac{\Omega_{1}^{2}}{R^{2}}+I_{H_{1}}}\Gamma^{\dagger},\text{ for }%
-\infty<t<\infty
\end{equation}
is defined only as a distribution,%
\begin{multline}
\int_{-\infty}^{\infty}a_{e}(t)v(t)\,dt=\lim_{R\rightarrow\infty}\int
_{-\infty}^{\infty}\Gamma\frac{e^{-\mathrm{i}\Omega_{1}t}}{\frac{\Omega
_{1}^{2}}{R^{2}}+I_{H_{1}}}\Gamma^{\dagger}v(t)dt,\\ \text{ for
}v\in C_{c}\left(  \mathbb{R},H_{0}\right)  ,
\end{multline}
and may not in fact be a function. In particular, there may be non-vanishing
instantaneous friction. However $a_{e}(t)$ is a relatively tame distribution;
it may be expressed as a second order differential operator applied to a
(strongly) continuous function
\begin{equation}
a_{e}(t)=\left(  -\frac{d^{2}}{dt^{2}}+1\right)  Se^{-\mathrm{i}\Omega_{1}%
t}S^{\dagger}\text{ with }S=\Gamma\left(  \Omega_{1}+\mathrm{i}I_{H_{1}%
}\right)  ^{-1}.
\end{equation}

In a key example, $H_{1}=L^{2}\left(  \mathbb{R},H_{0}\right)  $, the space of
square integrable $H_{0}$-valued functions on the real line, and $\Omega_{1}$
is multiplication by the independent variable, $\Omega_{1}\psi\left(
x\right)  =x\psi\left(  x\right)  $. Given any positive operator
$\alpha_{\infty}$, say bounded, on $H_{0}$ we define
\begin{equation}
\left[  \Gamma\psi\right]  (x)=\frac{1}{\sqrt{\pi}}\int_{-\infty}^{\infty
}\sqrt{\alpha_{\infty}}\psi\left(  x\right)  dx,
\end{equation}
which is $\Omega_{1}$ bounded since%
\begin{equation}
\left[  S\psi\right]  (x)=\frac{1}{\sqrt{\pi}}\int_{-\infty}^{\infty}%
\frac{1}{x+\mathrm{i}}\sqrt{\alpha_{\infty}}\psi\left(  x\right)  dx
\end{equation}
is bounded. Thus%
\begin{equation}
Se^{-\mathrm{i}\Omega_{1}t}S^{\dagger}=\frac{1}{\pi}\int_{-\infty}^{\infty
}dx\frac{e^{-\mathrm{i}xt}}{x^{2}+1}\alpha_{\infty}=e^{-\left|  t\right|
}\alpha_{\infty},
\end{equation}
and%
\begin{equation}
a_{e}(t)=\left(  -\frac{d^{2}}{dt^{2}}+1\right)  Se^{-\mathrm{i}\Omega_{1}%
t}S^{\dagger}=2\alpha_{\infty}\delta\left(  t\right)  ,
\end{equation}
purely instantaneous friction. We shall return to this example in our
discussion of the damped oscillator below.

\subsection{Linear response and the admittance operator}

In the linear response theory, e.g. ref. \cite[Section 3]{KuboToda2}, a system
is often characterized by its \emph{admittance operator} $\mathfrak{A}\left(
\omega\right)  :H_{0}\rightarrow H_{0}$, defined by the relation%
\begin{equation}
\hat{v}\left(  \omega\right)  =\mathfrak{A}\left(  \omega\right)  \hat
{f}\left(  \omega\right)  ,\ \hat{v}\left(  \omega\right)  ,\hat{f}\left(
\omega\right)  \in H_{0} \label{ax2b}%
\end{equation}
at real frequencies $\omega$. Here we discuss the admittance formulation of
dissipative dispersive systems satisfying the power dissipation condition
(\ref{ava4}) and truncated conservative systems described by (\ref{MAT}). For
this purpose, it is useful to recast the linear response equation (\ref{ax2b})
in the domain of complex frequencies $\zeta=\omega+\mathrm{i}\eta$, $\eta>0$,
which corresponds to replacing the Fourier transform by a Fourier-Laplace transform.

We assume that the system governed by (\ref{ax3})\ is at rest for all negative
times, i.e.%
\begin{equation}
v\left(  t\right)  =0,\ f\left(  t\right)  =0,\ t\leq0, \label{vf1}%
\end{equation}
and define the Laplace transforms%
\begin{align}
\hat{v}\left(  \zeta\right)   &  =\int_{0}^{\infty}e^{\mathrm{i}\zeta
t}v(t)dt,\text{ }\hat{f}\left(  \zeta\right)  =\int_{0}^{\infty}%
e^{\mathrm{i}\zeta t}f(t)dt,\label{azf1}\\
\hat{a}\left(  \zeta\right)   &  =\int_{\left[  0,\infty\right)
}e^{\mathrm{i}\zeta t}a\left(  t\right)  \,dt=\alpha_{\infty}+\int_{\left[
0,\infty\right)  }e^{\mathrm{i}\zeta t}\alpha\left(  t\right)
\,dt,\ \label{azf1a}%
\end{align}
on the domain $\left\{  \operatorname{Im}\zeta>0\right\}  $. It is convenient
to assume that $\alpha\left(  t\right)  $ is bounded as $t\rightarrow\infty$
so that $\hat{a}\left(  \zeta\right)  $ is well defined (see Condition
\ref{condfric} below).\ Under the Laplace transform, the evolution equation
(\ref{ax3}) is transformed into the following identity
\begin{equation}
\zeta m\hat{v}\left(  \zeta\right)  =\left[  A-\mathrm{i}\hat{a}\left(
\zeta\right)  \right]  \hat{v}\left(  \zeta\right)  +\mathrm{i}\hat{f}\left(
\zeta\right)  ,\ \zeta=\omega+\mathrm{i}\eta,\ \eta=\operatorname{Im}\zeta>0.
\label{vf2}%
\end{equation}

We have $\operatorname{Re}\hat{a}\left(  \zeta\right)  \geq0$ (for
$\operatorname{Im}\zeta>0$), since \
\begin{multline}
\operatorname{Re}\left(  v,\hat{a}\left(  \zeta\right)  v\right)
\\ =\frac{\operatorname{Im}\zeta}{2}\int_{\left[  0,\infty\right)
}\int_{\left[ 0,\infty\right)  }\left(  v,a_{e}\left(
t-\tau\right)  v\right)
e^{\mathrm{i}\zeta t}e^{-\mathrm{i}\bar{\zeta}\tau}dtd\tau\geq0 \label{azf3}%
\end{multline}
for any $v\in H_{0}$ by the power dissipation condition (\ref{ava4}). Hence,
the operator $\zeta m-A+\mathrm{i}\hat{a}\left(  \zeta\right)  $ is
invertible, and
\begin{gather}
\hat{v}\left(  \zeta\right)  =\mathfrak{A}_{m,A,a}\left(  \zeta\right)
\hat{f}\left(  \zeta\right)  ,\ \zeta=\omega+\mathrm{i}\eta,\ \eta
=\operatorname{Im}\zeta>0,\label{vt2}\\
\mathfrak{A}_{m,A,a}\left(  \zeta\right)  =\mathrm{i}\left[  \zeta
m-A+\mathrm{i}\hat{a}\left(  \zeta\right)  \right]  ^{-1}. \label{vt2a}%
\end{gather}
The equation (\ref{vt2}) generalizes (\ref{ax2b}) to a certain extent since it
is an identity for analytic functions in the upper half plane. Note that%
\begin{equation}
\begin{split}
  \operatorname{Re}\left\{  \mathfrak{A}_{m,A,a}\left(
\zeta\right)  \right\}
=& \ \frac{1}{2}\left[  \mathfrak{A}_{m,A,a}\left(  \zeta\right)  +\mathfrak{A}%
_{m,A,a}^{\dagger}\left(  \zeta\right)  \right] \\
=& \ \mathfrak{A}_{m,A,a}\left(  \zeta\right)  \left\{
\operatorname{Im}\zeta
m+\operatorname{Re}\hat{a}\left(  \zeta\right)  \right\}  \mathfrak{A}%
_{m,A,a}^{\dagger}\left(  \zeta\right)  \ \geq \ 0.
\end{split}\label{rea1}
\end{equation}
which expresses the power dissipation condition (\ref{ava4}) in
terms of the admittance operator $\mathfrak{A}_{m,A,a}$.

The \emph{admittance equation} (\ref{vt2}) provides an essentially equivalent
description of the\ system (\ref{ax3}). In particular, the various operators
in (\ref{ax3}) can be readily recovered from $\mathfrak{A}_{m,A,a}\left(
\omega\right)  $ by the relations%
\begin{multline}
      m^{-1}=-\lim_{\eta\rightarrow\infty}\eta\mathfrak{A}_{m,A,a}\left(
\mathrm{i}\eta\right)  ,\ A=-\lim_{\eta\rightarrow\infty}\sqrt{m}%
\operatorname{Im}\mathfrak{A}_{m,A,a}^{-1}\left(
\mathrm{i}\eta\right)
\sqrt{m},\\
\hat{a}\left(  \zeta\right)  =\mathrm{i}\left(  \zeta m-A\right)
+\left[ \mathfrak{A}_{m,A,a}\left(  \zeta\right)  \right] ^{-1}.
\label{ax4c}
\end{multline}
Hence, the admittance operator $\mathfrak{A}_{m,A,a}\left(  \omega\right)  $
carries all the information about the system initially described by the
triplet $\left\{  m,A,\hat{a}\left(  \zeta\right)  \right\}  $. Very often the
admittance equation (\ref{ax2b}) is a preferred form, since the admittance
$\mathfrak{A}\left(  \omega\right)  $ may be measured experimentally more
readily than $m$, $A$ or $\hat{a}\left(  \omega\right)  $.

There are also several technical advantages to the admittance
formulation (\ref{vt2}). First, the quantities $\hat{v}\left(
\zeta\right)  $, $\hat{a}\left(  \zeta\right)  $,
$\mathfrak{A}\left(  \zeta\right)  $, $\hat{f}\left(  \zeta\right)
$ are analytic functions in the upper half-plane
$\operatorname{Im}\zeta>0$, whereas their time counterparts may be
more singular functions. In addition (\ref{vt2}) has the advantage
that auxiliary operators which appear in the analysis are
\emph{bounded}. In particular, we shall see that for
$\alpha_{\infty}\neq0$ the admittance formulation permits us to
avoid the subtleties required for an unbounded coupling $\Gamma$.

Consider now a system which is the truncation of a general conservative system
of the form (\ref{MAT}). Under the Laplace transform (\ref{MAT}) is
transformed into%
\begin{equation}
-\mathrm{i}\zeta\mathcal{M}\hat{V}\left(  \zeta\right)  =-\mathrm{i}%
\mathcal{A}\hat{V}\left(  \zeta\right)  +T^{\dag}\hat{f}\left(  \zeta\right)
,\text{ }\operatorname{Im}\zeta>0 \label{MAT1a}%
\end{equation}
which is easily solved for $\hat{V}$,
\begin{equation}
\hat{V}\left(  \zeta\right)  =\mathrm{i}\left(  \zeta\mathcal{M}%
-\mathcal{A}\right)  ^{-1}T^{\dag}\hat{f}\left(  \zeta\right)  ,\text{
}\operatorname{Im}\zeta>0. \label{MAT1b}%
\end{equation}
Multiplying of the both sides of (\ref{MAT1b}) by the isometric truncation
$T:\mathcal{H}\rightarrow H_{0}$ yields%
\begin{align}
\hat{v}\left(  \zeta\right)   &  =\mathfrak{A}\left(  \zeta\right)  \hat
{f}\left(  \zeta\right)  ,\ \zeta=\omega+\mathrm{i}\eta,\ \eta
=\operatorname{Im}\zeta>0,\label{MAT2}\\
\ \mathfrak{A}\left(  \zeta\right)   &  =\mathrm{i}T\left(  \zeta
\mathcal{M}-\mathcal{A}\right)  ^{-1}T^{\dag}. \label{MAT2a}%
\end{align}
Notice also that the admittance operator as defined in (\ref{MAT2}) satisfies
\begin{multline}
\operatorname{Re}\left\{  \mathfrak{A}\left(  \zeta\right)
\right\} =\frac{\mathfrak{A}\left(  \zeta\right)
+\mathfrak{A}^{\dagger}\left( \zeta\right)  }{2}\\=\left(
\operatorname{Im}\zeta\right)  T\left(
\zeta\mathcal{M}-\mathcal{A}\right)  ^{-1}\mathcal{M}\left[
T\left(
\zeta\mathcal{M}-\mathcal{A}\right)  ^{-1}\right]  ^{\dag}. \label{MAT3}%
\end{multline}
This identity together with $\mathcal{M}>0$\ implies
\begin{equation}
\operatorname{Re}\left\{  \mathfrak{A}\left(  \zeta\right)  \right\}  \geq0.
\label{MAT3a}%
\end{equation}
\qquad\

For operators $\mathcal{M},\mathcal{A}$ in the block-matrix form (\ref{xfa3})
we have%
\begin{equation}
\mathfrak{A}\left(  \zeta\right)  =\mathrm{i}\left(  \zeta m-A-\Gamma\left(
\zeta I_{H_{1}}-\Omega_{1}\right)  ^{-1}\Gamma^{\dag}\right)  ^{-1},
\label{MAT4}%
\end{equation}
as may be easily verified. Comparing (\ref{MAT4}) and (\ref{vt2a}) gives the
following formula for the Laplace transform $\hat{a}\left(  \zeta\right)  $ of
the friction function for a truncated conservative system%
\begin{equation}
\hat{a}\left(  \zeta\right)  =\mathrm{i}\Gamma\left(  \zeta I_{H_{1}}%
-\Omega_{1}\right)  ^{-1}\Gamma^{\dag}, \label{MAT5}%
\end{equation}
which could also be verified by directly transforming
(\ref{xfa9a}).

A second approach to the central construction of this work is based upon the
observation that the Hilbert space $\mathcal{H}$ and the operator triple
$\mathcal{M}$, $\mathcal{A}$, $T$ in eq. (\ref{MAT2a}) can be reconstructed
from the admittance $\mathfrak{A}\left(  \zeta\right)  $. \ Alternatively
$H_{1}$, $\Gamma$, and $\Omega_{1}$ appearing in (\ref{MAT5}) can be
reconstructed from $\hat{a}\left(  \zeta\right)  $.

\section{Bases for a conservative extension}

We now describe how, given a dispersive system in the form (\ref{ax3}) or
(\ref{ax4}), one can reconstruct the hidden degrees of freedom. Of course, the
resulting mathematical reconstruction is initially devoid of physical
interpretation. However, the reader should bear in mind that one usually knows
a given evolution equation of the form (\ref{ax3}) involves ``hidden'' degrees
of freedom with a natural physical interpretation. Generally, the abstract
extension may be interpreted therefore in a physically concrete way.

For example in a classical dielectric medium, the time dispersion comes from
the material relation between the electric displacement $\mathbf{D}\left(
\mathbf{r},t\right)  $ at a point $\mathbf{r}$ and the electric field
$\mathbf{E}\left(  \mathbf{r},t\right)  $, namely%
\begin{equation}
\mathbf{D}\left(  \mathbf{r},t\right)  =\mathbf{E}\left(  \mathbf{r},t\right)
+4\pi\mathbf{P}\left(  \mathbf{r},t\right)  ,\text{ where }\mathbf{\hat{P}%
}\left(  \mathbf{r},\omega\right)  =\mathbf{\hat{\chi}}\left(  \omega\right)
\mathbf{\hat{E}}\left(  \mathbf{r},\omega\right)  \label{ax5}%
\end{equation}
with $\mathbf{\hat{P}}$ and $\mathbf{\hat{E}}$ the time Fourier transforms of
$\mathbf{P}$ and $\mathbf{E}$ and $\mathbf{\hat{\chi}}\left(  \omega\right)  $
the frequency dependent electric susceptibility tensor. The relations
(\ref{ax5}) can be viewed as a macroscopic picture of the interactions between
the electromagnetic field and the material medium, and we can naturally
interpret the hidden variables that arise in the spectral theory of this
equation as a model for the material medium. For the time being, however, we
ignore such considerations (which we take up in section 5 wherein we discuss
specific models). Herein we focus on the abstract, mathematical, structure of
the hidden variables.

Thus given a dissipative system, defined by either the evolution equation
(\ref{ax3}) or its frequency counterpart (\ref{ax4}), our main problem is to
find a larger conservative system, governed respectively by the equation
(\ref{ax1}) or its frequency counterpart (\ref{ax1a}), that reduces
correspondingly to (\ref{ax3}) or (\ref{ax4}) upon integrating out the hidden
variables. Clearly the power dissipation condition in the form (\ref{ava4})
for the friction function $a\left(  t\right)  $ or in the form (\ref{MAT3a})
for the admittance function $\mathfrak{A}\left(  \omega\right)  $ is a
necessary condition for the existence of such an extension, as indicated by
the relations (\ref{xfa10}-\ref{xfa11}) and (\ref{MAT2a}-\ref{MAT3}) obtained
for the truncation of a conservative system. Remarkably, we shall see it is
also a sufficient condition.

The equality (\ref{xfa10}) is a possible base for the construction of such a
conservative extension. Thus, we pose the following problem: given a friction
function $a\left(  t\right)  $ satisfying the dissipation condition
(\ref{ava4}) find an operator $\Gamma$ and a self-adjoint operator $\Omega
_{1}$ for which the equality (\ref{xfa10}) holds. Then the desired
conservative system is (\ref{xfa1})-(\ref{xfa4}). Alternatively, we may start
with the relation (\ref{MAT2a}) and ask, given the admittance $\mathfrak{A}%
\left(  \zeta\right)  $ with $\operatorname{Re}\mathfrak{A}\left(
\zeta\right)  \geq0$, whether we can find an isometric truncation
$T:\mathcal{H}\rightarrow H$ along with self-adjoint
$\mathcal{M}>0$ and $\mathcal{A}$ for which (\ref{MAT2a}) holds.
These two closely related approaches each lead to constructions of
an extension, based respectively on operator versions of the
following fundamental results: Bochner's Theorem, \cite[Section
60]{AkhGlaz}, \cite[Section XI, 13, Theorem 2]{Yosida}, and the
Herglotz-Nevanlinna Theorems, \cite[Section 59]{AkhGlaz},
\cite{KK}, \cite[Section 32.1, Theorem 2, Theorem 3]{Lax}. In the
second approach, the Naimark Theorem on positive operator valued
measures plays a key role, \cite[Vol II, Appendix I, Section
I]{AkhGlaz}, \cite[Appendix, Section 2, Theorem I]{RiNa}.

\subsection{Approach via Bochner's Theorem}

We begin by recalling, \cite[Section 60]{AkhGlaz}, \cite[Section XI, 13,
Theorem 2]{Yosida} and \cite[Theorem IX.9]{RS1}:

\begin{theorem}
[Bochner]\label{tBoch}A complex-valued continuous function $s(t)
$ of $-\infty<t<\infty$, is representable as%
\begin{equation}
s\left(  t\right)  =\int_{-\infty}^{\infty}e^{-\mathrm{i}t\sigma}\,dN\left(
\sigma\right)  \label{hg1}%
\end{equation}
with a non-decreasing, right-continuous bounded function $N\left(
\sigma\right)  $ if and only if $s\left(  t\right)  $ is positive-definite in
the following sense%
\begin{equation}
\int_{-\infty}^{\infty}\int_{-\infty}^{\infty}s\left(  t-\tau\right)
\overline{\varphi\left(  t\right)  }\varphi\left(  \tau\right)  \,dtd\tau\geq0
\label{hg2}%
\end{equation}
for every continuous function $\varphi\left(  t\right)  $ with compact support.
\end{theorem}

The measure $dN\left(  \sigma\right)  $ may be realized as the spectral
measure associated to the vector $\chi_{\mathbb{R}}$ and the operator
$\Omega_{1}$ of multiplication by $\sigma$ on $L^{2}(\mathbb{R},dN)$. Thus%
\begin{equation}
s(t)=\Gamma e^{-\mathrm{i}t\Omega_{1}}\Gamma^{\dagger} \label{hg1a}%
\end{equation}
with $\Gamma$ the following rank one operator%
\begin{equation}\Gamma\psi=\int\psi\left(  \sigma\right)  dN\left(
\sigma\right) :L^{2}(\mathbb{R},dN)\rightarrow\mathbb{C}.
\end{equation}
Hence, Bochner's Theorem may be seen as a special case of the following
result, which we prove below in Section 6.

\begin{theorem}
\label{tOperatorBoch}Let $\mathcal{B}\left(  H_{0}\right)  $ be the space of
all bounded linear operators in $H_{0}$. Then a strongly continuous
$\mathcal{B}\left(  H_{0}\right)  $-valued function $a_{e}\left(  t\right)  $,
$-\infty<t<\infty$, is representable as%
\begin{equation}
a_{e}\left(  t\right)  =\Gamma e^{-\mathrm{i}t\Omega_{1}}\Gamma^{\dagger},
\label{hg1b}%
\end{equation}
with $\Omega_{1}$ a self-adjoint operator on a Hilbert space $H_{1}$ and
$\Gamma:$ $H_{0}\rightarrow H_{1}$ a bounded linear map, if and only if
$a_e\left(  t\right)  $ satisfies the dissipation condition (\ref{ava4}) for
every continuous $H_{0}$ valued function $v(t)$ with compact support. If the
space $H_{1}$ is minimal -- in the sense that the linear span%
\begin{equation}
\left\langle g\left(  \Omega_{1}\right)  \Gamma^{\dagger}v:g\in C_{c}\left(
\mathbb{R}\right)  ,\text{ }v\in H_{0}\right\rangle
\end{equation}
is dense in $H_{1}$ -- then the triplet $\left\{  H_{1},\Omega_{1}%
,\Gamma\right\}  $ is determined uniquely up to an isomorphism.
\end{theorem}

\begin{remark}
In fact, it is sufficient to assume that $a_{e}\left(  t\right)  $ is locally
bounded and strongly measurable, strong continuity then follows from
(\ref{hg1b}).
\end{remark}

\begin{remark}
$\left\langle g\left(  \Omega_{1}\right)  \Gamma^{\dagger}v:\ g\in
C_{c}\left(  \mathbb{R}\right)  ,\text{ }v\in H_{0}\right\rangle $ denotes the
linear span, i.e., the subspace of linear combinations of finitely many
elements of the form $g\left(  \Omega_{1}\right)  \Gamma^{\dagger}v$.
\end{remark}

Our proof of this theorem is a very elementary generalization of the proof of
Bochner's Theorem given in \cite{RS1}. Nonetheless we are not aware that
Theorem \ref{tOperatorBoch} has appeared previously in the literature.

Theorem \ref{tOperatorBoch} provides a basis for constructing a conservative
extension of a dispersive system without instantaneous friction: given a
system described by a vector $v$ in a Hilbert space $H_{0}$ and governed by a
dissipative evolution (\ref{ax1}) with a strongly continuous friction
function, we simply represent it as the restriction of a conservative system
in the block-matrix form (\ref{xfa3}) with $\Omega_{1}$ and $\Gamma$ the
operators obtained from Theorem \ref{tOperatorBoch}.

The construction afforded by Theorem \ref{tOperatorBoch} is sufficiently
general to describe most systems of interest, excepting those with
instantaneous friction. For such a system we must admit a friction
``function''\ $a\left(  t\right)  $ which is not strongly continuous. For most
cases of interest though it is sufficient to assume the following condition

\begin{condition}
[friction function]\label{condfric}The friction function $a\left(  t\right)  $
is of the form
\begin{equation}
a\left(  t\right)  =\alpha_{\infty}\delta\left(  t\right)  +\alpha\left(
t\right)  \label{cf1}%
\end{equation}
where $\alpha_{\infty}$ is a bounded non-negative operator in $H_{0}$ and
$\alpha\left(  t\right)  $ a strongly continuous and bounded $\mathcal{B}%
\left(  H_{0}\right)  $-valued function for $t\geq0$, i.e.%
\begin{equation}
0\leq\alpha_{\infty}\leq CI_{H_{0}},\ C<\infty;\ \sup_{t\geq0}\left\|
\alpha\left(  t\right)  \right\|  <\infty. \label{cf2}%
\end{equation}
The extension $a_{e}\left(  t\right)  =2\alpha_{\infty}\delta\left(  t\right)
+\alpha_{e}\left(  t\right)  $, $-\infty<t<\infty$ of the function $a\left(
t\right)  $ is defined by the formula (\ref{ava2}).
\end{condition}

The following result for a friction function satisfying Condition
\ref{condfric} will be useful. We give its proof below in Section 6.

\begin{theorem}
\label{tGenOperatorBoch}Suppose that the friction function $a\left(  t\right)
$ satisfies Condition \ref{condfric}. Then its extension $a_{e}\left(
t\right)  =2\alpha_{\infty}\delta\left(  t\right)  +\alpha_{e}(t)$,
$-\infty<t<\infty$, is representable as%
\begin{equation}
a_{e}\left(  t\right)  =\operatorname*{Dlim}_{R\rightarrow\infty}\Gamma
e^{-\mathrm{i}t\Omega_{1}}\left(  \Gamma\Phi_{R}^{2}\right)  ^{\dagger}%
,\ \Phi_{R}^{2}=\left(  \frac{\Omega_{1}^{2}}{R^{2}}+I_{H_{1}}\right)  ^{-1},
\label{hg1c}%
\end{equation}
with $\Omega_{1}$ a self-adjoint operator on $H_{1}$ and $\Gamma
:\mathcal{D}\left(  \Omega_{1}\right)  \rightarrow H_{0}$ an $\Omega_{1}%
$-bounded linear map, if and only if $a_e\left(  t\right)  $ satisfies the
dissipation condition (\ref{ava4}) for every continuous $H_{0}$ valued function
$v(t)$ with compact support. If the space $H_{1}$ is minimal -- in
the sense that%
\begin{equation}
\left\langle \left(  \Gamma g\left(  \Omega_{1}\right)  \right)  ^{\dagger
}v:\ g\in C_{c}\left(  \mathbb{R}\right)  ,\text{ }v\in H_{0}\right\rangle
\end{equation}
is dense in $H_{1}$ -- then the triplet $\left\{  H_{1},\Omega_{1}%
,\Gamma\right\}  $ is determined uniquely up to an isomorphism.
\end{theorem}

\begin{remark}
Here $\operatorname*{Dlim}$ indicates the distributional limit, i.e.%
\begin{equation}
\int_{-\infty}^{\infty}a_{e}\left(  t\right)  v\left(  t\right)
dt=\lim_{R\rightarrow\infty}\int_{-\infty}^{\infty}\Gamma e^{-\mathrm{i}%
t\Omega_{1}}\Phi_{R}^{2}\Gamma^{\dagger}v\left(  t\right)  dt
\end{equation}
for every smooth $H_{0}$ valued function $v\left(  t\right)  $ with compact
support. In essence we have $a_{e}\left(  t\right)  =\Gamma e^{-\mathrm{i}%
t\Omega_{1}}\Gamma^{\dagger}$, but this expression is ambiguous so we
introduce a sort of principle value by regularizing with $\Phi_{R}^{2}$.
\end{remark}

\subsection{Approach via Herglotz-Nevanlinna Theorems}

The approach via Bochner's Theorem just outlined is quite straightforward and
adequate for many purposes, however there is an equally useful method which
works in the frequency domain and makes use of analytic function theory. This
second approach is based on an alternative description of the friction
function $a\left(  t\right)  $ through its Laplace transform\emph{
}$\hat{a}\left(  \zeta\right)  $ as defined above in eq. (\ref{azf1a}).

Condition \ref{condfric} for the friction function readily implies that%
\begin{equation}
\hat{a}\left(  \zeta\right)  =\alpha_{\infty}+\hat{\alpha}\left(
\zeta\right)  ,\ \left\|  \hat{\alpha}\left(  \zeta\right)  \right\|
\leq\frac{\sup_{t\geq0}\left\|  \alpha\left(  t\right)  \right\|
}{\operatorname{Im}\zeta},\ \operatorname{Im}\zeta>0. \label{aa1}%
\end{equation}
One advantage of $\hat{a}\left(  \zeta\right)  $ over $a\left(  t\right)  $ is
that it is an analytic function, even if $a\left(  t\right)  $ has the
singular term $\alpha_{\infty}\delta\left(  t\right)  $. Formally the Laplace
Transform $\hat{a}\left(  \zeta\right)  $ becomes the Fourier Transform
$\hat{a}\left(  \omega\right)  $ for $\zeta=\omega$ with real $\omega$.

The power dissipation condition (\ref{ava4}) implies that $\operatorname{Im}%
\hat{a}\left(  \zeta\right)  \geq0$ for $\operatorname{Im}\zeta>0$, as we have
seen in (\ref{azf3}). Therefore for each $v\in H$, $\zeta\mapsto\left(
v,\mathrm{i}\hat{a}\left(  \zeta\right)  v\right)  $ is an analytic map of the
upper half plane into itself. There is a classical representation theory for
such maps, \cite[Section 59]{AkhGlaz}, \cite{KK}, \cite[Section 32.1, Theorems
2, 3]{Lax}, which provides a tool for the construction of a conservative
extension governed by (\ref{xfa7a}) and (\ref{xfa7b}).

\begin{theorem}
[Nevanlinna]\label{tNev2}Every analytic function $g\left(
\zeta\right)  $ in the upper half-plane $\operatorname{Im}\zeta>0$
whose imaginary part is
everywhere non-negative and which satisfies the growth condition%
\begin{equation}
\limsup_{\eta\rightarrow+\infty}\eta\left\vert g\left(  i\eta\right)
\right\vert <\infty\label{hg4}%
\end{equation}
can be expressed uniquely in the form%
\begin{equation}
g\left(  \zeta\right)  =\int_{-\infty}^{\infty}\frac{dN\left(  \sigma\right)
}{\sigma-\zeta}, \label{hg5}%
\end{equation}
where $N\left(  \sigma\right)  $ is a non-decreasing, right-continuous,
bounded function such that%
\begin{equation}
\int_{-\infty}^{\infty}dN\left(  \sigma\right)  =\limsup_{\eta\rightarrow
+\infty}\eta\operatorname{Im}\left\{  g\left(  i\eta\right)  \right\}
<\infty.
\end{equation}
\end{theorem}

In fact, Theorem \ref{tNev2} is a special case of the following result.

\begin{theorem}
\label{tNev1}Every analytic function $g\left(  \zeta\right)  $ in
the half-plane $\left\{  \operatorname{Im}\zeta>0\right\}  $ whose
imaginary part
is everywhere non-negative can be expressed uniquely in the form%
\begin{equation}
g\left(  \zeta\right)  =\xi+\rho\zeta+\int_{-\infty}^{\infty}\frac{1+\sigma
\zeta}{\sigma-\zeta}\,d\tilde{N}\left(  \sigma\right)  \label{hg3}%
\end{equation}
where $\tilde{N}\left(  \sigma\right)  $ is a non-decreasing,
right-continuous, bounded function, $\xi$ is real, and $\rho\geq0$.
\end{theorem}

\begin{remark}
The measures from Theorem \ref{tNev1} and Theorem \ref{tNev2} are related by
the identity $dN\left(  \sigma\right)  =\left(  1+\sigma^{2}\right)
d\tilde{N}\left(  \sigma\right)  $, and the former result is obtained from the
later by noting that (\ref{hg4}) implies that $\rho=0$ and that $\left(
1+\sigma^{2}\right)  d\tilde{N}\left(  \sigma\right)  $ is a finite measure. \
\end{remark}

Returning to the analysis of $a\left(  t\right)  $, suppose first that the
instantaneous friction vanishes, $\alpha_{\infty}=0$.\ Then in view of eq.
(\ref{aa1}) we have%
\begin{equation}
\hat{a}\left(  \zeta\right)  =\hat{\alpha}\left(  \zeta\right)  ,\text{ and
}\left\|  \hat{a}\left(  \zeta\right)  \right\|  \leq\frac{\sup_{t}\left\|
\alpha\left(  t\right)  \right\|  }{\operatorname{Im}\zeta}.
\end{equation}
Hence, given $v\in H_{0}$, the function $\left(  v,\mathrm{i}\hat{a}\left(
\zeta\right)  v\right)  $ satisfies the hypotheses of Theorem \ref{tNev2} and
consequently there is a finite Borel measure $dN_{v,v}\left(  \sigma\right)  $
such that%
\begin{equation}
\left(  v,\mathrm{i}\hat{a}\left(  \zeta\right)  v\right)  =\int_{-\infty
}^{\infty}\frac{dN_{v,v}\left(  \sigma\right)  }{\sigma-\zeta}.
\end{equation}
For each pair $v,w\in H_{0}$, we define the ``off-diagonal''\ measures
$dN_{v,w}\left(  \sigma\right)  $ via polarization,
\begin{multline}
dN_{v,w}\left(  \sigma\right)  =\frac{1}{4}\left[  dN_{v+w,v+w}\left(
\sigma\right)  -dN_{v-w,v-w}\left(  \sigma\right) \right . \\
\left . -\mathrm{i}dN_{v+\mathrm{i}%
w,v+\mathrm{i}w}\left(  \sigma\right)  +\mathrm{i}dN_{v-\mathrm{i}%
w,v-\mathrm{i}w}\left(  \sigma\right)  \right]  ,
\end{multline}
so that%
\begin{equation}
\left(  v,\mathrm{i}\hat{a}\left(  \zeta\right)  w\right)  =\int_{-\infty
}^{\infty}\frac{dN_{v,w}\left(  \sigma\right)  }{\sigma-\zeta}\text{ for all
}v,w\in H_{0}.
\end{equation}

Using the measures $dN_{v,w}$ we define, for each $\sigma\in\mathbb{R}$, a
non-negative quadratic form
\begin{equation}
Q_{\sigma}\left(  v,w\right)  =\int_{\left(  -\infty,\sigma\right]  }%
dN_{v,w}\left(  \sigma^{\prime}\right)  .
\end{equation}
Because%
\begin{multline}
Q_{\sigma}\left(  v,v\right)
\leq\int_{-\infty}^{\infty}dN_{v,v}\left( \sigma\right)
\\ =\lim\sup_{\eta\rightarrow\infty}\eta\operatorname{Im}\left\{
\left(  v,\mathrm{i}\hat{a}\left(  \mathrm{i}\eta\right)  v\right) \right\}
\leq\left(  \sup_{t}\left\|  \alpha\left(  t\right) \right\|  \right) \left\|
v\right\|^2  ,
\end{multline}
we see that these forms are bounded. \ Thus for each $\sigma\in\mathbb{R}$
there is a non-negative bounded operator $K\left(  \sigma\right)  $ which
satisfies $Q_{\sigma}\left(  v,w\right)  =\left(  v,K\left(  \sigma\right)
w\right)  $. It is easy to see that this \emph{generalized spectral family} of
operators satisfies the following condition.

\begin{condition}
[generalized spectral family]\label{condgsp}\

\begin{enumerate}
\item $K\left(  \sigma\right)  $, $\sigma\in\mathbb{R}$ are bounded
non-negative operators in $H_{0}$.

\item $K\left(  \sigma\right)  \leq K\left(  \lambda\right)  $ for
$\sigma<\lambda$; $K\left(  \sigma+0\right)  =K\left(  \sigma\right)  $;
$\operatorname*{stlim}_{\sigma\rightarrow-\infty}K\left(  \sigma\right)  =0$;
$K\left( +\infty\right)
=\operatorname*{stlim}_{\sigma\rightarrow\infty}K\left( \sigma\right)  $ exists
and is bounded.
\end{enumerate}
\end{condition}

There is a fundamental result due to Naimark, \cite[in Russian]{Nai},
\cite[Vol. II, Appendix I, Section I]{AkhGlaz}, \cite[Appendix, Section 2,
Theorem I, includes uniqueness]{RiNa}, which provides a canonical
representation for generalized spectral families.

\begin{theorem}
[Naimark]\label{tNaim1}Let $K\left(  \sigma\right)  $, $\sigma\in\mathbb{R}$
be a generalized spectral family satisfying Condition \ref{condgsp}. Then
there exist a Hilbert space $H_{1}$, a bounded map $\Gamma:H_{1}\rightarrow
H_{0}$, and a resolution of the identity $E\left(  \sigma\right)  $,
$\sigma\in\mathbb{R}$ of $H_{1}$ such that:%
\begin{equation}
K\left(  \sigma\right)  =\Gamma E\left(  \sigma\right)  \Gamma^{\dag}%
,\sigma\in\mathbb{R}.
\end{equation}
If the space $H_{1}$ is minimal -- in the sense that%
\begin{equation}
\left\langle E\left(  \sigma\right)  \Gamma^{\dag}v:\sigma\in\mathbb{R},\text{
}v\in H_{0}\right\rangle
\end{equation}
is dense in $H_{1}$ -- then the triplet $\left\{  H_{1},\left\{  E\left(
\sigma\right)  ,\sigma\in\mathbb{R}\right\}  ,\Gamma\right\}  $ is determined
uniquely up to an isomorphism.
\end{theorem}

The construction of a conservative extension from these results proceeds as
follows. Given $\hat{a}\left(  \zeta\right)  $ with $\alpha_{\infty}=0$, we
obtain $K\left(  \sigma\right)  $ from Theorem \ref{tNev2} and thence $\Gamma$
and $E\left(  \sigma\right)  $ from Theorem \ref{tNaim1}. Letting $\Omega_{1}$
be the self-adjoint operator $\int_{-\infty}^{\infty}\sigma dE\left(
\sigma\right)  $, we obtain%
\begin{multline}
\hat{a}\left(  \zeta\right)  =-\mathrm{i}\int_{-\infty}^{\infty}%
\frac{1}{\sigma-\zeta}\Gamma dE\left(  \sigma\right)  \Gamma^{\dag}%
\\ =\mathrm{i}\Gamma\left[  \zeta I_{H_{1}}-\Omega_{1}\right]
^{-1}\Gamma^{\dag },\ \Omega_{1}=\int_{-\infty}^{\infty}\sigma
dE\left(  \sigma\right)  ,
\end{multline}
which is the desired representation for $\hat{a}\left(
\zeta\right)  $ -- compare with eq. (\ref{MAT5}). This
construction is summarized by the following operator
generalization of the Nevanlinna Theorem.

\begin{theorem}
\label{tOperatorNev2}Every $\mathcal{B}\left(  H_{0}\right)  $-valued analytic
function $G\left(  \zeta\right)  $ of the upper half plane $\operatorname{Im}%
\zeta>0$ with $\operatorname{Im}G\left(  \zeta\right)  $ everywhere
a non-negative operator, and which obeys the growth condition%
\begin{equation}
\limsup_{\eta\rightarrow+\infty}\eta\left\|  G\left(  \mathrm{i}\eta\right)
\right\|  <\infty, \label{hg4b}%
\end{equation}
can be expressed in the form
\begin{equation}
G(\zeta)=\Gamma\left[  \Omega_{1}-\zeta I_{H_{1}}\right]  ^{-1}\Gamma
^{\dagger} \label{hg4bb}%
\end{equation}
with $\Omega_{1}$ a self adjoint operator on a Hilbert space $H_{1}$ and
$\Gamma:H_{1}\rightarrow H_{0}$ a bounded map such that%
\begin{equation}
\Gamma\Gamma^{\dagger}v=\lim_{\eta\rightarrow+\infty}\eta G\left(
\mathrm{i}\eta\right)  v\text{ for every }v\in H_{0}. \label{gaga1}%
\end{equation}
If the space $H_{1}$ is minimal -- in the sense that%
\begin{equation}
\left\langle f\left(  \Omega_{1}\right)  \Gamma^{\dagger}v:\ f\in C_{c}\left(
\mathbb{R}\right)  ,\text{ }v\in H_{0}\right\rangle \label{gaga2}%
\end{equation}
is dense in $H_{1}$ -- then the triplet $\left\{  H_{1},\Omega_{1}%
,\Gamma\right\}  $ is determined uniquely up to an isomorphism.

\begin{remark}
Theorem \ref{tOperatorNev2} and operator generalizations of the
full Herglotz-Nevanlinna Theorem \ref{tNev1} are certainly known
to experts, and have seen application, for example, in the theory
of self-adjoint extensions of symmetric operators -- see, e.g.,
\cite{GeTs}.
\end{remark}
\end{theorem}

\begin{remark}
In the scalar case, the function $g\left(  \zeta\right)  $ in (\ref{hg5}) may
be expressed in terms of the resolvent of the self-adjoint operator
$\Omega_{1}\phi\left(  \sigma\right)  =\sigma\phi\left(  \sigma\right)  $ on
$L^{2}\left(  dN\right)  $,%
\begin{equation}
g\left(  \zeta\right)  =\Gamma\left[  \Omega_{1}-\zeta I_{H_{1}}\right]
^{-1}\Gamma^{\dagger},\
\end{equation}
with%
\begin{equation}
\Gamma\psi=\int\psi\left(  \sigma\right)  dN\left(  \sigma\right)  ,\text{
}\Gamma:L^{2}(\mathbb{R},dN)\rightarrow\mathbb{C}%
\end{equation}
the rank one operator as in the discussion of Bochner's Theorem above. \ Thus
Theorem \ref{tNev2} may be seen as a special case of Theorem
\ref{tOperatorNev2}. (Although the latter is in fact a consequence of the
former.)
\end{remark}

For systems with instantaneous friction ($\alpha_{\infty}\neq0$), we shall make
use of the following result, which is derived from Theorem \ref{tNev1} and the
Naimark representation (Theorem \ref{tNaim1}) in the same way as Theorem
\ref{tOperatorNev2}.\ We discuss the proof of this result, and related
constructions, in Section 7 below.

\begin{theorem}
\label{tGenOperatorNev2} Let $G\left(  \zeta\right)  $ be a $\mathcal{B}%
\left(  H_{0}\right)  $-valued analytic function of the upper half plane
$\operatorname{Im}\zeta>0$ with $\operatorname{Im}G\left(  \zeta\right)
\geq0$ everywhere. If%
\begin{equation}
\lim_{R\rightarrow\infty}G\left(  \mathrm{i}R\right)  v=\mathrm{i}g_{\infty}v
\label{Hh1}%
\end{equation}
exists for all $v\in H_{0}$ with $g_{\infty}\geq0$ a non-negative operator,
then $G(\zeta)$ can be expressed in the form%
\begin{equation}
G(\zeta)v=\lim_{R\rightarrow\infty}\Gamma\left[  \Omega_{1}-\zeta I_{H_{1}%
}\right]  ^{-1}\left(  \Gamma\frac{R^{2}}{\Omega_{1}^{2}+R^{2}I_{H_{1}}%
}\right)  ^{\dagger}v,\text{ for all }v\in H_{1} \label{Hh2}%
\end{equation}
with $\Omega_{1}$ a self-adjoint operator on $H_{1}$, and $\Gamma
:\mathcal{D}\left(  \Omega_{1}\right)  \rightarrow H_{0}$ a $\Omega_{1}%
$-bounded linear map. If the space $H_{1}$ is minimal -- in the sense that%
\begin{equation}
\left\langle \left(  \Gamma f\left(  \Omega_{1}\right)  \right)  ^{\dagger
}v:\ f\in C_{c}\left(  \mathbb{R}\right)  ,\text{ }v\in H_{0}\right\rangle
\end{equation}
is dense in $H_{1}$ -- then the triplet $\left\{  H_{1},\Omega_{1}%
,\Gamma\right\}  $ is determined uniquely up to an isomorphism.
\end{theorem}

For a friction function $a\left(  t\right)  $ satisfying Condition
\ref{condfric} and, consequently, $\hat{a}\left(  \zeta\right)  $
satisfying (\ref{aa1}) the function $G\left(  \zeta\right)
=\mathrm{i}\hat{a}\left( \zeta\right)  $ satisfies the hypotheses
of Theorem \ref{tGenOperatorNev2} with $g_{\infty}=$
$\alpha_{\infty}$. Hence, we can apply Theorem
\ref{tGenOperatorNev2} to obtain operators $\Gamma$ and
$\Omega_{1}$ which provide a basis for a conservative extension
via the abstract model of the previous section. The restriction
that $g_{\infty}$ be self-adjoint is natural in this context,
since any anti-Hermitian contribution to $\lim_{R\rightarrow
\infty}\hat{a}\left(  \mathrm{i}R\right)  $ can be absorbed in
$-\mathrm{i}A$. In the present context, the operator $G$ appearing
in Prop. \ref{pSelfAdjoint} is $G=\hat{a}\left(  \mathrm{i}\right)
$, which is in fact bounded.

\subsection{Spectral representation of\ the admittance operator}

The admittance operator $\mathfrak{A}\left(  \zeta\right)
=\mathrm{i}\left[ \zeta m-A+\mathrm{i}\hat{a}\left(  \zeta\right)
\right]  ^{-1}$ associated to the dispersive system (\ref{ax3}) is
an analytic function on the domain $\left\{
\operatorname{Im}\zeta>0\right\}  $, and the power dissipation
condition takes the form $\operatorname{Re}\mathfrak{A}\left(
\zeta\right) \geq0$ when expressed in terms of $\mathfrak{A}$ (see
eq. (\ref{MAT3a}).)
Furthermore, under condition \ref{condgsp}, we see that%
\begin{equation}
\lim_{\eta\rightarrow\infty}\eta\mathfrak{A}\left(  \mathrm{i}\eta\right)
=m^{-1}.
\end{equation}
Thus $\mathrm{i}\mathfrak{A}\left(  \zeta\right)  $ satisfies the conditions
of Theorem \ref{tOperatorNev2} and therefore has the following representation%
\begin{equation}
\mathfrak{A}\left(  \zeta\right)  =\mathrm{i}\Gamma\left(  \zeta
-\Omega\right)  ^{-1}\Gamma^{\dag},\text{ } \label{at2}%
\end{equation}
with $\Omega$ self-adjoint on a Hilbert space $\mathcal{H}$ and $\Gamma
:\mathcal{H}\rightarrow H_{0}$ bounded. In this section we develop a basis for
a slightly different representation of the admittance operator, namely%
\begin{multline}
\mathfrak{A}\left(  \zeta\right)  =\mathrm{i}T\left(  \zeta\mathcal{M}%
-\mathcal{A}\right)  ^{-1}T^{\dag} \label{at1} \\ \text{ with
}T:\mathcal{H}\rightarrow H_{0}\text{ an isometric truncation.} %
\end{multline}
In fact, eq. (\ref{at1}) is a consequence of eq. (\ref{at2}) and the polar
decomposition $\Gamma=T\mathcal{M}^{-\frac{1}{2}}$where $T:\mathcal{H}%
\rightarrow H_{0}$ is an isometric truncation and $0<\delta\leq\mathcal{M}$.
\ The relationship between eq. (\ref{at1}) and eq. (\ref{at2}) is expressed
through the identity $\mathcal{A}=\mathcal{M}^{\frac{1}{2}}\Omega
\mathcal{M}^{\frac{1}{2}}$. However, there is some flexibility in the choice
of $\mathcal{M}$, which we discuss below. Since $\mathcal{M}^{\frac{1}{2}%
}\Omega\mathcal{M}^{\frac{1}{2}}$ may not be well defined for unbounded
$\mathcal{M}$, we shall require throughout that $\mathcal{M}$ be bounded.

Let us first fix some notation. Given two Hilbert spaces $\mathcal{H},H_{0}$
and a bounded linear operator $L:\mathcal{H}\rightarrow H_{0}$ we denote%
\begin{equation}
\operatorname*{Ker}L=\left\{  V\in\mathcal{H}:LV=0\right\}
,\ \operatorname*{Ran}L=\left\{  LV:V\in\mathcal{H}\right\}  .
\end{equation}
We denote the closure of a subset $S$ in a Hilbert space by
$\overline{S}$, and the restriction of $L$ to $S$ by $\left.
L\right|  _{S}$. We need the following elementary facts
\begin{equation}
\operatorname*{Ker}L=\left[  \operatorname*{Ran}L^{\dag}\right]  ^{\bot
},\ \overline{\operatorname*{Ran}L}=\left[  \operatorname*{Ker}L^{\dag
}\right]  ^{\bot}, \label{at3}%
\end{equation}
where $\left[  \cdot\right]  ^{\bot}$ denotes the orthogonal complement in the
relevant Hilbert space. We refer to the orthogonal direct sum decomposition
\begin{equation}
\mathcal{H}=\overline{\operatorname*{Ran}L^{\dag}}\oplus\operatorname*{Ker}%
L\text{ where }\overline{\operatorname*{Ran}L^{\dag}}=\left[
\operatorname*{Ker}L\right]  ^{\bot}, \label{hh1}%
\end{equation}
as the $L$-decomposition of the Hilbert space $\mathcal{H}$.

Theorem \ref{tOperatorNev2} readily implies the following statements regarding
the decomposition associated to an operator function $G\left(  \zeta\right)  $
of the type considered there.

\begin{corollary}
\label{cred}Let $G(\zeta):H_{0}\rightarrow H_{0}$ satisfy all the conditions
of Theorem \ref{tOperatorNev2}, so $G\left(  \zeta\right)  =\Gamma\left(
\Omega_{1}-\zeta I_{\mathcal{H}}\right)  ^{-1}\Gamma^{\dagger}$. Then we have
\begin{equation}
\operatorname*{Ker}G(\zeta)=\operatorname*{Ker}G^{\dagger}(\zeta
)=\operatorname*{Ker}\Gamma^{\dagger}\text{ for all }\operatorname{Im}\zeta>0.
\label{gg1}%
\end{equation}
Furthermore, the $\Gamma^{\dagger}$-decomposition of the Hilbert space $H_{0}%
$, i.e.%
\begin{equation}
H_{0}=\tilde{H}_{0}\oplus\tilde{H}_{0}^{\bot},\ \tilde{H}_{0}=\left[
\operatorname*{Ker}\Gamma^{\dagger}\right]  ^{\bot}=\overline
{\operatorname*{Ran}\Gamma} \label{gg2}%
\end{equation}
reduces $G(\zeta)$ in the sense that $G(\zeta)$ has the following block form
under (\ref{gg2}):%
\begin{multline}
G(\zeta) =G_{\tilde{H}_{0}}(\zeta)\oplus0=\left[
\begin{array}
[c]{cc}%
G_{\tilde{H}_{0}}(\zeta) & 0\\
0 & 0
\end{array}
\right]  \label{gg3},\\ \text{ where }
G_{\tilde{H}_{0}}(\zeta)  =\tilde{\Gamma}\left[  \Omega_{1}-\zeta I_{H_{1}%
}\right]  ^{-1}\tilde{\Gamma}^{\dagger}:\tilde{H}_{0}\rightarrow\tilde{H}%
_{0},\ \tilde{\Gamma}=P_{\tilde{H}_{0}}\Gamma.
\end{multline}
In addition%
\begin{equation}
\operatorname*{Ker}G_{\tilde{H}_{0}}(\zeta)=\operatorname*{Ker}\tilde{\Gamma
}^{\dagger}=\left\{  0\right\}  ,\ \overline{\operatorname*{Ran}\tilde{\Gamma
}}=\tilde{H}_{0},\ \operatorname{Im}\zeta>0. \label{gg4}%
\end{equation}
\end{corollary}

Corollary \ref{cred} allows us to extract from the operator function
$G(\zeta)$ its nontrivial component and motivates the following definition.

\begin{definition}
\label{dred}For any function $G(\zeta):H_{0}\rightarrow H_{0}$ satisfying all
the conditions of Theorem \ref{tOperatorNev2} the pair $\left\{  \tilde{H}%
_{0},G_{\tilde{H}_{0}}(\zeta)\right\}  $ defined in Corollary \ref{cred} is
called its \emph{reduced representation.}
\end{definition}

\noindent For the representation eq. (\ref{at1}) to hold we shall of course
require that the function $G\left(  \zeta\right)  =-\mathrm{i}\mathfrak{A}%
\left(  \zeta\right)  $ is in reduced form. However, to obtain a bounded mass
operator, we must strengthen this requirement by assuming there is
$\varepsilon>0$ such that%
\begin{equation}
\operatorname{Im}G\left(  \mathrm{i}\eta\right)  \geq\frac{\varepsilon}{\eta
},\text{ }\eta>1, \label{gg5}%
\end{equation}
which implies that $\Gamma\Gamma^{\dagger}\geq\varepsilon$. Under this
additional assumption, we have the following theorem, which provides a basis
for a spectral representation of the admittance operator.

\begin{theorem}
\label{tadmittance1}Let $G\left(  \zeta\right)  $ be a $\mathcal{B}\left(
H_{0}\right)  $-valued analytic function of the upper half plane
$\operatorname{Im}\zeta>0$ with $\operatorname{Im}G\left(  \zeta\right)
\geq0$ for every $\zeta$ and assume that $G\left(  \zeta\right)  $ obeys the
growth conditions eq. (\ref{gg5}) and%
\begin{equation}
\limsup_{\eta\rightarrow+\infty}\eta\left\|  G\left(  \mathrm{i}\eta\right)
\right\|  <\infty. \label{gh2}%
\end{equation}
Then there exists a Hilbert space $\mathcal{H}$ such that%
\begin{multline}
G(\zeta)=\Gamma\left[  \Omega-\zeta I_{\mathcal{H}}\right]  ^{-1}%
\Gamma^{\dagger},\ \text{where }\Omega\text{ is self-adjoint in }%
\mathcal{H}\text{, }\label{gh4}\\
\Gamma:\mathcal{H}\rightarrow H_{0}\text{ is bounded, and
}\Gamma\Gamma^{\dag }=\lim_{\eta\rightarrow+\infty}\eta G\left(
\mathrm{i}\eta\right) \geq\varepsilon.
\end{multline}
Furthermore, if $T$ denotes the isometric truncation $T:=\left[  \Gamma
\Gamma^{\dag}\right]  ^{-\frac{1}{2}}\Gamma$, then $T=UP_{H_{0}^{\prime}}$
with $U=\left.  T\right|  _{H_{0}^{\prime}}$ a unitary map from $H_{0}%
^{\prime}=\left[  \operatorname*{Ker}\Gamma\right]  ^{\bot}$ to $H_{0}$, and
the following representation holds:
\begin{equation}
G(\zeta)=T\left[  \mathcal{A}_{\mathcal{M}}-\zeta\mathcal{M}\right]
^{-1}T^{\dagger}\text{ , } \label{gh8}%
\end{equation}
where $\mathcal{A}_{\mathcal{M}}=\mathcal{M}^{\frac{1}{2}}\Omega
\mathcal{M}^{\frac{1}{2}}$ is self adjoint in $\mathcal{H}$ and $\mathcal{M}%
:\mathcal{H}\rightarrow\mathcal{H}$ $\ $is any operator of the form
\begin{equation}
\mathcal{M}=m_{G}\oplus m_{1}, \label{gh7}%
\end{equation}
with $m_{1}:\operatorname*{Ker}\Gamma\rightarrow\operatorname*{Ker}\Gamma$
bounded and strictly positive ($m_{1}\geq\delta I_{\operatorname*{Ker}\Gamma}%
$, $\delta>0$) and
\begin{equation}
m_{G}=U^{-1}\left[  \Gamma\Gamma^{\dag}\right]  ^{-\frac{1}{2}}U.
\end{equation}
Note that $\mathcal{M}$ is bounded and strictly positive since $m_{G}%
\geq\left\|  \Gamma\Gamma^{\dag}\right\|  ^{-\frac{1}{2}}I_{H_{0}^{\prime}}$
and $\left\|  m_{G}\right\|  \leq\frac{1}{\varepsilon}$. If the space
$\mathcal{H}$ is minimal -- in the sense that%
\begin{equation}
\left\langle f\left(  \Omega\right)  \Gamma^{\dag}v:\ f\in C_{c}\left(
\mathbb{R}\right)  ,\text{ }v\in H_{0}\right\rangle \label{gh9}%
\end{equation}
is dense in $\mathcal{H}$ -- then the triplet $\left\{  \mathcal{H}%
,\mathcal{M},\mathcal{A}\right\}  $ is determined uniquely up to an
isomorphism and the choice of $m_{1}$.
\end{theorem}

As the reader may easily verify, Theorem \ref{tadmittance1} follows from
Theorem \ref{tOperatorNev2} and the \emph{polar decomposition }(see, e.g.,
ref. \cite[Section VI.7]{Kato}), summarized here:

\begin{theorem}
\label{tpolar2}Let $\Gamma:$ $\mathcal{H}\rightarrow H_{0}$ be a bounded
linear operator with $\operatorname*{Ker}\Gamma^{\dag}=\left\{  0\right\}  $
and let%
\begin{equation}
\mathcal{H}=H_{0}^{\prime}\oplus H_{1},\ H_{0}^{\prime}=\left[
\operatorname*{Ker}\Gamma\right]  ^{\bot},\
H_{1}=\operatorname*{Ker}\Gamma
\end{equation}
be the $\Gamma$-decomposition of $\mathcal{H}$. Then%
\begin{equation}
\Gamma V\neq0\text{ for any nonzero }V\in H_{0}^{\prime}\text{ and }%
\overline{\Gamma H_{0}^{\prime}}=H_{0},
\end{equation}%
\begin{equation}
\Gamma^{\dag}\Gamma=KP_{H_{0}^{\prime}}\text{ where }K=\left.  \Gamma^{\dag
}\Gamma\right|  _{H_{0}^{\prime}}\text{ }>0\text{,}%
\end{equation}
and the following ``polar decomposition'' holds%
\begin{equation}
\Gamma=UK^{\frac{1}{2}}P_{H_{0}^{\prime}}\text{ where }U:H_{0}^{\prime
}\rightarrow H_{0}\text{ is unitary,} \label{hh3}%
\end{equation}
indicating, in particular, that $H_{0}^{\prime}$ is an isometric copy of
$H_{0}$. In addition,
\begin{align}
\Gamma\Gamma^{\dag}  &  =UKU^{-1},\ K=\left.  \Gamma^{\dag}\Gamma\right|
_{H_{0}^{\prime}}=U^{-1}\Gamma\Gamma^{\dag}U,\label{hh3a}\\
\Gamma &  =\left[  \Gamma\Gamma^{\dag}\right]  ^{\frac{1}{2}}UP_{H_{0}%
^{\prime}}, \label{hh3b}%
\end{align}
and $\Gamma$ is an isometric truncation if and only if $K=I_{H_{0}^{\prime}}$,
in which case%
\begin{equation}
\Gamma=UP_{H_{0}^{\prime}}\text{ .} \label{hh2}%
\end{equation}
Furthermore, let $\mathcal{K}$ be any operator in $\mathcal{H}$ of the form%
\begin{equation}
\mathcal{K}=K\oplus K_{1}\text{ with }K_{1}:H_{1}\rightarrow H_{1}\text{
non-negative.} \label{hh4}%
\end{equation}
Then the following ``generalized polar decomposition'' holds%
\begin{equation}
\Gamma=T\mathcal{K}^{\frac{1}{2}}=T\left[  K^{\frac{1}{2}}\oplus
K_{1}^{\frac{1}{2}}\right]  \text{ ,} \label{hh5}%
\end{equation}
where $T=UP_{H_{0}^{\prime}}:\mathcal{H}\rightarrow H_{0}$ is an isometric
truncation. Notice that $T$ and $K$ are uniquely determined by $\Gamma$, but
$\mathcal{K}$ depends on the choice of $K_{1}$.
\end{theorem}

\section{Final schemes of the construction of the conservative extension.}

In this section we summarize in a concise form the main consequences for time
dispersive systems of the analysis of preceding sections. There are two
intimately related ways to construct a conservative extension of a dispersive
system. The first, the ``friction function scheme,'' is based on the
operator-valued friction function $a\left(  t\right)  $ or its Laplace
transform $\hat{a}\left(  \zeta\right)  $, whereas the second, the
``admittance operator scheme,'' is based on the admittance operator
$\mathfrak{A}\left(  \zeta\right)  $.

\subsection{Friction function scheme}

The original time dispersive system is described by an evolution equation%
\begin{equation}
m\partial_{t}v\left(  t\right)  =-\mathrm{i}Av\left(  t\right)  -\int
_{0}^{\infty}a\left(  \tau\right)  v\left(  t-\tau\right)  \,d\tau+f\left(
t\right)  ,\ v\left(  t\right)  \in H_{0}, \label{vv1}%
\end{equation}
with $H_{0}$ the space of system states $v$ describing \textquotedblleft
observable\textquotedblright\ variables, $A$ a self-adjoint operator in
$H_{0}$ describing the internal dynamics, $a\left(  \tau\right)
:H_{0}\rightarrow H_{0}$ an operator-valued friction function accounting for
time dispersion and losses, and $f\left(  t\right)  \in H_{0}$ a time
dependent external force. Based on the notion that the friction function
$a\left(  \tau\right)  $ is a result of a coupling between the
\textquotedblleft observable\textquotedblright\ variables $v\in H_{0}$ and
some \textquotedblleft hidden\textquotedblright\ variables described by a
vector $w$ belonging to a Hilbert space $H_{1}$, \emph{we seek a conservative
(not time dispersive) extension of eq. (\ref{vv1}) in the form}
\begin{align}
m\partial_{t}v\left(  t\right)   &  =-\mathrm{i}Av\left(  t\right)
-\mathrm{i}\Gamma w\left(  t\right)  +f\left(  t\right)  ,\label{vv2}\\
\partial_{t}w\left(  t\right)   &  =-\mathrm{i}\Gamma^{\dagger}v\left(
t\right)  -\mathrm{i}\Omega_{1}w\left(  t\right)  , \label{vv3}%
\end{align}
where $\Omega_{1}$ is a self-adjoint operator in $H_{1}$ describing the
internal dynamics of the \textquotedblleft hidden\textquotedblright\ variables
and $\Gamma:H_{1}\rightarrow H_{0}$ is a coupling operator between the
\textquotedblleft hidden\textquotedblright\ and \textquotedblleft
observable\textquotedblright\ variables. The system (\ref{vv2})-(\ref{vv3}),
of course, is expected to reduce to the original system (\ref{vv1}) after the
\textquotedblleft hidden\textquotedblright\ variables $w\left(  t\right)  $
are eliminated by solving eq. (\ref{vv3}) and plugging the result into\ eq.
(\ref{vv2}).

Now, the question is how to construct the conservative extension of the form
(\ref{vv2})-(\ref{vv3}) based on the original dispersive system (\ref{vv1})?
As we have seen, a necessary and sufficient condition for such an extension to
exist is the \emph{power dissipation condition}%
\begin{equation}
\int_{-\infty}^{\infty}\int_{-\infty}^{\infty}\left(  v\left(  t\right)
,a_{e}\left(  t-\tau\right)  v\left(  \tau\right)  \right)  \,dtd\tau
\geq0,\text{ for every }v(t). \label{vv7}%
\end{equation}
A first possibility is to obtain the triplet $\left\{  H_{1},\Gamma,\Omega
_{1}\right\}  $ from the function $a\left(  t\right)  $ via the operator
version of the Bochner Theorem \ref{tOperatorBoch}, which represents $a\left(
t\right)  $ as%
\begin{equation}
a\left(  t\right)  =\Gamma e^{-\mathrm{i}t\Omega_{1}}\Gamma^{\dagger}.
\end{equation}
However, in practice most systems are specified in the frequency domain, and
it is more convenient to carry out the construction in that setting.

Taking the Fourier-Laplace transforms of (\ref{vv1}) and (\ref{vv2}%
)-(\ref{vv3}) with respect to $t$ yields%
\begin{equation}
m\zeta\hat{v}\left(  \zeta\right)  =\left[  A-\mathrm{i}\hat{a}\left(
\zeta\right)  \right]  \hat{v}\left(  \zeta\right)  +\mathrm{i}\hat{f}\left(
\zeta\right)  ,\ \zeta=\omega+\mathrm{i}\eta,\ \eta=\operatorname{Im}\zeta>0,
\label{vv4}%
\end{equation}
and the conservative system, recast in block matrix form,%
\begin{gather}
\zeta\mathcal{M}\hat{V}\left(  \zeta\right)  =\mathcal{A}\hat{V}\left(
\zeta\right)  +\mathrm{i}\hat{F}\left(  \zeta\right)  ,\ \zeta=\omega
+\mathrm{i}\eta,\ \eta=\operatorname{Im}\zeta>0,\label{vv5}\\
\hat{V}\left(  \zeta\right)  =\left[
\begin{array}
[c]{c}%
\hat{v}\left(  \zeta\right) \\
\hat{w}\left(  \zeta\right)
\end{array}
\right]  ,\ \hat{F}\left(  \zeta\right)  =\left[
\begin{array}
[c]{c}%
\hat{f}\left(  \zeta\right) \\
0
\end{array}
\right]  ,\nonumber\\
\mathcal{A}=\left[
\begin{array}
[c]{cc}%
A & \Gamma\\
\Gamma^{\dagger} & \Omega_{1}%
\end{array}
\right]  ,\ \mathcal{M}=\left[
\begin{array}
[c]{cc}%
m & 0\\
0 & I_{H_{1}}%
\end{array}
\right]  ,\ \hat{v}\left(  \zeta\right)  =T\hat{V}\left(  \zeta\right)
,\ T=\left[
\begin{array}
[c]{cc}%
I_{H_{0}} & 0
\end{array}
\right]  .\nonumber
\end{gather}
Note that the operator $T$ defined in (\ref{vv5}) is an isometric
truncation of the extended space $H_{0}\oplus H_{1}$ onto the
space of observable variables $H_{0}$. After the Laplace
transform, the power dissipation
condition (\ref{vv7}) becomes%
\begin{equation}
\operatorname{Re}\hat{a}\left(  \zeta\right)  \geq0\text{ for }%
\operatorname{Im}\zeta>0, \label{vv8}%
\end{equation}
which is equivalent to (\ref{vv7}).

The scheme of the construction is as follows. Given an
operator-valued friction function $\hat{a}\left(  \zeta\right)  $
which satisfies the power dissipation condition (\ref{vv8}), we
apply the operator version of the Nevanlinna Theorem formulated in
Theorems \ref{tOperatorNev2} and \ref{tGenOperatorNev2} to
construct a triplet $\left\{  H_{1},\Gamma
,\Omega_{1}\right\}  $ giving the representation%
\begin{align}
\hat{a}\left(  \zeta\right)   &  =\mathrm{i}\Gamma\left(  \zeta I_{H_{1}%
}-\Omega_{1}\right)  ^{-1}\Gamma^{\dagger},\ \label{av2}\\
\Gamma &  :H_{1}\rightarrow H_{0},\ \Omega_{1}:H_{1}\rightarrow H_{1}\text{ is
self-adjoint,}\nonumber
\end{align}
if $\left\|  \hat{a}\left(  \mathrm{i}\eta\right)  \right\|  =O(\eta^{-1})$,
or
\begin{equation}
\hat{a}\left(  \zeta\right)  =\operatorname*{stlim}_{R\rightarrow\infty
}\mathrm{i}\Gamma\left(  \zeta I_{H_{1}}-\Omega_{1}\right)  ^{-1}\left(
\Gamma\frac{R^{2}}{\Omega_{1}^{2}+R^{2}I_{H_{1}}}\right)  ^{\dagger},
\label{av2aa}%
\end{equation}
if $\operatorname*{stlim}_{\eta\rightarrow\infty}\hat{a}\left(  \mathrm{i}%
\eta\right)  $ exists.

In more detail, assuming\ $\left\|  \hat{a}\left(
\mathrm{i}\eta\right) \right\|  =O(\eta^{-1})$, we first obtain
the operator version of the Nevanlinna representation
\begin{equation}
\hat{a}\left(  \zeta\right)  =\mathrm{i}\int_{-\infty}^{\infty}\frac{1}%
{\zeta-\sigma}dK\left(  \sigma\right)  , \label{av2a}%
\end{equation}
where $dK\left(  \sigma\right)  :H_{0}\rightarrow H_{0}$ is an non-negative
operator-valued measure over $\mathbb{R}$. Notice that $dK\left(
\sigma\right)  $ is not necessarily a resolution of identity, and for
different intervals $\Delta_{1}$ and $\Delta_{2}$ in $\mathbb{R}$ the
operators $K\left(  \Delta_{1}\right)  \geq0$ and $K\left(  \Delta_{2}\right)
\geq0$ may not commute. However, having found $dK\left(  \sigma\right)  $ we
apply the Naimark Theorem \ref{tNaim1} and get (i) a Hilbert space $H_{1}$;
(ii) a resolution of identity $dE\left(  \sigma\right)  $; (iii) an operator
$\Gamma:H_{1}\rightarrow H_{0}$ such that%
\begin{equation}
dK\left(  \sigma\right)  =\Gamma dE\left(  \sigma\right)  \Gamma^{\dagger
}.\ \label{av2b}%
\end{equation}
To complete the construction of the triplet $\left\{
H_{1},\Gamma,\Omega _{1}\right\}  $, we define \begin{equation}
\Omega_{1}:=\int_{-\infty}^{\infty}\sigma dE\left(  \sigma\right)
.
\end{equation}
\emph{It is when applying the Naimark Theorem \ref{tNaim1} that we get the
desired triplet }$\left\{  H_{1},\Gamma,\Omega_{1}\right\}  $, \emph{which is
the central point of the construction of a conservative extension for a time
dispersive system satisfying the power dissipation condition (\ref{vv7}),
(\ref{vv8}). }

We may present a more constructive picture of the triplet the
triplet $\left\{  H_{1},\Gamma,\Omega_{1}\right\}  $ under the
additional assumptions
that%
\begin{equation}
\ \operatorname{Re}\hat{a}\left(  \sigma+\mathrm{i}0\right)
=\operatorname*{stlim}_{\eta\rightarrow0}\operatorname{Re}\hat{a}\left(
\sigma+\mathrm{i}\eta\right)  \label{as1}%
\end{equation}
exists for almost every $\sigma$, and
\begin{equation}
\lim_{\eta\rightarrow0}\int_{I}\left\|  \operatorname{Re}\hat{a}\left(
\sigma+\mathrm{i}\eta\right)  -\operatorname{Re}\hat{a}\left(  \sigma
+\mathrm{i}0\right)  \right\|  d\sigma=0 \label{as2}%
\end{equation}
for any finite interval $I$. With these assumptions, the Stieltjes-Inversion
formula\ -- discussed in Appendix A.1 below -- provides the following explicit
formula for $dK$:%
\begin{equation}
dK\left(  \sigma\right)  =\frac{1}{\pi}\operatorname{Re}\hat{a}\left(
\sigma+\mathrm{i}0\right)  d\sigma. \label{av2c}%
\end{equation}
Let us define $N\left(  \sigma\right)  =\pi^{-1}\operatorname{Re}%
\hat{a}\left(  \sigma+\mathrm{i}0\right)  $, and note that $N\left(
\sigma\right)  $ is a non-negative operator for almost every $\sigma$. We
choose the Hilbert space $H_{1}=L^{2}\left(  \mathbb{R},H_{0}\right)  $, which
is the space of square integrable functions from $\mathbb{R}$ to $H_{0}$ with
inner product%
\begin{equation}
\left\langle w_{1},w_{2}\right\rangle _{H_{1}}=\int_{-\infty}^{\infty
}\left\langle w_{1}\left(  \sigma\right)  ,w_{2}\left(  \sigma\right)
\right\rangle _{H_{0}}d\sigma.
\end{equation}
Choosing for $\Omega_{1}$ the operator
\begin{equation}
\Omega_{1}w\left(  \sigma\right)  =\sigma w\left(  \sigma\right)  ,
\end{equation}
and for $\Gamma$ the map%
\begin{equation}
\Gamma w=\int_{\mathbb{-\infty}}^{\infty}\sqrt{N\left(  \sigma\right)
}w\left(  \sigma\right)  d\sigma,
\end{equation}
it is easy to see that%
\begin{equation}
\hat{a}\left(  \zeta\right)  =\mathrm{i}\int_{-\infty}^{\infty}\frac{1}%
{\zeta-\sigma}N\left(  \sigma\right)  d\sigma=\mathrm{i}\Gamma\frac{1}%
{\zeta-\Omega_{1}}\Gamma^{\dagger}, \label{av2d}%
\end{equation}
which is the desired representation.

The above representation on $H_{1}=L^{2}\left(  \mathbb{R},H_{0}\right)  $ may
not be minimal if, for $\sigma$ from a set of positive measure, $N\left(
\sigma\right)  $ has a non-trivial kernel. In that case it is more useful to
consider the Hilbert space $H_{1}=L^{2}\left(  dK\right)  $ defined to be the
space%
\begin{equation}
\left\{  w:\mathbb{R}\rightarrow H_{0}\text{ measurable }\left|  \int
_{-\infty}^{\infty}\left\langle w\left(  \sigma\right)  ,N\left(
\sigma\right)  w\left(  \sigma\right)  \right\rangle d\sigma<\infty\right.
\right\}  ,
\end{equation}
modulo null functions with $N\left(  \sigma\right)  w\left(  \sigma\right)
=0$ for almost every $\sigma$. \ In fact the proof of the Naimark theorem
(without assumptions (\ref{as1})-(\ref{as2})) proceeds by constructing the
space $L^{2}\left(  dK\right)  $. \ A sketch of this construction is given in
Appendix A.2.

Finally, having found the triplet $\left\{
H_{1},\Gamma,\Omega_{1}\right\} $, we construct the conservative
extension as the system of equations (\ref{vv2})-(\ref{vv3}).
Using the matrix form (\ref{vv5})\ of the conservative system
(\ref{vv2})-(\ref{vv3}) we also get the following representation
for
the admittance operator%
\begin{equation}
\mathfrak{A}\left(  \zeta\right)  =\mathrm{i}T\left(  \zeta\mathcal{M}%
-\mathcal{A}\right)  ^{-1}T^{\dagger}=\left\{  \zeta m-\left[  A-\mathrm{i}%
\hat{a}\left(  \zeta\right)  \right]  \right\}  ^{-1}. \label{av4}%
\end{equation}
\emph{The above construction of the triplet }$\left\{  H_{1},\Gamma,\Omega
_{1}\right\}  $\emph{, including the Hilbert space }$H_{1}$\emph{ of ``hidden
variables'', is essentially independent of the operator }$A$\emph{ which
describes the internal dynamics of the ``observable'' variables.}

\subsection{Admittance operator scheme}

Suppose the original time dispersive system is given by its
admittance operator $\mathfrak{A}\left(  \zeta\right)  $, acting
in the Hilbert space of \ ``observable'' variables $H_{0}$ by the
following equation
\begin{equation}
\hat{v}\left(  \zeta\right)  =\mathfrak{A}\left(  \zeta\right)  \hat{f}\left(
\zeta\right)  ,\ \zeta=\omega+\mathrm{i}\eta,\ \eta=\operatorname{Im}\zeta>0,
\label{vu1}%
\end{equation}
relating the generalized velocity $\hat{v}\left(  \zeta\right)  $ and the
generalized force $\hat{f}\left(  \zeta\right)  $ in the complex frequency
domain. For real $\zeta=\omega$ the equation (\ref{vu1}) reduces to the
familiar real frequency form%
\begin{equation}
\hat{v}\left(  \omega\right)  =\mathfrak{A}\left(  \omega\right)  \hat
{f}\left(  \omega\right)  . \label{vu2}%
\end{equation}
We assume that the Hilbert space $H_{0}$ and the admittance operator
$\mathfrak{A}\left( \zeta\right)  $ are already reduced in the sense of
Definition \ref{dred}, i.e. $H_{0}=\tilde{H}_{0}$ and
$\mathfrak{A}_{\tilde{H}_{0}}\left( \zeta\right) =\mathfrak{A}\left(
\zeta\right)  $, and that $\mathfrak{A}\left( \zeta\right)  $
satisfies the power dissipation and the growth conditions%
\begin{equation}
\operatorname{Re}\mathfrak{A}\left(  \zeta\right)  \geq0\text{ for
}\operatorname{Im}\zeta>0,\ \limsup_{\eta\rightarrow+\infty}\eta\left\|
\mathfrak{A}\left(  \mathrm{i}\eta\right)  \right\|  <\infty. \label{vu2a}%
\end{equation}

\emph{We seek a conservative, non time-dispersive extension, of
eq. (\ref{vu1}) via the following representation for the
admittance operator}
\begin{equation}
\mathfrak{A}\left(  \zeta\right)  =\mathrm{i}T\left(  \zeta\mathcal{M}%
-\mathcal{A}\right)  ^{-1}T^{\dagger}, \label{vu5}%
\end{equation}
with $T$ an isometric truncation from a Hilbert space
$\mathcal{H}$ to $H_{0}$, $\mathcal{M}\geq\delta I_{\mathcal{H}}$,
$\delta>0$ on $\mathcal{H}$, and $\mathcal{A}$ self-adjoint in
$\mathcal{H}$, corresponding to the following evolution equation
for $v$,
\begin{equation}
\mathcal{M}\dot{V}\left(  t\right)  =-\mathrm{i}\mathcal{A}V\left(  t\right)
+T^{\dagger}f\left(  t\right)  ,\text{ }v\left(  t\right)  =TV\left(
t\right)  . \label{vu4}%
\end{equation}
To construct the representation (\ref{vu5}), we first construct the Hilbert
space $\mathcal{H}$ and the representation
\begin{equation}
\mathfrak{A}\left(  \zeta\right)  =\mathrm{i}\Gamma_{\mathfrak{A}}\left(
\zeta\mathcal{I}-\Omega_{\mathfrak{A}}\right)  ^{-1}\Gamma_{\mathfrak{A}%
}^{\dagger}, \label{vu8}%
\end{equation}
following the argument which led to (\ref{av2}) for the friction function in
the previous section -- applying Theorem \ref{tadmittance1} to $G(\zeta
)=\mathrm{i}\mathfrak{A}\left(  \zeta\right)  $.

Having obtained (\ref{vu8}), we use Theorem \ref{tpolar2} to get the following
polar decomposition%
\begin{gather}
\Gamma_{\mathfrak{A}}=Um_{\mathfrak{A}}^{-\frac{1}{2}}P_{H_{0}^{\prime}}\text{
where }U:H_{0}^{\prime}\rightarrow H_{0}\text{ in unitary,}\label{vu9}\\
m_{\mathfrak{A}}=U^{-1}\left[  \Gamma_{\mathfrak{A}}\Gamma_{\mathfrak{A}%
}^{\dag}\right]  ^{-1}U\geq\left\|  \Gamma_{\mathfrak{A}}\Gamma_{\mathfrak{A}%
}^{\dag}\right\|  ^{-1}I_{H_{0}^{\prime}},\ \Gamma_{\mathfrak{A}}%
\Gamma_{\mathfrak{A}}^{\dag}=\lim_{\eta\rightarrow+\infty}\mathrm{i}%
\eta\mathfrak{A}\left(  \mathrm{i}\eta\right)  >0.\nonumber
\end{gather}
Then we introduce the mass operator
\begin{equation}
\mathcal{M}=m_{\mathfrak{A}}\oplus m_{1}\text{ where }m_{1}:H_{1}\rightarrow
H_{1}\geq\delta I_{H_{1}},\ \delta>0, \label{vu9a}%
\end{equation}
with $m_{1}$ being chosen as we please. \ In view of (\ref{vu9})%
\begin{equation}
\Gamma_{\mathfrak{A}}=UP_{H_{0}^{\prime}}\mathcal{M}^{-\frac{1}{2}}%
=UP_{H_{0}^{\prime}}\left(  m_{\mathfrak{A}}^{-\frac{1}{2}}\oplus
m_{1}^{-\frac{1}{2}}\right)  , \label{vu9b}%
\end{equation}
and we get\ the desired representation
\begin{multline}
\mathfrak{A}\left(  \zeta\right)  =\mathrm{i}T\left(  \zeta\mathcal{M}%
-\mathcal{A}\right)  ^{-1}T^{\dagger}, \\ \text{ where
}T=UP_{H_{0}^{\prime}}\text{
and }\mathcal{A}=\mathcal{M}^{\frac{1}{2}}\Omega_{\mathfrak{A}}\mathcal{M}%
^{\frac{1}{2}}. \label{vu10}%
\end{multline}
Note that the Hilbert space of ``hidden variables'' is $H_{1}=\mathcal{H}%
\circleddash H_{0}^{\prime}$, which is $\ker\Gamma_{\mathfrak{A}}$, i.e.%
\begin{equation}
H_{1}=\left\{  \psi\in\mathcal{H}:\Gamma_{\mathfrak{A}}\psi=0\right\}  .
\label{vu11}%
\end{equation}

\section{Examples of the construction of conservative systems}

In this section we apply the general scheme for the construction of
conservative extensions, as described in Sections 3 and 4, to a few well known
classical systems: a damped oscillator, a general scalar dispersive
dissipative system, and a classical dielectric medium. Related constructions
have appeared elsewhere in the literature, e.g., Lamb's representation of a
damped oscillator as a mass attached to an infinitely long tense string
\cite{Lamb}. Interesting examples and very detailed studies of relations
between admittance operators and spectral measures for loaded strings (as
described by Krein-Feller operators) are offered in the second paper in ref.
\cite{KK}, wherein dispersion is introduced via boundary conditions.

\subsection{Damped oscillator}

We consider a damped oscillator with mass $m_{o}>0$, real frequency
$\Omega_{o}$, and friction coefficient $\gamma_{o}>0$, described by a complex
variable $v$ which evolves according to the following equation%
\begin{equation}
m_{o}\partial_{t}v=-\mathrm{i}m_{o}\Omega_{o}v-\gamma_{o}v+f\left(  t\right)
,\ \label{dvv1}%
\end{equation}
where $f\left(  t\right)  $, the external force, is a complex-valued function.
Evidently, the equation (\ref{dvv1}) is a particular case of a system of the
form (\ref{ax3})
with%
\begin{multline}
H_{0}=\mathbb{C},\ A=m_{o}\Omega_{o},\\ a\left(  t\right)
=a_{o}\left( t\right)  =\gamma_{o}\delta\left(  t\right),  \text{
and }\hat{a}_{o}\left(
\zeta\right)  =\gamma_{o}\text{ for }\operatorname{Im}\zeta\geq0. \label{dvv2}%
\end{multline}

Following the friction function scheme described in Section 3, we look at eq.\
(\ref{dvv1}) as a consequence of a larger conservative system with hidden
degrees of freedom of the form (\ref{vv2})-(\ref{vv3}), and proceed with the
construction of the triplet $\left\{  H_{1},\Omega_{1},\Gamma\right\}  $ based
on Theorem \ref{tGenOperatorNev2}. Notice that $G\left(  \zeta\right)
=\mathrm{i}\hat{a}_{o}\left(  \zeta\right)  =\mathrm{i}\gamma_{o}$ evidently
satisfies the hypothesis of Theorem \ref{tGenOperatorNev2}. Hence, we must
have%
\begin{equation}
\gamma_{o}=\mathrm{i}\lim_{R\rightarrow\infty}\Gamma\frac{1}{\zeta I_{H_{1}%
}-\Omega_{1}}\left(  \Gamma\frac{R^{2}}{\Omega_{1}^{2}+R^{2}I_{H_{1}}}\right)
^{\dagger},\ \operatorname{Im}\zeta\geq0. \label{dvv3}%
\end{equation}

The discussion following eq. (\ref{av2c}) -- in particular eq. (\ref{av2d})
with $N\left(  \sigma\right)  =\gamma_{0}/\pi$ -- suggests that we take
$H_{1}=L^{2}\left(  \mathbb{R}\right)  $,%
\begin{equation}
\Omega_{1}\psi\left(  \sigma\right)  =\sigma\psi\left(  \sigma\right)
,\ \sigma\in\mathbb{R},\ \psi\left(  \sigma\right)  \in L_{2}\left(
\mathbb{R}\right)  , \label{dvv6}%
\end{equation}
and%
\begin{multline}
\Gamma\psi=\sqrt{\frac{\gamma_{o}}{\pi}}\lim_{R\rightarrow\infty}\int_{-R}%
^{R}\psi\left(  \sigma\right)  =\sqrt{\frac{\gamma_{o}}{\pi}}\left\langle
\mathbf{1},\psi\left(  \sigma\right)  \right\rangle ,\\ \text{ where }%
\mathbf{1}=\mathbf{1}\left(  \sigma\right)  =1,\ \sigma\in\mathbb{R}.
\label{dvv7}%
\end{multline}
Although $\Gamma$ is not bounded, we note that $\Gamma$ is an $\Omega_{1}%
$-bounded map from $\mathcal{D}\left(  \Omega_{1}\right)  \rightarrow
\mathbb{C}$, since%
\begin{equation}
\left|  \Gamma\psi\right|  \leq\sqrt{\frac{\gamma_{o}}{\pi}}\lim
_{R\rightarrow\infty}\int_{-R}^{R}\left|  \frac{1}{\sigma+\mathrm{i}}\right|
\left|  \left(  \sigma+\mathrm{i}\right)  \psi\left(  \sigma\right)  \right|
d\sigma\leq\sqrt{\gamma_{o}}\left\|  \left(  \Omega_{1}+\mathrm{i}\right)
\psi\right\|  .
\end{equation}
Formally,%
\begin{equation}
\left(  \Gamma^{\dagger}v\right)  \left(  \sigma\right)  =\sqrt{\frac{\gamma
_{o}}{\pi}}v,\ \sigma\in\mathbb{R}. \label{dvv8}%
\end{equation}

The generator of the dynamics in the conservative system is the self-adjoint
operator $\mathcal{A}$ acting in the Hilbert space $\mathcal{H}=\mathbb{C}%
\oplus L_{2}\left(  \mathbb{R},d\sigma\right)  $, defined by
\begin{equation}
\begin{split}
\mathcal{A}\left[
\begin{array}
[c]{c}%
v\\
\psi\left(  \sigma\right)
\end{array}
\right]   &  =\left[
\begin{array}
[c]{cc}%
m_{o}\Omega_{o} & \sqrt{\frac{\gamma_{o}}{\pi}}\left\langle \mathbf{1}%
,\cdot\right\rangle \\
\sqrt{\frac{\gamma_{o}}{\pi}}\mathbf{1} & \sigma
\end{array}
\right]  \left[
\begin{array}
[c]{c}%
v\\
\psi\left(  \sigma\right)
\end{array}
\right] \\
&  =\left[
\begin{array}
[c]{c}%
m_{o}\Omega_{o}v+\sqrt{\frac{\gamma_{o}}{\pi}}\left\langle \mathbf{1}%
,\psi\left(  \sigma\right)  \right\rangle \\
\sqrt{\frac{\gamma_{o}}{\pi}}v+\sigma\psi\left(  \sigma\right)
\end{array}
\right]  ,\ \sigma\in\mathbb{R},
\end{split}\label{dvv9}
\end{equation}
with the domain%
\begin{equation}
D\left(  \mathcal{A}\right)  =\left\{  \left[
\begin{array}
[c]{c}%
v\\
\psi\left(  \sigma\right)
\end{array}
\right]  ,\ \sigma\in\mathbb{R}:\sqrt{\frac{\gamma_{o}}{\pi}}v+\sigma
\psi\left(  \sigma\right)  \in L_{2}\left(  \mathbb{R},d\sigma\right)
\right\}  . \label{dvv10}%
\end{equation}
One can show that for any vector in $D\left(  \mathcal{A}\right)  $ the
expression $\left\langle \mathbf{1},\psi\left(  \sigma\right)  \right\rangle $
is well defined as the following limit
\begin{equation}
\left\langle \mathbf{1},\psi\left(  \sigma\right)  \right\rangle
:=\lim_{R\rightarrow\infty}\int_{-R}^{R}\psi\left(  \sigma\right)  \,d\sigma,
\label{dvv11}%
\end{equation}
and by Prop. \ref{pSelfAdjoint}, we see that $\mathcal{A}$ is self adjoint.

Finally the mass operator here takes the form%
\begin{equation}
\mathcal{M}=\left[
\begin{array}
[c]{cc}%
m_{o} & 0\\
0 & I_{H_{1}}%
\end{array}
\right]  ,
\end{equation}
and the desired extended conservative system for the damped oscillator is%
\begin{equation}
\begin{split}
m_{o}\partial_{t}v=&-\mathrm{i}m_{o}\Omega_{o}v- \mathrm{i} \sqrt{\frac{\gamma_{o}}{\pi}%
}\int_{-\infty}^{\infty}\psi\left(  \sigma\right)  \,d\sigma+f\left(  t\right)
\\
\partial_{t}\psi\left(  \sigma\right)  =&-\mathrm{i}\sigma\psi\left(  \sigma\right)
+\mathrm{i}\sqrt{\frac{\gamma_{o}}{\pi}}v
\end{split}\label{dvv12}\end{equation}
for $\left(  v,\psi\right)  \in\ \mathcal{H}=\mathbb{C}\oplus
L_{2}\left( \mathbb{R},d\sigma\right)  $.

\subsection{General scalar dispersive dissipative system}

A number of classical dispersive systems, including homogeneous dielectrics
and an oscillator with a retarded friction, can be described by a scalar
complex variable $v$ governed by an evolution equation%
\begin{equation}
m_{\operatorname*{s}}\partial_{t}v\left(  t\right)  =-\mathrm{i}%
m_{\operatorname*{s}}\Omega_{s}v\left(  t\right)  -\int_{0}^{\infty
}a_{\operatorname*{s}}\left(  \tau\right)  v\left(  t-\tau\right)
\,d\tau+f\left(  t\right)  , \label{sds1}%
\end{equation}
with $a_{\operatorname*{s}}\left(  \tau\right)  $ complex-valued
functions, $m_{\operatorname*{s}}$ a positive number and
$\Omega_{s}$ a real number. The function $f$ is the external
force. The scalar friction function $a_{\operatorname*{s}}\left(
t\right)  $ is assumed to satisfy Condition \ref{condfric} and the
power dissipation condition (\ref{ava4}), which is to say it is a
positive definite function as in the classical Bochner's theorem.
The system described (\ref{sds1}) has, according to (\ref{vt2a}),
the
following admittance form%
\begin{multline}
\hat{v}\left(  \zeta\right)  =\mathfrak{A}_{\operatorname*{s}}\left(
\zeta\right)  \hat{f}\left(  \zeta\right)  ,\\ \mathfrak{A}_{\operatorname*{s}%
}\left(  \zeta\right)  =\left\{  m_{\operatorname*{s}}\left(  \zeta-\Omega
_{s}\right)  +\mathrm{i}\hat{a}_{s}\left(  \zeta\right)  \right\}
^{-1},\text{ }\operatorname{Im}\zeta>0. \label{sds1b}%
\end{multline}

Let us consider here a general scalar dispersive dissipative system given in
admittance operator form%
\begin{equation}
\hat{v}\left(  \zeta\right)  =\mathfrak{A}_{\operatorname*{s}}\left(
\zeta\right)  \hat{f}\left(  \zeta\right)  ,\ \zeta=\omega+\mathrm{i}%
\eta,\ \eta=\operatorname{Im}\zeta>0, \label{sds0}%
\end{equation}
where the scalar admittance operator $\mathfrak{A}_{\operatorname*{s}}\left(
\zeta\right)  $ satisfies the power dissipation condition%
\begin{equation}
\operatorname{Re}\mathfrak{A}_{\operatorname*{s}}\left(  \zeta\right)
\geq0,\ \zeta=\omega+\mathrm{i}\eta,\ \eta=\operatorname{Im}\zeta>0,
\label{sds0a}%
\end{equation}
and
\begin{equation}
\lim\sup_{\eta\rightarrow\infty}\eta\left|  \mathfrak{A}_{\operatorname*{s}%
}\left(  \mathrm{i}\eta\right)  \right|  <\infty, \label{sds0b}%
\end{equation}
and $\mathfrak{A}_{\operatorname*{s}}\left(  \zeta\right)  \neq0$ at least for
one $\zeta.$

To find a conservative extension for (\ref{sds0}), we use the admittance
operator scheme from Section 4. In this case $H_{0}=\mathbb{C}$ and we first
seek a Hilbert space $\mathcal{H}$ and a self-adjoint operator
$\Omega_{\mathfrak{A}_{\operatorname*{s}}}$ in it such that
\begin{multline}
\mathfrak{A}_{\operatorname*{s}}\left(  \zeta\right)  =\mathrm{i}%
\Gamma_{\mathfrak{A}_{\operatorname*{s}}}\left(  \zeta I-\Omega_{\mathfrak{A}%
_{\operatorname*{s}}}\right)  ^{-1}\Gamma_{\mathfrak{A}_{\operatorname*{s}}%
}^{\dagger}\\ =-\mathrm{i}\int_{-\infty}^{\infty}\frac{\Gamma_{\mathfrak{A}%
_{\operatorname*{s}}}E_{\Omega_{\mathfrak{A}_{\operatorname*{s}}}}\left(
d\sigma\right)  \Gamma_{\mathfrak{A}_{\operatorname*{s}}}^{\dagger}}%
{\sigma-\zeta},\ \operatorname{Im}\zeta\geq0. \label{sds2}%
\end{multline}
We construct the spectral representation (\ref{sds2}) as follows.
In view of conditions (\ref{sds0a}) and (\ref{sds0b}), the
classical Nevanlinna Theorem
\ref{tNev2} gives a non-negative scalar measure $dN_{\mathfrak{A}%
_{\operatorname*{s}}}\left(  \sigma\right)  $ satisfying%
\begin{equation}
\mathrm{i}\mathfrak{A}_{\operatorname*{s}}\left(  \zeta\right)  =\int
_{-\infty}^{\infty}\frac{dN_{\mathfrak{A}_{\operatorname*{s}}}\left(
\sigma\right)  }{\sigma-\zeta},\ \operatorname{Im}\zeta\geq0,\text{
}dN_{\mathfrak{A}_{\operatorname*{s}}}\left(  \sigma\right)  \geq0.
\label{sds3a}%
\end{equation}
Let us define $m_{\mathfrak{A}_{\operatorname*{s}}}$ by%
\begin{equation}
\frac{1}{m_{\mathfrak{A}_{\operatorname*{s}}}}=\int_{-\infty}^{\infty
}dN_{\mathfrak{A}_{\operatorname*{s}}}\left(  \sigma\right)  =\lim
_{\eta\rightarrow\infty}\eta\mathfrak{A}_{\operatorname*{s}}\left(
\mathrm{i}\eta\right)  , \label{sds3b}%
\end{equation}
and note that if $\mathfrak{A}_{\operatorname*{s}}$ is given by (\ref{sds1b}),
$m_{\mathfrak{A}_{\operatorname*{s}}}=m_{\operatorname*{s}}$.

We set%
\begin{equation}
\mathcal{H}=L_{2}\left(  dN_{\mathfrak{A}_{\operatorname*{s}}}\left(
\sigma\right)  ,\mathbb{C}\right)  ,\ H_{0}^{\prime}=\left\{  \psi
\in\mathcal{H}:\psi\left(  \sigma\right)  =v\in\mathbb{C},\ \sigma
\in\mathbb{R}\right\}  . \label{sds4}%
\end{equation}
In other words,\emph{ }$H_{0}^{\prime}$\emph{ is defined as a set of constant
functions of }$\sigma$,\emph{ }which is, evidently, unitarily equivalent to the
set of complex numbers $\mathcal{C}$ through the following mapping
\begin{equation}
U:H_{0}^{\prime}\rightarrow\mathbb{C},\ U\left(  v\mathbf{1}\right)
=\frac{v}{\sqrt{m_{\mathfrak{A}_{\operatorname*{s}}}}}\text{, where
}\mathbf{1}=\mathbf{1}\left(  \sigma\right)  =1\text{ for }\sigma\in
\mathbb{R}. \label{sds4a}%
\end{equation}
We also define
\begin{multline}
\Omega_{\mathfrak{A}_{\operatorname*{s}}}\psi\left(  \sigma\right)
=\sigma\psi\left(  \sigma\right)  ,\ \psi\in\mathcal{H},\\ \left[
\Gamma_{\mathfrak{A}_{\operatorname*{s}}}\psi\right]  =\left(  \mathbf{1}%
,\psi\right)  =%
{\displaystyle\int}
\psi\left(  \sigma\right)  \,dN_{\mathfrak{A}_{\operatorname*{s}}}\left(
\sigma\right)  \in\mathbb{C}, \label{sds4b}%
\end{multline}
implying%
\begin{equation}
\Gamma_{\mathfrak{A}_{\operatorname*{s}}}^{\dag}v=v\mathbf{1},\ v\in
\mathbb{C}. \label{sds4c}%
\end{equation}

The representation (\ref{sds2}) readily follows from the definitions
(\ref{sds4})-(\ref{sds4c}). Also, from (\ref{sds4b}) and (\ref{sds4c}) it
follows that
\begin{equation}
\Gamma_{\mathfrak{A}_{\operatorname*{s}}}\Gamma_{\mathfrak{A}%
_{\operatorname*{s}}}^{\dag}=\int_{-\infty}^{\infty}dN_{\mathfrak{A}%
_{\operatorname*{s}}}\left(  \sigma\right)  =\frac{1}{m_{\mathfrak{A}%
_{\operatorname*{s}}}}:\mathbb{C}\rightarrow\mathbb{C}. \label{sds4d}%
\end{equation}
Observe that the orthogonal projection $P_{H_{0}^{\prime}}$ is%
\begin{equation}
\left[  P_{H_{0}^{\prime}}\psi\right]  \left(  \sigma\right)  =\left[
m_{\mathfrak{A}_{\operatorname*{s}}}\int_{-\infty}^{\infty}\psi\left(
\sigma\right)  \,dN_{\mathfrak{A}_{\operatorname*{s}}}\left(  \sigma\right)
\right]  \mathbf{1},\text{\ }\sigma\in\mathbb{R}. \label{sds4f}%
\end{equation}
We may choose the mass operator $\mathcal{M}$ to be just the scalar operator
$\mathcal{M}=m_{\mathfrak{A}_{\operatorname*{s}}}I_{\mathcal{H}}$, and,
consequently, get the desired components of an extended conservative system%
\begin{multline}
\mathcal{H}=L_{2}\left(  dN_{\mathfrak{A}_{\operatorname*{s}}}\left(
\sigma\right)  ,\mathbb{C}\right)  ,\ \mathcal{M}=m_{\mathfrak{A}%
_{\operatorname*{s}}}I_{\mathcal{H}},\\ \mathcal{A}\psi\left(
\sigma\right)
=m_{\mathfrak{A}_{\operatorname*{s}}}\sigma\psi\left(
\sigma\right) ,\ T=UP_{H_{0}^{\prime}},
\end{multline}
where $U$ is defined by (\ref{sds4a}).

The measure $dN_{\mathfrak{A}}\left(  \sigma\right)  $ can be recovered from
Stieltjes' formula -- see eq. (\ref{hg8}) below --%
\begin{multline}
\int_{-\infty}^{\infty}f\left(  \sigma\right)  dN_{\mathfrak{A}%
_{\operatorname*{s}}}\left(  \sigma\right)  \\ =\lim_{\eta\rightarrow+0}%
\frac{1}{\pi}\int_{-\infty}^{\infty}f\left(  \sigma\right)  \operatorname{Re}%
\left\{  \mathfrak{A}_{\operatorname*{s}}\left(  \sigma+\mathrm{i}\eta\right)
\right\}  \,d\sigma,\text{ for }f\in C_{0}\left(  \mathbb{R}\right)  .
\end{multline}
In particular, if $dN_{\mathfrak{A}_{\operatorname*{s}}}\left(  \sigma\right)
$ has a density $n_{\mathfrak{A}_{\operatorname*{s}}}\left(  \sigma\right)  $
with respect to the Lebesgue measure, i.e.%
\begin{equation}
dN_{\mathfrak{A}_{\operatorname*{s}}}\left(  \sigma\right)  =n_{\mathfrak{A}%
_{\operatorname*{s}}}\left(  \sigma\right)  \,d\sigma, \label{sds5}%
\end{equation}
then the density $n_{\mathfrak{A}_{\operatorname*{s}}}\left(  \sigma\right)  $
is the following pointwise limit%
\begin{equation}
n_{\mathfrak{A}_{\operatorname*{s}}}\left(  \sigma\right)  =\lim
_{\eta\rightarrow0}\frac{1}{\pi}\operatorname{Re}\left\{  \mathfrak{A}%
_{\operatorname*{s}}\left(  \sigma+\mathrm{i}\eta\right)  \right\}  ,
\label{sds7}%
\end{equation}
which, in view of (\ref{sds3b}), satisfies%
\begin{equation}
n_{\mathfrak{A}_{\operatorname*{s}}}\left(  \sigma\right)  \geq0\text{ and
}\int_{-\infty}^{\infty}n_{\mathfrak{A}_{\operatorname*{s}}}\left(
\sigma\right)  \,d\sigma=\lim_{\eta\rightarrow\infty}\eta\mathfrak{A}%
_{\operatorname*{s}}\left(  \mathrm{i}\eta\right)  =\frac{1}{m_{\mathfrak{A}%
_{\operatorname*{s}}}}. \label{sds7b}%
\end{equation}

When the measure $dN_{\mathfrak{A}_{\operatorname*{s}}}\left(  \sigma\right)
$ has a density $n_{\mathfrak{A}_{\operatorname*{s}}}\left(  \sigma\right)  $,
there is a slightly different description based on the Hilbert space
$\mathcal{H}=L_{2}\left(  \mathbb{R}\right)  $, namely:
\begin{gather}
\Omega_{\mathfrak{A}_{\operatorname*{s}}}\psi\left(  \sigma\right)
=\sigma\psi\left(  \sigma\right)  ,\ \psi\in L_{2}\left(
\mathbb{R}\right);\label{sds7a}\\ H_{0}^{\prime }=\left\{  \psi\in
L_{2}\left( \mathbb{C}\right)  :\psi\left(  \sigma\right)
=v\sqrt{n_{\mathfrak{A}_{\operatorname*{s}}}\left(  \sigma\right)  }%
,\ v\in\mathbb{C}\right\} \\
\left[  \Gamma_{\mathfrak{A}_{\operatorname*{s}}}\psi\right]  \left(
\sigma\right)  =\int_{-\infty}^{\infty}\sqrt{n_{\mathfrak{A}%
_{\operatorname*{s}}}\left(  \sigma\right)  }\psi\left(  \sigma\right)
\,d\sigma,\ \left[  \Gamma_{\mathfrak{A}_{\operatorname*{s}}}^{\dagger
}v\right]  \left(  \sigma\right)  =v\sqrt{n_{\mathfrak{A}_{\operatorname*{s}}%
}\left(  \sigma\right)  }.
\end{gather}
Then we get%
\begin{gather}
U:H_{0}^{\prime}\rightarrow\mathbb{C},\ U\left(  v\sqrt{n_{\mathfrak{A}%
_{\operatorname*{s}}}\left(  \sigma\right)  }\right)  =\frac{v}%
{m_{\mathfrak{A}_{\operatorname*{s}}}},\label{uh1}\\
\left[  P_{H_{0}^{\prime}}\psi\right]  \left(  \sigma\right)
=\left[
m_{\mathfrak{A}_{\operatorname*{s}}}\int_{-\infty}^{\infty}\sqrt
{n_{\mathfrak{A}_{\operatorname*{s}}}\left(  \sigma^{\prime}\right)  }%
\psi\left(  \sigma^{\prime}\right)  \,d\sigma^{\prime}\right]  \sqrt
{n_{\mathfrak{A}_{\operatorname*{s}}}\left(  \sigma\right)  },\text{\ }%
\sigma\in\mathbb{R},
\end{gather}
and, consequently,%
\begin{equation}
\mathcal{H}=L_{2}\left(  \mathbb{C}\right)  ,\ \mathcal{M}=m_{\mathfrak{A}%
_{\operatorname*{s}}}I_{\mathcal{H}},\ \mathcal{A}\psi\left(  \sigma\right)
=m_{\mathfrak{A}_{\operatorname*{s}}}\sigma\psi\left(  \sigma\right)
,\ T=UP_{H_{0}^{\prime}}. \label{uh2}%
\end{equation}

Finally, the evolution of the extended conservative system, as described by
its state $\psi\left(  t,\sigma\right)  $, is governed by the following
equations
\begin{multline}
m_{\mathfrak{A}_{\operatorname*{s}}}\partial_{t}\psi\left(
t,\sigma\right)
\\=-\mathrm{i}m_{\mathfrak{A}_{\operatorname*{s}}}\sigma\psi\left(
t,\sigma\right)  +f\left(  t\right)  \sqrt{m_{\mathfrak{A}_{\operatorname*{s}%
}}n_{\mathfrak{A}_{\operatorname*{s}}}\left(  \sigma\right)  },\ \psi\left(
t,\sigma\right)  \in L_{2}\left(  \mathbb{C}\right)  , \label{sds8}%
\end{multline}
with $n_{\mathfrak{A}_{\operatorname*{s}}}\left(  \sigma\right)  =\lim
_{\eta\rightarrow0}\frac{1}{\pi}\operatorname{Re}\left\{  \mathfrak{A}%
_{\operatorname*{s}}\left(  \sigma+\mathrm{i}\eta\right)  \right\}
\geq0$ and $f\left(  t\right)  $ the external force. \emph{The
state }$v\left( t\right)  $\emph{ of the scalar dispersive system
(\ref{sds1}) is represented
by the following linear functional of }$\psi\left(  t,\sigma\right)  $%
\begin{equation}
v\left(  t\right)  =T\psi\left(  t\right)  =\int_{-\infty}^{\infty}%
\sqrt{m_{\mathfrak{A}_{\operatorname*{s}}}n_{\mathfrak{A}_{\operatorname*{s}}%
}\left(  \sigma\right)  }\psi\left(  t,\sigma\right)  \,d\sigma. \label{sds9}%
\end{equation}

It is of interest to note that under this construction, the extended system
described by $\psi$ is always governed by the canonical evolution equation
(\ref{sds8}) in the Hilbert space $\mathcal{H}=L_{2}\left(  \mathbb{C}\right)
$ with generator\ $\mathcal{A}\psi\left(  \sigma\right)  =\sigma\psi\left(
\sigma\right)  $. Consequently, the only feature which distinguishes different
scalar dispersive systems is the mass $m_{\mathfrak{A}_{\operatorname*{s}}}$
and the ``observable variable''\ $v\left(  t\right)  $ obtained by projecting
onto a one-dimensional Hilbert space $H_{0}$ spanned by the vector
$\sqrt{m_{\mathfrak{A}_{\operatorname*{s}}}n_{\mathfrak{A}_{\operatorname*{s}%
}}\left(  \sigma\right)  }$ in $L_{2}\left(  \mathbb{C}\right)  $. Observe
that the external force $f\left(  t\right)  \sqrt{m_{\mathfrak{A}%
_{\operatorname*{s}}}n_{\mathfrak{A}_{\operatorname*{s}}}\left(
\sigma\right)  }$ is in the space $H_{0}^{\prime}$.

\subsection{Maxwell equations for lossy and dispersive media}

In this section we construct a conservative extension of the
Maxwell equations for a homogeneous, lossy and dispersive medium
following the friction-admittance scheme from Section 3.3.

The classical Maxwell equations for a homogeneous, lossy and
dispersive medium are \cite[Section 1.1]{Born}
\begin{equation}
\nabla\times\mathbf{E}\left(  \mathbf{r},t\right)  =-\partial_{t}%
\mathbf{B}\left(  \mathbf{r},t\right)  -4\pi\mathbf{J}_{B}\left(
\mathbf{r},t\right)  \,,\ \nabla\cdot\mathbf{H}\left(  \mathbf{r},t\right)
=0\,, \label{max1}%
\end{equation}%
\begin{equation}
\nabla\times\mathbf{H}\left(  \mathbf{r},t\right)  =\partial_{t}%
\mathbf{D}\left(  \mathbf{r},t\right)  +4\pi\mathbf{J}\left(  \mathbf{r}%
,t\right)  \,,\ \nabla\cdot\mathbf{D}\left(  \mathbf{r},t\right)  =0\,,
\label{max2}%
\end{equation}
where $\mathbf{H}$, $\mathbf{E}$, $\mathbf{B}$ and $\mathbf{D}$ are
respectively the magnetic and electric fields, and magnetic and electric
inductions, and $\mathbf{J}$ and $\mathbf{J}_{B}$ are respectively the
external electric and magnetic currents. For simplicity, we consider here a
non-magnetic medium, which amounts to taking%
\begin{equation}
\mathbf{H}=\mathbf{B}\text{ and }\mathbf{J}_{B}=0, \label{max3}%
\end{equation}
in units such that $\mu_{0}=1$. We also assume there are no free charges,
which is the assumption that the current $\mathbf{J}\left(  \mathbf{r}%
,t\right)  $ is divergence free%
\begin{equation}
\nabla\cdot\mathbf{J}\left(  \mathbf{r},t\right)  =0. \label{max3a}%
\end{equation}

The dispersive properties of the medium come through the material
(constitutive) relations which, in the simplest case of a
homogeneous and isotropic medium, take the form
\begin{multline}
\mathbf{D}\left(  \mathbf{r},t\right)  =\mathbf{E}\left(
\mathbf{r},t\right) +4\pi\mathbf{P}\left(  \mathbf{r},t\right) ,
\\ \text{ where }\mathbf{P}\left(
\mathbf{r},t\right)  =\int_{0}^{\infty}\chi\left(  \tau\right)  \mathbf{E}%
\left(  \mathbf{r},t-\tau\right)  \,d\tau, \label{pet1}%
\end{multline}
and $\chi\left(  \tau\right)  $ is the scalar-valued \emph{response
(aftereffect) function}, \cite[Chapter 2]{Bohren}, \cite[Section 3]%
{KuboToda2}. In the frequency domain the relation (\ref{pet1})
between the polarization $\mathbf{P}\left(  \mathbf{r},t\right)  $
and the electric field
$\mathbf{E}\left(  \mathbf{r},t\right)  $ becomes%
\begin{equation}
\mathbf{\hat{P}}\left(  \mathbf{r},\omega\right)  =\hat{\chi}\left(
\omega\right)  \mathbf{\hat{E}}\left(  \mathbf{r},\omega\right)  ,\ \hat{\chi
}\left(  \omega\right)  =\int_{0}^{\infty}\chi\left(  t\right)  e^{\mathrm{i}%
\omega t}\,dt, \label{pet2}%
\end{equation}
where $\hat{\chi}\left(  \omega\right)  $ is the so-called \emph{frequency
dependent electric susceptibility}, which is a scalar-valued function for the
case we consider. Since the medium is homogeneous and isotropic, $\hat{\chi}$
does not depend on $\mathbf{r}$ and we have%
\begin{equation}
\nabla\cdot\mathbf{D}\left(  \mathbf{r},t\right)  =\nabla\cdot\mathbf{E}%
\left(  \mathbf{r},t\right)  =\nabla\cdot\mathbf{P}\left(  \mathbf{r}%
,t\right)  =0. \label{pet2a}%
\end{equation}

To construct a conservative extension we begin by recasting the Maxwell
equations for the dispersive medium in the general form\ from Section 3. The
``observable'' variables $v$ and the corresponding Hilbert space $H_{0}$ in
this case are
\begin{multline}
v\left(  t\right)     =\left[
\begin{array}
[c]{c}%
\mathbf{E}\left(  \mathbf{r},t\right) \\
\mathbf{B}\left(  \mathbf{r},t\right)
\end{array}
\right]  \in H_{0},\\
\text{where } H_{0}  =\left\{ v\in L^{2}\left(
\mathbb{C}^{6}\right) :\nabla\cdot\mathbf{E}\left(
\mathbf{r},t\right) =\nabla\cdot\mathbf{B}\left(
\mathbf{r},t\right)  =0\right\} . \label{vH1}
\end{multline}
In other words, $H_{0}$ consists of square-integrable
$6$-dimensional fields with the components $\mathbf{E}\left(
\mathbf{r},t\right)  $ and $\mathbf{B}\left(  \mathbf{r},t\right)
$ divergence free. Comparing equations (\ref{max1})-(\ref{pet2a})
with the general evolution equation (\ref{ax3}) we
set:%
\begin{multline}
m_{\operatorname*{M}}    =I_{H_{0}},\ A_{\operatorname*{M}}=\left[
\begin{array}
[c]{cc}%
0 & \mathrm{i}\nabla^{\times}\\
-\mathrm{i}\nabla^{\times} & 0
\end{array}
\right]  ,\ a_{\operatorname*{M}}\left(  t\right)  =a_{\operatorname*{s}%
}\left(  t\right)  \left[
\begin{array}
[c]{cc}%
I_{3} & 0\\
0 & 0
\end{array}
\right]  ,\\
\text{where }a_{\operatorname*{s}}\left(  t\right)
=4\pi\partial_{t}\chi\left( t\right)  ,\ I_{3}=\left[
\begin{array}
[c]{ccc}%
1 & 0 & 0\\
0 & 1 & 0\\
0 & 0 & 1
\end{array}
\right]  ,\label{vH2}
\end{multline}
or, in the complex frequency domain%
\begin{equation}
\hat{a}_{\operatorname*{M}}\left(  \zeta\right)  =\hat{a}_{s}\left(
\zeta\right)  \left[
\begin{array}
[c]{cc}%
I_{3} & 0\\
0 & 0
\end{array}
\right]  ,\ \hat{a}_{\operatorname*{s}}\left(  \zeta\right)  =-4\pi
\mathrm{i}\zeta\hat{\chi}\left(  \zeta\right)  ,\ \operatorname{Im}\zeta\geq0.
\label{vH3}%
\end{equation}
The reduced form of $\hat{a}_{\operatorname*{M}}\left(  \zeta\right)  $ -- as
in Definition \ref{dred} -- is%
\begin{equation}
\begin{split}
&\tilde{H}_{0}=\left\{  v\in L^{2}\left(  \mathbb{C}^{6}\right)
:\nabla \cdot\mathbf{E}\left(  \mathbf{r},t\right)  =0,\
\mathbf{B}\left(
\mathbf{r},t\right)  =0\right\}  ,\ \\
&\hat{a}_{\operatorname*{M},\tilde{H}_{0}}\left(  \zeta\right)  \mathbf{E}%
\left(  \mathbf{r}\right)  =\hat{a}_{s}\left(  \zeta\right)
\mathbf{E}\left( \mathbf{r}\right)  . \end{split} \label{vH3a}
\end{equation}

Notice that the friction function $a_{\operatorname*{M}}\left(  t\right)  $ is
related to the time derivative of the polarization $\partial_{t}\mathbf{P}$,
represented by the scalar function $\zeta\hat{\chi}\left(  \zeta\right)  $ in
the complex frequency domain. That is consistent with the physical fact that
the work per unit time done by the electric field $\mathbf{E}$ to produce the
polarization is given by $\partial_{t}\mathbf{P\cdot E}$. In view of the
simple structure of the operator friction function (\ref{vH3}) the power
dissipation condition turns here into the following condition for the scalar
function $\hat{a}_{s}\left(  \zeta\right)  =-4\pi\mathrm{i}\zeta\hat{\chi
}\left(  \zeta\right)  $
\begin{equation}
\operatorname{Re}\hat{a}_{\operatorname*{s}}\left(  \zeta\right)
=4\pi\operatorname{Im}\left\{  \zeta\hat{\chi}\left(  \zeta\right)  \right\}
\geq0,\ \operatorname{Im}\zeta\geq0. \label{vH4}%
\end{equation}
It is easy to see that the construction of a conservative extension is
essentially reduced to the construction of the conservative extension for the
scalar friction function $a_{\operatorname*{s}}\left(  t\right)  $. Thus, a
conservative extension\ for the electric polarization can be found using the
results of the previous section for a general scalar dispersive dissipative
system and we obtain, in particular, the evolution equations (\ref{sds8}) and
the representation (\ref{sds9}).

For simplicity, suppose that the following limit exists for every real
$\sigma$%
\begin{multline}
n_{\hat{\chi}}\left(  \sigma\right)  =\lim_{\eta\rightarrow+0}\frac{1}{\pi
}\operatorname{Re}\left\{  \hat{a}_{\operatorname*{s}}\left(  \sigma
+\mathrm{i}\eta\right)  \right\}  \\ =\lim_{\eta\rightarrow+0}4\operatorname{Im}%
\left\{  \left(  \sigma+\mathrm{i}\eta\right)  \hat{\chi}\left(
\sigma+\mathrm{i}\eta\right)  \right\}  , \label{pet4a}%
\end{multline}
and that%
\begin{equation}
\lim_{\eta\rightarrow\infty}\eta\operatorname{Re}\left\{
\hat{a}_{\operatorname*{s}}\left(  \mathrm{i}\eta\right)  \right\}
=\lim_{\eta\rightarrow\infty}4\eta\operatorname{Im}\left\{  \left(
\sigma+\mathrm{i}\eta\right)  \hat{\chi}\left(  \sigma+\mathrm{i}\eta\right)
\right\}  <\infty, \label{pet4aa}%
\end{equation}
conditions which are satisfied in many non-trivial examples. Then the desired
\emph{conservative extension of the original Maxwell equations (\ref{max1}%
)-(\ref{pet2a}) for a dispersive and dissipative dielectric medium} takes the
form%
\begin{equation}
\begin{split}
  \partial_{t}\mathbf{H}\left(  \mathbf{r},t\right)   &  =-\nabla\times
\mathbf{E}\left(  \mathbf{r},t\right)  ,\\
\partial_{t}\mathbf{E}\left(  \mathbf{r},t\right)   &  =\nabla\times
\mathbf{H}\left(  \mathbf{r},t\right)
\\& \quad -\int_{-\infty}^{\infty}\sqrt {m_{\hat{\chi}}n_{\hat{\chi}}\left(
\sigma\right)  }\mathbf{\Psi}\left(
\mathbf{r},t,\sigma\right)  \,d\sigma-4\pi\mathbf{J}\left(  \mathbf{r}%
,t\right)  ,\\
m_{\hat{\chi}}\partial_{t}\mathbf{\Psi}\left(
\mathbf{r},t,\sigma\right)   &
=-\mathrm{i}m_{\hat{\chi}}\sigma\mathbf{\Psi}\left(
\mathbf{r},t,\sigma
\right)  +\sqrt{m_{\hat{\chi}}n_{\hat{\chi}}\left(  \sigma\right)  }%
\mathbf{E}\left(  \mathbf{r},t\right)  ,
\end{split}\label{cEH1}
\end{equation}
where
\begin{multline}
n_{\hat{\chi}}\left(  \sigma\right)  =-4\lim_{\eta\rightarrow+0}%
\operatorname{Im}\left\{  \left(  \sigma+\mathrm{i}\eta\right)  \hat{\chi
}\left(  \sigma+\mathrm{i}\eta\right)  \right\}  \geq0,\ m_{\hat{\chi}}%
^{-1}=\int_{-\infty}^{\infty}n_{\hat{\chi}}\left(  \sigma\right)
\,d\sigma,\\
4\pi\zeta\hat{\chi}\left(  \zeta\right)  =\int_{-\infty}^{\infty
}\frac{n_{\hat{\chi}}\left(  \sigma\right) }{\sigma-\zeta}\,d\sigma ,\
\operatorname{Im}\zeta>0, \ \mathbf{\Psi}\left(  \mathbf{r},t,\sigma\right)
\in L_{2}\left( \mathbb{R};\mathbb{C}^{3}\right)  ,\label{cEH2}
\end{multline}
and the fields $\mathbf{H}\left(  \mathbf{r},t\right)  $, $\mathbf{D}\left(
\mathbf{r},t\right)  $ are divergence free, i.e.%
\begin{multline}
\nabla\cdot\mathbf{H}\left(  \mathbf{r},t\right)  =0,\\
\nabla\cdot \mathbf{E}\left(  \mathbf{r},t\right)
+4\pi\int_{-\infty}^{t}\int_{-\infty
}^{\infty}\sqrt{m_{\hat{\chi}}n_{\hat{\chi}}\left(  \sigma\right)  }%
\nabla\cdot\mathbf{\Psi}\left(  \mathbf{r},\tau,\sigma\right)  \,d\sigma
d\tau=0. \label{cEH3}%
\end{multline}
The electric polarization $\mathbf{P}\left(  \mathbf{r},t\right)  $, its time
derivative $\partial_{t}\mathbf{P}\left(  \mathbf{r},t\right)  $ and the
electric induction $\mathbf{D}\left(  \mathbf{r},t\right)  $ are now defined
by
\begin{equation}
\begin{split}
\mathbf{P}\left(  \mathbf{r},t\right)   &  =\frac{1}{4\pi}\int_{-\infty}%
^{t}\int_{-\infty}^{\infty}\sqrt{m_{\hat{\chi}}n_{\hat{\chi}}\left(
\sigma\right)  }\mathbf{\Psi}\left(  \mathbf{r},\tau,\sigma\right)  \,d\sigma
d\tau\mathbf{,}\\
\partial_{t}\mathbf{P}\left(  \mathbf{r},t\right)   &  =\frac{1}{4\pi}%
\int_{-\infty}^{\infty}\sqrt{m_{\hat{\chi}}n_{\hat{\chi}}\left(
\sigma\right)  }\mathbf{\Psi}\left(  \mathbf{r},t,\sigma\right)
\,d\sigma,\\
\mathbf{D}\left(  \mathbf{r},t\right)   &  =\mathbf{E}\left(  \mathbf{r}%
,t\right)  +\int_{-\infty}^{t}\int_{-\infty}^{\infty}\sqrt{m_{\hat{\chi}%
}n_{\hat{\chi}}\left(  \sigma\right)  }\mathbf{\Psi}\left(  \mathbf{r}%
,\tau,\sigma\right)  \,d\sigma d\tau\mathbf{.}
\end{split}\label{cEH4}
\end{equation}

We note that according to (\ref{cEH1}) the vector $\frac{1}{4\pi}\int
_{-\infty}^{t}\mathbf{\Psi}\left(  \mathbf{r},\tau,\sigma\right)  d\tau$ in
(\ref{cEH4}) evidently represents a time dependent microscopic dipole of mass
$m_{\hat{\chi}}$, localized at $\mathbf{r}$, which oscillates with natural
frequency $\sigma$. The total polarization $\mathbf{P}\left(  \mathbf{r}%
,t\right)  $ of the medium at point $\mathbf{r}$ is a superposition of a
(continuum) number of microscopic dipoles localized at $\mathbf{r}$. The
natural definition for the energy of the ``hidden'' medium is%
\begin{equation}
\frac{1}{2}\int_{\mathbb{R}^{3}}\int_{-\infty}^{\infty}\left|  \mathbf{\Psi
}\left(  \mathbf{r},t,\sigma\right)  \right|  ^{2}\,d\sigma d\mathbf{r,}%
\end{equation}
consistent with the evolution equation and the assumption that
$\int _{\mathbb{R}^{3}}\mathbf{E}( \mathbf{r},t) \cdot
\partial_{t}\mathbf{P}( \mathbf{r},t)
 d\mathbf{r}$ is the instantaneous
rate of work done by the electromagnetic field.

In a forthcoming paper, we shall discuss in greater detail conservative models
for dispersion in dielectric media -- including scattering theory and
inhomogeneous media -- however, to complete the discussion here it may be
useful to consider one physically relevant non-trivial example. \ For this
purpose, let us take the so-called Lorentz medium in which \cite[Section
3.5]{Scaife}, \cite[Section 9.1]{Bohren}%
\begin{align}
\chi\left(  t\right)   &  =\chi_{\operatorname*{L}}\left(  t\right)
=\omega_{\operatorname*{p}}^{2}\exp\left\{  -\frac{\gamma}{2}t\right\}
\frac{\sin\xi t}{\xi},\ \xi=\sqrt{\omega_{0}^{2}-\frac{\gamma^{2}}{4}}%
,\ t\geq0;\label{pet5}\\
\hat{\chi}\left(  \zeta\right)   &
=\hat{\chi}_{\operatorname*{L}}\left( \zeta\right)
=\frac{\omega_{\operatorname*{p}}^{2}}{\omega_{0}^{2}-\zeta
^{2}-\mathrm{i}\gamma\zeta},
\end{align}
with $\omega_{\operatorname*{p}}$, $\omega_{0}$ and $\gamma$ positive
parameters. Observe that%
\begin{equation}
\operatorname{Im}\left\{  \zeta\hat{\chi}_{\operatorname*{L}}\left(
\zeta\right)  \right\}  =\frac{\omega_{\operatorname*{p}}^{2}\left[
\eta\left(  \omega_{0}^{2}+\omega^{2}+\eta^{2}+\gamma\eta\right)  +\omega
^{2}\gamma\right]  }{\left(  \omega_{0}^{2}-\omega^{2}+\eta^{2}+\gamma
\eta\right)  ^{2}+\omega^{2}\left(  2\eta+\gamma\right)  ^{2}},\ \zeta
=\omega+\mathrm{i}\eta\label{pet6}%
\end{equation}
and hence%
\begin{equation}
\operatorname{Im}\left\{  \zeta\hat{\chi}_{\operatorname*{L}}\left(
\zeta\right)  \right\}  \geq0,\ \operatorname{Im}\zeta=\eta\geq0, \label{pet7}%
\end{equation}
in full compliance with the power dissipation condition (\ref{vH4}). In
addition, from (\ref{sds7}), (\ref{sds7b}) we see that%
\begin{equation}
n_{\hat{\chi}_{\operatorname*{L}}}\left(  \sigma\right)  =4\lim_{\eta
\rightarrow+0}\operatorname{Im}\left\{  \left(  \sigma+\mathrm{i}\eta\right)
\hat{\chi}_{\operatorname*{L}}\left(  \sigma+\mathrm{i}\eta\right)  \right\}
=4\frac{\omega_{\operatorname*{p}}^{2}\sigma^{2}\gamma}{\left(  \omega_{0}%
^{2}-\sigma^{2}\right)  ^{2}+\sigma^{2}\gamma^{2}} \label{pet8}%
\end{equation}
exists pointwise, and%
\begin{equation}
m_{\hat{\chi}_{\operatorname*{L}}}^{-1}=\int_{-\infty}^{\infty}n_{\hat{\chi}%
}\left(  \sigma\right)  \,d\sigma=\lim_{\eta\rightarrow+\infty}\eta^{2}%
\hat{\chi}_{\operatorname*{L}}\left(  \mathrm{i}\eta\right)  =4\pi
\omega_{\operatorname*{p}}^{2}. \label{pet8a}%
\end{equation}
Plugging in the above values $n_{\hat{\chi}_{\operatorname*{L}}}\left(
\sigma\right)  $ and $m_{\hat{\chi}_{\operatorname*{L}}}$ into extended
Maxwell equations (\ref{cEH1})-(\ref{cEH4}) we get the desired conservative
description of the Lorentz medium.

\section{Dissipation and continuity of the spectrum}

It may seem startling that dissipation, i.e., losses, can arise when we
truncate a \emph{unitary} evolution. However, such results are familiar from
the theory of unitary dilations of contractive semi-groups \cite{Pav}, which is
the theory of solutions to (\ref{ax3}) for friction without retardation, i.e.,
$a(t)=\alpha_{\infty}\delta\left(  t\right)  $. Furthermore, the mechanism at
work is physically very natural: there is energy transport to a large number of
``invisible'' degrees of freedom, i.e. to ``heat.''

Mathematically, a rigorous analysis of losses in very large but finite systems
is generally complicated by the fact that systems with a finite number of
degrees of freedom eventually (perhaps after an extremely long time) return
arbitrarily close to their starting configuration (Poincar\'{e} recurrence).
However, if we consider an idealization in which there are infinitely many
hidden degrees of freedom, resulting in infinite recurrence time, then the
Poincar\'{e} recurrence may not occur and there is hope of describing losses
in a cleaner and simpler way. \emph{As it turns out, a sufficient condition
for losses is strict positivity of }$\operatorname{Re}\hat{a}\left(
\zeta\right)  $\emph{, which implies absolute continuity of the spectral
measure for the generator }$\mathcal{A}$ \emph{of the dynamics of a
conservative extension.}

To state a quite general condition, we use the notion of \emph{non-tangential
boundedness}. Given $\omega_{0}\in\mathbb{R}$ let the cone of aperture $\theta
\in(0,\pi)$ at $\omega_{0}$, denoted $\Gamma_{\theta}(\omega_{0})$, be the set%
\begin{equation}
\Gamma_{\theta}(\omega_{0})=\left\{  \omega+\mathrm{i}\eta:\eta<1\text{ and }|\omega-\omega_{0}%
|<\eta\tan\frac{\theta}{2}\right\}  .
\end{equation}
A function $F:\mathbb{C}_{+}\rightarrow\mathbb{C}$, with $\mathbb{C}_+ = \{
\zeta : \mathrm{Im} \zeta > 0\}$, is said to be
\emph{non-tangentially bounded at }$\omega_{0}\in\mathbb{R}$ if%
\begin{equation}
\sup_{\zeta\in\Gamma_{\theta}(\omega_{0})}\left|  F(\omega)\right|  <\infty,
\end{equation}
for some $\theta\in(0,\pi)$. We shall use the following well known theorem
from harmonic analysis regarding boundary values of non-tangentially bounded
harmonic functions, see e.g. \cite{St}.

\begin{theorem}
\label{thm:nontangentiallimits}Let $F:\mathbb{C}_{+}\rightarrow\mathbb{C}$ be
a harmonic function and suppose that $F$ is non-tangentially bounded at every
point of a set $E\subset\mathbb{R}$. \ Then
\begin{equation}
F(\omega+\mathrm{i}0)=\lim_{\eta \rightarrow 0}F(\omega+\mathrm{i}\eta)
\end{equation}
exists for almost every $\omega\in E$.
\end{theorem}

\noindent\emph{Note:}\ In fact, the so-called non-tangential limits
$\lim_{\zeta\rightarrow \omega}F(\zeta)$, with $\zeta$ restricted to
$\Gamma_{\theta}(\omega)$, exist at almost every $\omega$, but we will not use
this fact.

Our principle theorem on losses is the following

\begin{theorem}
\label{dissipative dynamics}Suppose the mass operator $m$ is strictly positive,
$m \ge \delta >0$, and the Laplace transform $\hat{a}\left( \zeta\right)  $ of
the friction function satisfies the following strengthened form of the
power-dissipation condition (\ref{azf3}): there are measurable
functions $\theta:\mathbb{R}\rightarrow(0,\pi)$, $\gamma:\mathbb{R}%
\rightarrow(0,\infty)$, with $\gamma(\cdot)\in L_{loc}^{1}(\mathbb{R})$ such
that for almost every $\omega\in\mathbb{R}$
\begin{equation}
\operatorname{Re}\left(  v,\hat{a}\left(  \zeta\right)  v\right)
\,\geq\frac{1}{\gamma(\omega)} \| v\|^2 ,\text{ for all
}\zeta\in\Gamma_{\theta(\omega)}(\omega)\text{
and }v\in H_{1}. \label{strictpositivity}%
\end{equation}
Then for any compactly supported generalized force $f\in L_{c}^{1}%
(\mathbb{R};H_{0})$, the solution $v_{f}\left(  t\right)  $ to the evolution
equation (\ref{ax3}) vanishes in the large $t$ limit:%
\begin{equation}
\lim_{t\rightarrow\infty}\left\|  v_{f}(t)\right\|  =0.
\end{equation}

\begin{remark}
The condition (\ref{strictpositivity}) holds in particular if
$\operatorname{Re}\hat{a}\left(  \zeta\right)  $ is strictly positive,
$\operatorname{Re}\hat{a}\left(  \zeta\right)  \geq\delta I_{H_{0}}$ for all
$\operatorname{Im}\zeta>0$. For example, in the Lorentz medium we have
$\hat{a}\left(  \zeta\right)  =\gamma+\mathrm{i}\frac{\omega_{o}^{2}}{\zeta}$,
so $\operatorname{Re}\hat{a}\left(  \zeta\right)  \geq\gamma I_{H_{0}}$.
\end{remark}
\end{theorem}

\begin{proof}
Note that the admittance $\mathfrak{A}\left(  \zeta\right)  =\mathrm{i}%
\left(  \zeta m-A+\mathrm{i}\hat{a}\left(  \zeta\right)  \right)  ^{-1}$ obeys%
\begin{multline}
\operatorname{Re}\mathfrak{A}\left(  \zeta\right)
=\mathfrak{A}\left( \zeta\right)  \left[  \operatorname{Im}\zeta
m+\operatorname{Re}\hat{a}\left(
\zeta\right)  \right]  \mathfrak{A}\left(  \zeta\right)  ^{\dagger}\\
\geq\frac{1}{\gamma(\omega)}\,\mathfrak{A}\left(  \zeta\right)  \mathfrak{A}%
\left(  \zeta\right)  ^{\dagger},\text{for
}\zeta\in\Gamma_{\theta(\omega)}(\omega).
\end{multline}
We conclude that
\begin{equation}
\left\|  \mathfrak{A}\left(  \zeta\right)  \right\|  \geq\left\|
\operatorname{Re}\mathfrak{A}\left(  \zeta\right)  \right\|  \geq
\frac{1}{\gamma(\omega)}\,\left\|  \mathfrak{A}\left(  \zeta\right)  \right\|
^{2},
\end{equation}
and thus%
\begin{equation}
\left\|  \mathfrak{A}\left(  \zeta\right)  \right\|
\leq\gamma(\omega),\text{for }\zeta\in\Gamma_{\theta(\omega)}(\omega).
\end{equation}

Using Theorem \ref{tOperatorNev2}, we find a conservative extension consisting
of a self-adjoint operator $\mathcal{A}$ on a Hilbert space $\mathcal{H}$ and
an isometric imbedding $T^{\dagger}:H_{0}\rightarrow\mathcal{H}$ such that
$\mathfrak{A}\left(  \zeta\right)  =\mathrm{i}m^{-1/2} T\left(  \zeta-\mathcal{A}%
\right)  ^{-1}T^{\dagger} m^{-1/2} $. \ Furthermore the solution $v_{f}$ to
(\ref{ax3})
with generalized force $f$ is%
\begin{equation}
v_{f}\left(  t\right)  =\int_{-\infty}^{t}m^{-1/2} Te^{\mathrm{i}\left(
s-t\right) \mathcal{A}}T^{\dagger}m^{-1/2}f\left(  s\right)  ds.
\end{equation}
The theorem now follows from Theorem \ref{selfadjointdissipation} below via
dominated convergence.
\end{proof}

\begin{theorem}
\label{selfadjointdissipation}Let $\mathcal{A}$ be a self adjoint operator on
a separable Hilbert space $\mathcal{H}$, and let $T:\mathcal{H}\rightarrow
H_{0}$ be a bounded map from $\mathcal{H}$ into a separable Hilbert space
$H_{0}$. \ Suppose there are measurable functions $\gamma:\mathbb{R}%
\rightarrow(0,\infty)$ and $\theta:\mathbb{R}\rightarrow(0,\pi)$ with
$\gamma(\cdot)\in L_{loc}^{1}(\mathbb{R})$ such that for almost every
$\omega\in\mathbb{R}$%
\begin{equation}
\sup_{\zeta\in\Gamma_{\theta(\omega)}(\omega)}\left\|  \operatorname{Im}T(\zeta
-\mathcal{A})^{-1}T^{\dagger}\right\|  \leq\gamma(\omega). \label{eq:upperbound}%
\end{equation}
Then
\begin{equation}
\lim_{t\rightarrow\pm\infty}Te^{-\mathrm{i}t\mathcal{A}}T^{\dagger}v=0\,
\label{eq:zerostlim}%
\end{equation}
for every $v\in H_{0}$.
\end{theorem}

\begin{remark}
(1)\ As will be clear from the proof, if $\gamma(\cdot)\in L^{1}$ (instead of
$L_{loc}^{1}$), then eq. (\ref{eq:zerostlim}) holds in \emph{norm}. \ (2)\ Eq.
(\ref{eq:upperbound}) implies that the spectral measure associated to
$\mathcal{A}$ and any $\psi\in\operatorname*{ran}T^\dagger$ is purely
absolutely
continuous, with density%
\begin{equation}
-\lim_{\varepsilon\rightarrow+0}\left\langle \psi,\operatorname{Im}%
(\omega+\mathrm{i}\eta-\mathcal{A})^{-1}\psi\right\rangle .
\end{equation}
\end{remark}

\begin{proof}
We begin by showing that for each bounded open interval $I\subset\mathbb{R}$%
\begin{equation}
\left\Vert Te^{-\mathrm{i}t\mathcal{A}}E_{\mathcal{A}}(I)T^{\dagger
}\right\Vert \rightarrow0:\text{as}:t\rightarrow\pm\infty,
\end{equation}
where $E_{\mathcal{A}}(\cdot)$ is the spectral resolution of $\mathcal{A}$.

Let us define%
\begin{equation}
F(\zeta)=\operatorname{Im}T^{\dagger}(\zeta-\mathcal{A})^{-1}T\text{ for
}\operatorname{Im}\zeta\neq0. \end{equation} By the spectral theorem and the
observation that the spectral measures of $\mathcal{A}$, $\mathrm{d} \|
E_{\mathcal{A}}(\lambda) \psi \|^2$, are purely absolutely continuous, we have
\begin{equation}\label{dynamics}
Te^{-\mathrm{i}t\mathcal{A}}E_{\mathcal{A}}(I)T^{\dagger}v=-\frac{1}{\pi}%
\lim_{\eta\downarrow0}\int_{I}d\omega\,e^{-\mathrm{i}t\omega}F(\omega+\mathrm{i}%
\eta)\,v,
\end{equation}
for each finite $t$ and every $v\in H_{0}$

Since $H_{0}$ is separable, it has a countable basis $\phi_{j}$, $j=1,....,$.
Using Theorem \ref{thm:nontangentiallimits} we conclude that there is
$I^{\prime}\subset I$ with $m(I\backslash I^{\prime})=0$ such that
\begin{equation}
\lim_{\varepsilon\rightarrow0}\langle\phi_{i},F(\omega+\mathrm{i}\eta
)\phi_{j}\rangle
\end{equation}
exists for every $\omega\in I^{\prime}$ and every pair of basis vectors
$\phi_{i}$, $\phi_{j}$. \ Let $\mathcal{S}$ denote the subspace of linear
combinations of finitely many basis vectors. Given $\psi\in\mathcal{S}$,
\begin{multline}
\sum_{i=1}^{\infty}\lim_{\eta\rightarrow0}\left\vert \langle\phi
_{i},F(\omega+\mathrm{i}\eta)\psi\rangle\right\vert ^{2}   \leq
\liminf_{\eta\rightarrow0}\sum_{i=1}^{\infty}\left\vert \langle\phi
_{i},F(\omega+\mathrm{i}\eta)\psi\rangle\right\vert ^{2}\\
  =\liminf_{\eta\rightarrow0}\left\Vert F(\omega+\mathrm{i}\eta
)\psi\right\Vert ^{2}\leq\gamma(\omega)^{2} \|\psi \|^2.
\end{multline}
We conclude that, for every $\omega\in I^{\prime}$, the limit
\begin{equation} \operatorname*{wk-lim}_{\eta
}F(\omega+\mathrm{i}\eta)\psi=\sum\lim_{\eta}\langle\phi
_{i},F(\omega+\mathrm{i}\eta)\psi\rangle\phi_{i}
\end{equation} exists, and the map%
\begin{equation}
\psi\mapsto\sum_{i=1}^{\infty}\left(  \lim_{\eta\rightarrow0}%
\langle\phi_{i},F(\omega+\mathrm{i}\eta)\psi\rangle\right)  \phi_{i}%
\end{equation}
is bounded from $\mathcal{S}$ into $H_{0}$. Since $\mathcal{S}$ is dense, this
map may be extended to a unique\ bounded linear map $F(\omega+\mathrm{i}%
0):H_{0}\rightarrow H_{0}$ with $\left\Vert F(\omega+\mathrm{i}0)\right\Vert
\leq\gamma(u)$. It is elementary to see that,
\begin{equation}
\left\langle \psi_{1},F(\omega+\mathrm{i}0)\psi_{2}\right\rangle =\lim
_{\eta\rightarrow0}\left\langle \psi_{1},F(\omega+\mathrm{i}\eta
)\psi_{2}\right\rangle \text{ for any }\psi_{1},\psi_{2}\in H_{0},
\end{equation}
i.e.%
\begin{equation}
F(\omega+\mathrm{i}0)=\operatorname*{wklim}_{\eta}F(\omega+\mathrm{i}%
\eta)\text{ for }\omega\in I^{\prime}.
\end{equation}

By dominated convergence we find from \eqref{dynamics}
\begin{equation}
Te^{-\mathrm{i}t\mathcal{A}}E_{\mathcal{A}}(I)T^{\dagger}v=-\frac{1}{\pi}%
\int_{I}d\omega e^{-\mathrm{i}t\omega}\,F(\omega+\mathrm{i}0)\,v.
\end{equation}
for every $v\in H_{0}$. Since $\left\Vert F(\omega+\mathrm{i}0)\right\Vert
\leq\gamma(\omega)\in L^{1}(I)$, we have
\begin{equation}
\lim_{t\rightarrow\pm\infty}\left\Vert Te^{-\mathrm{i}t\mathcal{A}%
}E_{\mathcal{A}}(I)T^{\dagger}\right\Vert =0
\end{equation}
by the Riemann-Lebesgue lemma -- the extension of this result to operator
valued functions is elementary.

To complete the proof, we note that given $\varepsilon>0$ and $\psi\in H_{0}$
we can find a finite interval $I_{\varepsilon}\subset\mathbb{R}$ such that
$\left\|  (1-E_{\mathcal{A}}(I_{\varepsilon}))T^{\dagger}\psi\right\|
\leq\varepsilon$. Thus
\begin{equation}
\limsup_{t\rightarrow\pm\infty}\left\|  Te^{-\mathrm{i}t\mathcal{A}}%
T^{\dagger}\psi\right\|  \leq\limsup_{t\rightarrow\pm\infty}\left\|
Te^{-\mathrm{i}t\mathcal{A}}E_{\mathcal{A}}(I_{\varepsilon})T^{\dagger}%
\psi\right\|  +\varepsilon=\varepsilon.
\end{equation}
Since $\varepsilon$ is arbitrary, we see that (\ref{eq:zerostlim}) holds.
\end{proof}

\section{Operator versions of classical spectral theorems}

In this section we discuss the proofs of operator versions of Bochner's Theorem
\ref{tBoch} -- Theorems \ref{tOperatorBoch} and \ref{tGenOperatorBoch} above --
and operator versions of the Herglotz-Nevanlinna Theorems \ref{tNev1},
\ref{tNev2} -- Theorem \ref{tOperatorNev2} and \ref{tGenOperatorNev2}. For
properties of operator-valued functions holomorphic in a half-plane and their
boundary values see \cite{RoRo}.

\subsection{Bochner's Theorem}

The two operator valued generalizations of Bochner's theorem stated above --
Theorems \ref{tOperatorBoch} and \ref{tGenOperatorBoch} -- are combined in the
following statement.

\begin{theorem}
The friction function $a_{e}\left(  t\right)  =2\alpha_{\infty}\delta\left(
t\right)  +\alpha_{e}(t)$, $-\infty<t<\infty$, with $\alpha_{e}(t)$ a strongly
continuous $\mathcal{B}\left(  H_{0}\right)  $ valued function, and
$\alpha_{\infty}$ a bounded non-negative operator is representable as%
\begin{equation}
a_{e}\left(  t\right)  =\operatorname*{Dlim}_{R\rightarrow\infty}\Gamma
e^{-\mathrm{i}t\Omega_{1}}\left(  \Gamma\Phi_{R}^{2}\right)  ^{\dagger}%
,\ \Phi_{R}^{2}=\left(  \frac{\Omega_{1}^{2}}{R^{2}}+I_{H_{1}}\right)  ^{-1},
\label{hg1c2}%
\end{equation}
with $\Omega_{1}$ a self-adjoint operator on $H_{1}$ and $\Gamma:D\left(
\Omega_{1}\right)  \rightarrow H_{0}$ a $\Omega_{1}$-bounded linear map, if
and only if $a\left(  t\right)  $ satisfies the dissipation condition
(\ref{ava4}) for every continuous $H_{0}$ valued function $v(t)$ with compact
support. The operator $\Gamma$ is bounded if and only if $\alpha_{\infty}=0$,
in which case%
\begin{equation}
a_{e}\left(  t\right)  =\Gamma e^{-\mathrm{i}t\Omega_{1}}\Gamma^{\dagger}.
\label{hg1b2}%
\end{equation}
If the space $H_{1}$ is minimal -- in the sense that
\begin{equation}
\left\langle \left( \Gamma f\left(  \Omega_{1}\right) \right)
^{\dagger}v:\ f\in C_{c}\left( \mathbb{R}\right)  ,\text{ }v\in
H_{0}\right\rangle \end{equation} is dense in $H_{1}$ -- then the
triplet $\left\{  H_{1},\Omega_{1},\Gamma\right\}  $ is determined
uniquely up to an isomorphism.
\end{theorem}

\begin{proof}
Let us start by defining the Hilbert space $H_{1}$, which we take to be an
extension of the Banach space $V=L^{1}\left(  \mathbb{R},H_{0}\right)  \cap
L^{2}\left(  \mathbb{R},H_{0}\right)  $ of measurable $H_{0}$ valued functions
$\phi$ with
\begin{equation}
\left\Vert \phi\right\Vert _{V}:=\left(  \int_{-\infty}^{\infty}\left\Vert
\phi\left(  t\right)  \right\Vert ^{2}\,dt\right)  ^{1/2}+\int_{-\infty
}^{\infty}\left\Vert \phi\left(  t\right)  \right\Vert \,dt<\infty.
\end{equation}
Let $\left(  \cdot,\cdot\right)_{H_0}  $ denote the inner product on $H_{0}$
and
define on $V$ a quadratic form%
\begin{align}
\left\langle \phi,\psi\right\rangle  &  :=\int_{-\infty}^{\infty}\int
_{-\infty}^{\infty}\left(  \phi\left(  t\right)  ,a_{e}\left(  t-\tau\right)
\psi\left(  \tau\right)  \right)_{H_0}  \,dtd\tau\label{preip}\\
&  =2\int_{-\infty}^{\infty}\left(  \phi\left(  t\right)  ,\alpha_{\infty}%
\psi\left(  t\right)  \right)_{H_0}  \,dt
\\ & \quad +\int_{-\infty}^{\infty}\int_{-\infty }^{\infty}\left(  \phi\left(  t\right)
,\alpha_{e}\left(  t-\tau\right) \psi\left(  \tau\right)  \right)_{H_0}
\,dtd\tau,\nonumber
\end{align}
which is positive semi-definite by virtue of the power dissipation condition. \
However, there may be null vectors in $V$, that is vectors $\phi$ with
$\left\langle \phi,\phi\right\rangle =0$. \ Let $N$ denote the set of null
vectors and define $H_{1}=\overline{V/N}$, where \textquotedblleft \
$\overline{\cdot}$ \textquotedblright\ denotes closure in the norm inherited
from the inner product. Then $H_{1}$ is a Hilbert space, whose inner product we
also denote by $\left\langle \cdot,\cdot\right\rangle $.

Any $\phi\in V$ defines a unique element $\left[  \phi\right]  \in V/N\subset
H_{1}$ with $\left\Vert \left[  \phi\right]  \right\Vert _{H_{1}}%
^{2}=\left\langle \phi,\phi\right\rangle $. In particular, $\left[
\phi\right]  $ is zero if and only if $\left\langle \phi,\phi\right\rangle
=0$. Furthermore, the map $\phi\mapsto\left[  \phi\right]  $ is bounded from
$V$ into $H_{1}$ since%
\begin{multline}
\left\Vert \left[  \phi\right]  \right\Vert
_{H_{1}}^{2}\leq2\left\Vert \alpha_{\infty}\right\Vert
\int_{-\infty}^{\infty}\left\Vert \phi\left( t\right)  \right\Vert
^{2}\,dt+\sup_{\tau}\left\Vert \alpha_{e}\left( \tau\right)
\right\Vert \left(  \int_{-\infty}^{\infty}\left\Vert \phi\left(
t\right)  \right\Vert \,dt\right)  ^{2}\\ \lesssim\left\Vert
\phi\right\Vert _{V}^{2}.
\end{multline}
Thus a convergent sequence $\phi_{j}$ in $V$ gives rise to a convergent
sequence $\left[  \phi_{j}\right]  $ in $H_{1}$ and
\begin{equation}
\lim_{j}\left[  \phi_{j}\right]  =\left[  \lim_{j}\phi_{j}\right]  .
\end{equation}

We define the operator $\Omega_{1}$ to be the self adjoint generator of the
one parameter unitary group of time translations. Specifically, we note that
the transformations $T_{s}$ of $V$ into itself given by%
\begin{equation}
T_{s}\phi\left(  t\right)  =\phi\left(  t-s\right)  ,
\end{equation}
form a group which preserves the pre-inner product (\ref{preip}). Therefore,
$s\mapsto T_{s}$ extends to a one parameter unitary group $s\mapsto U_{s}$ on
$H_{1}$, which is in fact strongly continuous (as follows from strong
continuity of $T_{s}$ on $V$). The Stone--von\ Neumann theorem implies there is
a unique self adjoint operator $\Omega_{1}$ on $H_{1}$ with $U_{s}%
=e^{\mathrm{i}s\Omega_{1}}$. \

Clearly there is a connection between $\Omega_{1}$ and the operation of
differentiation on $V$. \ To understand this, note that given $\phi\in V$ we
have $\left[  \phi\right]  \in\mathcal{D}\left(  \Omega_{1}\right)  $ if and
only if%
\begin{equation}
\Omega_{1}\left[  \phi\right]  =\lim_{s\rightarrow0}\frac{U_{s}\left[
\phi\right]  -\left[  \phi\right]  }{\mathrm{i}s}=-\mathrm{i}\lim
_{s\rightarrow0}\left[  \frac{T_{s}\phi-\phi}{s}\right]
\end{equation}
exists. \ Thus a sufficient condition for $\left[  \phi\right]  $ to be in
$\mathcal{D}\left(  \Omega_{1}\right)  $ is for $s^{-1}\left(  T_{s}\phi
-\phi\right)  $ to converge in $V$, which holds if and only if $\phi
\in\mathcal{D}\left(  \partial_{t}\right)  $, in which case%
\begin{equation}
\lim_{s}\frac{T_{s}\phi-\phi}{s}=-\partial_{t}\phi.
\end{equation}
Therefore we have%
\begin{equation}
\Omega_{1}\left[  \phi\right]  =\mathrm{i}\left[  \partial_{t}\phi\right]
\text{ for }\phi\in\mathcal{D}\left(  \partial_{t}\right)  .
\label{differential}%
\end{equation}
Note, however, that $\left[  \phi\right]  \in\mathcal{D}\left(  \Omega
_{1}\right)  $ does not necessarily imply $\phi\in\mathcal{D}\left(
\partial_{t}\right)  $.

If we formally define $\Gamma^{\dagger}$ to be the map $\Gamma^{\dagger
}v=\left[  v\delta\left(  \cdot\right)  \right]  $, ignoring for the moment
that $v\delta\left(  \cdot\right)  \notin V$, we may calculate that%
\begin{equation}
\Gamma\left[  \phi\right]  =\int_{-\infty}^{\infty}a_{e}\left(  -t\right)
\phi\left(  t\right)  \,dt, \label{Gammadefn}%
\end{equation}
and therefore
\begin{equation}
\Gamma e^{-\mathrm{i}t\Omega_{1}}\Gamma^{\dagger}v=\int_{-\infty}^{\infty
}a_{e}\left(  -s\right)  v\delta\left(  s+t\right)  \,dt=a_{e}(t)v.
\end{equation}
If $\alpha_{\infty}=0$ then in fact (\ref{Gammadefn}) defines a bounded
operator, and the above calculation may be justified.\ In that case, we could
have started with a space $V$ including point measures so that $v\delta\left(
t\right)  $ for $v\in H_{0}$ would be in $V$ and we would have $\Gamma
^{\dagger}v=\left[  v\delta\left(  \cdot\right)  \right]  $.

However, to consider also $\alpha_{\infty}\neq0$, we work indirectly by
defining the bounded map $S^{\dagger}:H_{0}\rightarrow H_{1}$,%
\begin{equation}
S^{\dagger}v:=\left[  vG\right]  ;\text{ }G\left(  t\right)  =\mathrm{i}%
e^{t}\begin{cases} 1 & t < 0 \; , \\ 0 & t  > 0 \; . \end{cases}
\end{equation}
Note that formally, $\left(  \Omega_{1}-\mathrm{i}I_{H_{1}}\right)
S^{\dagger}v=\left[  v\delta\left(  \cdot\right)  \right]  $, since $\left\{
\mathrm{i}\partial_{t}-\mathrm{i}\right\}  G\left(  t\right)  =\delta\left(
t\right)  $. To proceed rigorously, let us compute $S:=\left(  S^{\dagger
}\right)  ^{\dagger}$. Given $\phi\in V$%
\begin{equation}
\left(  v,S\left[  \phi\right]  \right)  =\left\langle S^{\dagger}v,\left[
\phi\right]  \right\rangle =\int_{-\infty}^{\infty}\int_{-\infty}^{\infty
}\overline{G}\left(  t\right)  \left(  v,a_{e}\left(  t-\tau\right)
\phi\left(  \tau\right)  \right)  \,dtd\tau,
\end{equation}
so%
\begin{multline}
S\left[  \phi\right]     =\int_{-\infty}^{\infty}\left(  -\mathrm{i}%
\int_{-\infty}^{0}e^{t}a_{e}\left(  t-\tau\right)  dt\right)  \phi\left(
\tau\right)  \,d\tau\\
  =\int_{-\infty}^{\infty}\left(  -\mathrm{i}\int_{\tau}^{\infty}e^{\tau
-t}a_{e}\left(  -t\right)  dt\right)  \phi\left(  \tau\right)
\,d\tau .
\end{multline}
If we define $\Gamma:\mathcal{D}\left(  \Omega_{1}\right)  \rightarrow H_{0}$
by%
\begin{equation}
\text{ }\Gamma:=S\left(  \Omega_{1}+\mathrm{i}I_{H_{1}}\right)  , \label{sg1}%
\end{equation}
then we recover (\ref{Gammadefn}) for $\phi\in\mathcal{D}\left(  \partial
_{t}\right)  $,
\begin{align}
\Gamma\left[  \phi\right]   &  =\int_{-\infty}^{\infty}\left(  \int_{\tau
}^{\infty}e^{\tau-t}a_{e}\left(  -t\right)  \,dt\right)  \left\{
\partial_{\tau}+1\right\}  \phi\left(  \tau\right)  \,d\tau\nonumber\\
&  =\int_{-\infty}^{\infty}a_{e}\left(  -\tau\right)  \phi\left(  \tau\right)
\,d\tau=2\alpha_{\infty}\phi\left(  0\right)  +\int_{-\infty}^{\infty}%
\alpha_{e}\left(  -\tau\right)  \phi\left(  \tau\right)  \,d\tau,
\end{align}
since $\phi\in$ $\mathcal{D}\left(  \partial_{t}\right)  $ implies $\phi$ is
continuous (so $\phi\left(  0\right)  $ is unambiguous).

One may easily verify that%
\begin{multline}\label{secondderivS}
Se^{-\mathrm{i}t\Omega_{1}}S^{\dagger}=\frac{1}{2}\int_{-\infty}^{\infty
}e^{-\left\vert t+s\right\vert }a_{e}\left(  s\right)  \,ds
\\ =e^{-\left\vert t\right\vert
}\alpha_{\infty}+\frac{1}{2}\int_{-\infty}^{\infty}e^{-\left\vert
t+s\right\vert }\alpha_{e}\left(  s\right)  \,ds,
\end{multline}
and therefore%
\begin{equation}
\left(  -\frac{d^{2}}{dt^{2}}+1\right)  Se^{-\mathrm{i}t\Omega_{1}}S^{\dagger
}=2\alpha_{\infty}\delta\left(  t\right)  +\alpha_{e}\left(  t\right)
=a_{e}\left(  t\right)  .
\end{equation}
Thus, with $\Phi_{R}$ defined by (\ref{hg1c2}) and using (\ref{sg1}) and (\ref{secondderivS}) we get%
\begin{align}
\Gamma e^{-\mathrm{i}t\Omega_{1}}\left(  \Gamma\Phi_{R}^{2}\right)
^{\dagger}  &  =\left(  -\frac{d^{2}}{dt^{2}}+1\right)  \Gamma\frac{1}%
{\Omega_{1}^{2}+I_{H_{1}}}e^{-\mathrm{i}t\Omega_{1}}\left(  \Gamma\Phi_{R}%
^{2}\right)  ^{\dagger}\nonumber\\
&  =\left(  -\frac{d^{2}}{dt^{2}}+1\right)  Se^{-\mathrm{i}t\Omega_{1}}%
\Phi_{R}^{2}S^{\dagger}\overset{\mathcal{D}}{\rightarrow}a_{e}\left(
t\right)  \text{ as }R\rightarrow\infty\text{,}%
\end{align}
where $\overset{\mathcal{D}}{\rightarrow}$ denotes limit in the sense of
distributions.

The uniqueness up to isomorphism can be understood as follows. Let $\left\{
H_{1},\Omega_{1},\Gamma\right\}  $ and $\left\{  H_{1}^{\prime},\Omega
_{1}^{\prime},\Gamma^{\prime}\right\}  $ be distinct representations, and
suppose that%
\begin{equation}
\mathcal{S}=\left\langle \left(  \Gamma f\left(  \Omega_{1}\right)  \right)
^{\dagger}v:\ f\in C_{c}\left(  \mathbb{R}\right)  ,\text{ }v\in
H_{0}\right\rangle
\end{equation}
is dense in $H_{1}$. \ We denote by $\hat{f}$ the Fourier transform of $f\in
C_{c}\left(  \mathbb{R}\right)  $, so that%
\begin{equation}
f\left(  \Omega_{1}\right)  =\int_{-\infty}^{\infty}\hat{f}\left(  t\right)
e^{\mathrm{i}t\Omega_{1}}\,dt.
\end{equation}
Then, given $f,g\in C_{c}\left(  \mathbb{R}\right)  $ and $v,w\in H_{0}$, we
see that%
\begin{multline}
\left\langle \left(  \Gamma g\left(  \Omega_{1}\right)  \right)
^{\dagger }w,\left(  \Gamma f\left(  \Omega_{1}\right)  \right)
^{\dagger }v\right\rangle _{H_{1}}\\
\begin{aligned}
=&\lim_{R\rightarrow\infty}\left(  w,\Gamma g\left(
\Omega_{1}\right)  f\left(  \Omega_{1}\right)  ^{\dagger}\left(
\Gamma
\Phi_{R}^{2}\right)  ^{\dagger}v\right) \\
=&\lim_{R\rightarrow\infty}\int_{-\infty}^{\infty}\int_{-\infty}^{\infty
}\hat{g}\left(  t\right)  \hat{f}^{\ast}\left(  \tau\right)
\left(  w,\Gamma
e^{\mathrm{i}\left(  \tau-t\right)  \Omega_{1}}\left(  \Gamma\Phi_{R}%
^{2}\right)  ^{\dagger}v\right)  \,dtd\tau\\
=&\int_{-\infty}^{\infty}\int_{-\infty}^{\infty}\hat{g}\left(
t\right) \hat{f}^{\ast}\left(  \tau\right)  \left(  w,a_{e}\left(
t-\tau\right) v\right)  \,dtd\tau \\
=&\left\langle \left( \Gamma^{\prime}g\left(  \Omega
_{1}^{\prime}\right)  \right) ^{\dagger}w,\left(
\Gamma^{\prime}f\left( \Omega_{1}^{\prime}\right)  \right)
^{\dagger}v\right\rangle _{H_{1}^{\prime }}.
\end{aligned}
\end{multline}
Thus, defining%
\begin{equation}
T\left(  \Gamma f\left(  \Omega_{1}\right)  \right)  ^{\dagger}v=\left(
\Gamma^{\prime}f\left(  \Omega_{1}^{\prime}\right)  \right)  ^{\dagger}v
\end{equation}
and extending $T$ to $\mathcal{S}$ by linearity, we produce a well defined
isometry $T:\mathcal{S}\hookrightarrow H_{1}^{\prime}$. \ The closure of this
map, also denoted $T$, is an isometric imbedding $T:H_{1}\hookrightarrow
H_{1}^{\prime}$. \ It is now easy to verify that
\begin{equation}
f\left(  \Omega_{1}\right)  =T^{\dagger}f\left(  \Omega_{1}^{\prime}\right)
T,\text{ for }f\in C_{c}\left(  \mathbb{R}\right)  ;\Gamma^{\prime}=\Gamma
T^{\dagger},
\end{equation}
and thus the representation $\left\{  H_{1},\Omega_{1},\Gamma\right\}  $ is
isomorphic to the restriction of $\left\{  H_{1}^{\prime},\Omega_{1}^{\prime
},\Gamma^{\prime}\right\}  $ to the closure of%
\begin{equation}
\left\langle \left(  \Gamma^{\prime}f\left(  \Omega_{1}^{\prime}\right)
\right)  ^{\dagger}v:\ f\in C_{c}\left(  \mathbb{R}\right)  ,\text{ }v\in
H_{0}\right\rangle .
\end{equation}
\end{proof}

\subsection{Herglotz-Nevanlinna Theorems}

In Section 2, we have already presented a more or less complete proof of
Theorem \ref{tOperatorNev2} based on the classical Nevanlinna Theorem
\ref{tNev2} and the Naimark representation \ref{tNaim1} of generalized spectral
families. Since Theorem \ref{tGenOperatorNev2} is proved in a very similar
fashion, we present only a somewhat streamlined proof here.

\begin{proof}
[Proof of Theorem \ref{tGenOperatorNev2}]By the
Herglotz-Nevanlinna theorem, for each $v\in H_{0}$ there is a
finite measure $d\tilde{N}_{v,v}$ and a real
number $\xi_{v,v}$ such that%
\begin{equation}
\left(  v,G\left(  \zeta\right)  v\right)  =\xi_{v,v}+\int_{-\infty}^{\infty
}\frac{1+\sigma\zeta}{\sigma-\zeta}d\tilde{N}_{v,v}\left(  \sigma\right)  .
\end{equation}
The $\theta$ term drops out of the representation because $\zeta^{-1}G\left(
\zeta\right)  \rightarrow0$ as $\zeta\rightarrow\infty$. \ In fact, $\xi
_{v,v}=\operatorname{Re}\left(  v,G\left(  \mathrm{i}\right)  v\right)  $. \ As
in the derivation of Theorem \ref{tOperatorNev2}, we define \textquotedblleft
off-diagonal\textquotedblright\ measures $d\tilde{N}_{v,w}$ for each pair
$v,w\in H_{0}$ by polarization so that
\begin{equation}
\left(  v,G\left(  \zeta\right)  w\right)  =\left(  v,\operatorname{Re}%
G\left(  \mathrm{i}\right)  w\right)  +\int_{-\infty}^{\infty}\frac{1+\sigma
\zeta}{\sigma-\zeta}d\tilde{N}_{v,w}\left(  \sigma\right)  .
\end{equation}
\ \qquad

As above there is a \textquotedblleft generalized spectral
family\textquotedblright\ $K\left(  \sigma\right)  $ satisfying the hypothesis
of the Naimark Theorem \ref{tNaim1} such that%
\begin{equation}
\left(  v,K\left(  \sigma\right)  w\right)  =\int_{\left(  -\infty
,\sigma\right]  }d\tilde{N}_{v,w}\left(  \sigma\right)  .
\end{equation}
We denote by $H_{1}$ and $E\left(  \sigma\right)  $ the Hilbert space and
resolution of the identity guaranteed by the Naimark Theorem, letting $T$
denote the associated mapping $T:$ $H_{1}\rightarrow H_{0}$. Thus, we have%
\begin{equation}
G\left(  \zeta\right)  =\operatorname{Re}G\left(  \mathrm{i}\right)
+\int_{-\infty}^{\infty}\frac{1+\sigma\zeta}{\sigma-\zeta}TdE\left(
\sigma\right)  T^{\dagger},
\end{equation}
or%
\begin{equation}
G\left(  \zeta\right)  =\operatorname{Re}G\left(  \mathrm{i}\right)
+T\frac{I_{H_{1}}+\Omega_{1}\zeta}{\Omega_{1}-\zeta I_{H_{1}}}T^{\dagger
},\ \Omega_{1}=\int_{-\infty}^{\infty}\sigma dE\left(  \sigma\right)  .
\end{equation}
From this we may easily compute the following formula%
\begin{equation}
G\left(  \zeta\right)  =\operatorname{Re}G\left(  \mathrm{i}R\right)
+T\frac{R^{2}I_{H_{1}}+\Omega_{1}\zeta}{\Omega_{1}-\zeta I_{H_{1}}%
}\frac{I_{H_{1}}+\Omega_{1}^{2}}{R^{2}I_{H_{1}}+\Omega_{1}^{2}}T^{\dagger}.
\end{equation}

\ Therefore, given $v\in H_{0}$,%
\begin{equation}
G\left(  \zeta\right)  v=\lim_{R\rightarrow\infty}\Gamma\frac{1}{\Omega
_{1}-\zeta I_{H_{1}}}\left(  \Gamma\frac{R^{2}}{R^{2}I_{H_{1}}+\Omega_{1}^{2}%
}\right)  ^{\dagger},\ \Gamma=T\sqrt{I_{H_{1}}+\Omega_{1}^{2}},
\end{equation}
because we have
\begin{equation}
\lim_{R\rightarrow\infty}\left\{  \operatorname{Re}G\left(  \mathrm{i}%
R\right)  \right\}  v=0
\end{equation}
by assumption, and
\begin{equation}
\lim_{R\rightarrow\infty}T\frac{\Omega_{1}\zeta}{\Omega_{1}-\zeta I_{H_{1}}%
}\frac{I_{H_{1}}+\Omega_{1}^{2}}{R^{2}I_{H_{1}}+\Omega_{1}^{2}}T^{\dagger}v=0
\end{equation}
for each $\zeta$ in the upper half plane, since $\frac{I_{H_{1}}+\Omega
_{1}^{2}}{R^{2}I_{H_{1}}+\Omega_{1}^{2}}\rightarrow0$ strongly.
\end{proof}

\appendix

\section{Appendix: Stieltjes Inversion Formula and Naimark's Theorem}

In Section 3 -- and also in the examples of Section 4 -- we
discussed the construction of conservative extensions. Reference
was made there to two classical constructions -- the Stieltjes
Inversion formula and Naimark's construction for generalized
spectral measures -- which quite generally provide an explicit
description of the Hilbert space $H_{1}$ in the operator versions
of the Herglotz-Nevanlinna theorems. For completeness we include a
discussion of those results here.

\subsection{Stieltjes Inversion Formula}

The Nevanlinna Theorem \ref{tNev2} -- in particular the relation
(\ref{hg5}) -- suggests the introduction of the so-called
\emph{Cauchy transform} defined for complex-valued measures of
finite variation on $\mathbb{R}$:
\begin{equation}
\tilde{N}\left(  \zeta\right)  =\int_{-\infty}^{\infty}\frac{dN\left(
\sigma\right)  }{\sigma-\zeta},\ \operatorname{Im}\zeta \neq 0. \label{hg6}%
\end{equation}
The Nevanlinna theorem states that the set of functions which are
Cauchy transforms of \emph{non-negative }finite measures is
exactly the class of analytic maps of the upper half plane into
itself which decay as
$\mathcal{O}\left(  1/\operatorname{Im}\zeta\right)  $ as $\operatorname{Im}%
\zeta\rightarrow\infty$. \ There does not seem to be such a simple description
of the set of Cauchy transforms of complex measures. \ Nonetheless, a complex
measure is uniquely determined by its Cauchy transform, \cite[(Section 32.1,
Lemma 4)]{Lax}, \cite{KK}.

\begin{proposition}
\label{pCauchy}The Cauchy transform (\ref{hg6}) is one-to-one, i.e. a complex
measure of finite variation is uniquely determined by its Cauchy
transformation. Furthermore, if $dN\left(  \sigma\right)  $ is a real (signed)
measure of finite variation it can be recovered from its Cauchy transform
$\tilde{N}\left( \zeta\right)  $ restricted to
$\{ \mathrm{Im} \zeta > 0\}$ by Stieltjes' formula :%
\begin{gather}
\frac{N\left(  \sigma_{1}+0\right)  +N\left(  \sigma_{1}-0\right)  }%
{2}-\frac{N\left(  \sigma_{0}+0\right)  +N\left(  \sigma_{0}-0\right)  }%
{2}=\label{hg7}\\
=\lim_{\eta\rightarrow+0}\frac{1}{\pi}\int_{\sigma_{0}}^{\sigma_{1}%
}\operatorname{Im}\tilde{N}\left(  \sigma+\mathrm{i}\eta\right)
\,d\sigma.\,\nonumber
\end{gather}
If $n\left(  \sigma\right)  d\sigma$ is the absolutely continuous component of
the measure $dN\left(  \sigma\right)  $, in particular if $dN\left(
\sigma\right)  =n\left(  \sigma\right)  d\sigma$, we also have%
\begin{equation}
n\left(  \sigma\right)  =\lim_{\eta\rightarrow+0}\frac{1}{\pi}%
\operatorname{Im}\tilde{N}\left(  \sigma+\mathrm{i}\eta\right)  \text{ for
Lebesgue almost every }\sigma. \label{hg8}%
\end{equation}
\end{proposition}

\begin{remark}
Another manifestation of (\ref{hg7}), is the weak convergence
\begin{equation}
\operatorname*{wklim}_{\eta\rightarrow0}\frac{1}{\pi}\operatorname{Im}%
\tilde{N}\left(  \sigma+\mathrm{i}\eta\right)  \,d\sigma=dN\left(
\sigma\right)  ,
\end{equation}
that is,%
\begin{multline}
\int_{-\infty}^{\infty}f\left(  \sigma\right)  dN\left(
\sigma\right)
\\ =\lim_{\eta\rightarrow0}\frac{1}{\pi}\int_{-\infty}^{\infty}f\left(
\sigma\right)  \operatorname{Im}\tilde{N}\left(
\sigma+\mathrm{i}\eta\right) \,d\sigma\text{ for all }f\in
C_{c}\left(  \mathbb{R}\right)  .
\end{multline}
\end{remark}

Thus a scalar Herglotz function $g\left(  \zeta\right)  $ which is
$\mathcal{O}\left(  1/\operatorname{Im}\zeta\right)  $ as $\operatorname{Im}%
\zeta\rightarrow\infty$ may be represented by the formula%
\begin{equation}
g\left(  \zeta\right)  =\lim_{\eta\rightarrow0}\frac{1}{\pi}\int_{-\infty
}^{\infty}\frac{1}{\sigma-\zeta}\operatorname{Im}g\left(  \sigma
+\mathrm{i}\eta\right)  \,d\sigma.
\end{equation}
In particular, if $\operatorname{Im}g\left(  \sigma+\mathrm{i}0\right)
=\lim_{\eta\rightarrow0}\operatorname{Im}g\left(  \sigma+\mathrm{i}%
\eta\right)  $ exists for almost every $\sigma$ and%
\begin{equation}
\lim_{\eta\rightarrow0}\int_{-\infty}^{\infty}\left\vert \operatorname{Im}%
g\left(  \sigma+\mathrm{i}\eta\right)  -\operatorname{Im}g\left(
\sigma+\mathrm{i}0\right)  \right\vert \,d\sigma=0,
\end{equation}
then%
\begin{equation}
g\left(  \zeta\right)  =\frac{1}{\pi}\int_{-\infty}^{\infty}\frac{1}%
{\sigma-\zeta}\operatorname{Im}g\left(  \sigma+\mathrm{i}0\right)  \,d\sigma.
\end{equation}
In general, however there may be a singular component to the measure $dN$ in
the representation (\ref{hg5}).

For an operator valued Herglotz function $G\left(  \zeta\right)  $, with
$\left\Vert G\left(  \zeta\right)  \right\Vert =\mathcal{O}\left(
1/\operatorname{Im}\zeta\right)  $, we have therefore%
\begin{equation}
G\left(  \zeta\right)  =\lim_{\eta\rightarrow0}\frac{1}{\pi}\int_{-\infty
}^{\infty}\frac{1}{\sigma-\zeta}\operatorname{Im}G\left(  \sigma
+\mathrm{i}\eta\right)  \,d\sigma,
\end{equation}
with the integral understood in the weak sense, i.e.%
\begin{multline}
\left(  v,G\left(  \zeta\right)  w\right)  \\ =\lim_{\eta\rightarrow0}%
\frac{1}{\pi}\int_{-\infty}^{\infty}\frac{1}{\sigma-\zeta}\left(
v,\operatorname{Im}G\left(  \sigma+\mathrm{i}\eta\right)  w\right)
\,d\sigma,\text{ for }v,w\in H_{0}.
\end{multline}
Thus the generalized spectral family $K\left(  \sigma\right)  $ associated to
$G\left(  \zeta\right)  $ can be expressed through the formula%
\begin{equation}
d\left(  v,K\left(  \sigma\right)  w\right)  =\operatorname*{wklim}%
_{\eta\rightarrow0}\frac{1}{\pi}\left(  v,\operatorname{Im}G\left(
\sigma+\mathrm{i}\eta\right)  w\right)  d\sigma. \label{dMwklim}%
\end{equation}

\subsection{Naimark's Theorem}

The Naimark construction for $K\left(  \sigma\right)  $, which leads to Theorem
\ref{tNaim1}, is most easily understood by realizing $H_{1}$ as the Hilbert
space $L^{2}\left(  dK\right)  $, where the latter space needs to be
appropriately defined. \ Formally $L^{2}\left(  dK\right)  $ should consist of
all $H_{0}$ valued functions $\Psi$ such that $\int_{-\infty}^{\infty}\left(
\Psi\left(  \sigma\right)  ,dK\left(  \sigma\right)  \Psi\left( \sigma\right)
\right)  $ is finite, modulo null functions (for which the integral is $0$). \
It is not always clear how to make sense of this integral however, and one must
turn to a more abstract definition of $L^{2}\left(  K\right)  $. However, if
$dK\left(  \sigma\right)  =m\left(  \sigma\right)  d\sigma$, with $m\left(
\sigma\right)  $ bounded for almost every $\sigma$, we can avoid the abstract
construction by defining $H_{1}=L^{2}\left(  dK\right)  $ to be the
space of $H_{0}$ valued functions $\Psi$ such that%
\begin{equation}
\left\|  \Psi\right\|  _{K}^{2}=\int_{-\infty}^{\infty}\left(  \Psi\left(
\sigma\right)  ,m\left(  \sigma\right)  \Psi\left(  \sigma\right)  \right)
\,d\sigma<\infty, \label{L2M}%
\end{equation}
modulo null functions for which $\left\|  \Psi\right\|  _{K}^{2}=0$. \ Note
that $m\left(  \sigma\right)  $ is necessarily a positive operator since%
\begin{equation}
\left(  v,m\left(  \sigma\right)  v\right)
=\lim_{\varepsilon\rightarrow
0}\frac{1}{2\varepsilon}\int_{\sigma-\varepsilon}^{\sigma+\varepsilon}d\left(
v,K\left(  \sigma\right)  v\right)  . \end{equation}

For more general measures $dK$ there are two options. We could
define the norm appearing in (\ref{L2M}) for $H_{0}$ valued
\emph{simple functions} -- functions taking a finite number of
values -- and let $H_{1}$ be the closure of the space of simple
functions under this norm. Alternatively, we can express $dK\left(
\sigma\right)  $ as the weak limit (\ref{dMwklim}) and define
$H_{1}=L^{2}\left(  dK\right)  $ as the space of functions $\Psi$
such
that%
\begin{equation}
\left\|  \Psi\right\|  _{K}^{2}=\lim_{\eta\rightarrow0}\frac{1}{\pi}%
\int_{-\infty}^{\infty}\left(  \Psi\left(  \sigma\right)  ,\operatorname{Im}%
G\left(  \sigma+\mathrm{i}\eta\right)  \Psi\left(  \sigma\right)  \right)
\,d\sigma
\end{equation}
exists and is finite, modulo null functions (as always.)

However it may be defined, once the space $H_{1}=L^{2}\left(  dK\right)  $ has
been constructed, there is a natural spectral measure\ $E\left(
\sigma\right)  $ given by
\begin{equation}
\left[  E\left(  \sigma\right)  \Psi\right]  \left(  \nu\right)  =\left\{
\begin{array}
[c]{cc}%
\Psi\left(  \nu\right)  & \text{if }\nu\leq\sigma\\
0\text{ } & \text{if }\nu>\sigma
\end{array}
\right.  ,
\end{equation}
that is $E\left(  \sigma\right)  $ corresponds to multiplication by the
characteristic function of $\left(  -\infty,\sigma\right]  $. \ The associated
self adjoint operator $\Omega_{1}=\int_{-\infty}^{\infty}\sigma dE\left(
\sigma\right)  $ is simply multiplication by the independent variable:%
\begin{equation}
\Omega_{1}\Psi\left(  \sigma\right)  =\sigma\Psi\left(  \sigma\right)  .
\end{equation}
Finally there is a natural map $\Gamma^{\dagger}:H_{0}\rightarrow H_{1}$ which
takes an element $\psi\in H_{0}$ to the constant function with value $\psi$:%
\begin{equation}
\left[  \Gamma^{\dagger}\psi\right]  \left(  \sigma\right)  =\psi\text{ for
every }\sigma\in\mathbb{R}.
\end{equation}
It is easy to verify that
\begin{equation}
K\left(  \sigma\right)  =\Gamma E\left(  \sigma\right)  \Gamma^{\dagger}.
\end{equation}


\textbf{Acknowledgment and Disclaimer:} The effort of A. Figotin
was sponsored by the Air Force Office of Scientific Research, Air
Force Materials Command, USAF, under grant number
F49620-01-1-0567. J. H. Schenker was supported in part by a
National Science Foundation post-doctoral fellowship and received
travel support under the aforementioned USAF grant. The US
Government is authorized to reproduce and distribute reprints for
governmental purposes notwithstanding any copyright notation
thereon. The views and conclusions contained herein are those of
the authors and should not be interpreted as necessarily
representing the official policies or endorsements, either
expressed or implied, of the Air Force Office of Scientific
Research or the US Government.


\begin{thebibliography}{9}                                                                                                %

\bibitem {AkhGlaz}Akhiezer, N. I. and Glazman, I. M., \textsl{Theory of Linear
Operators in Hilbert Space}, Dover, New York, 1993.

\bibitem {Born}Born, M. and Wolf, E., \textsl{Principles of Optics}, Pergamon
Press, 1993.

\bibitem {Bohren}Bohren, C. and Huffman, D. \textsl{Absorbtion and Scattering
of Light by Small Particles}, John Wiley \& Sons, 1983.

\bibitem {GeTs}Gesztesy, F. and Tsekanovskii, E., \textsl{On matrix-valued
Herglotz functions}, Math. Nachr. 218 (2000), 61--138.

\bibitem {GoKr}Gohberg, I. and Krein, M., \textsl{Introduction to the Theory
of Linear Nonselfadjoint Operators}, AMS, 1969.

\bibitem {Kato}Kato T., \textsl{Perturbation Theory for Linear Operators},
Springer, 1980.

\bibitem {KK}Kac, I. and Krein, M., \textsl{\textquotedblleft R-Functions -
Analytic Functions Mapping the Upper Half Plane into Itself\textquotedblright,
\textquotedblleft On the Spectral Functions of the String\textquotedblright, in
American Mathematical Translations}, Series 2, Vol. 103, 1974.

\bibitem {KuboToda2}Kubo, R., Toda, M. and Hashitsume, N., \textsl{Statistical
Physics II, Nonequilibrium Stastical Mechanics}, Second Edition,
Springer-Verlag, 1991.

\bibitem {Lamb}Lamb, H., \textsl{On a Pecularity of the Wave-System due the
Free Vibrations on a Nucleus in an Extented Medium}, Proc. of Lond. Math. Soc.
Vol. XXXII, No. 723, p. 208-211, 1900.

\bibitem {Lax}Lax, P., \textsl{Functional Analysis}, Wiley-Interscience, 2002.

\bibitem {Nai}Naimark, M. A. \textsl{On a representation of additive operator
set functions}, C. R. (Doklady) Acad. Sci. URSS (N.S.) 41 (1943), 359--361.

\bibitem {Pav}Pavlov, B.\ S. \textsl{Spectral analysis of a singular
Schr\"{o}dinger operator in terms of a functional model}. In \emph{Partial
differential equations, VIII}, pages 87-153, Springer-Verlag, Berlin, 1992.

\bibitem {RiNa}Riesz, F. and Sz.-Nagy, B., \textsl{Functional Analysis},
Dover, 1990.

\bibitem {RS1}Reed, M. and Simon, B., \textsl{Functional Analysis}, Vol. I,
Academic Press, 1980.

\bibitem {RoRo}Rosenblum, M. and Rovnyak J.,
\textsl{Hardy Classes and Operator Theory}, Oxford University Press, 1985.


\bibitem {St}Stein, E. M., \textsl{Singular integrals and differentiability
properties of functions}, Princeton University Press, 1970.

\bibitem {Scaife}Scaife, B., \textsl{Principles of Dielectrics}, Oxford Press, 1998.

\bibitem {Tip} Tip, A., \textsl{Linear absorptive dielectrics}, Phys.\ Rev.\ A,
57 (1998), 4818-4841.

\bibitem {Yosida}Yosida, K., \textsl{Functional Analysis}, Springer, 1995.
\end{thebibliography}
\end{document}